\documentclass[a4paper,12pt]{article}
\usepackage[utf8]{inputenc}
\usepackage{enumitem}
\usepackage[margin=1in]{geometry}
\usepackage{abstract}
\usepackage{setspace}
\usepackage{indentfirst}
\usepackage{amsmath}
\usepackage{amssymb}
\usepackage{mathtools}
\usepackage{graphicx}
\usepackage{csquotes}
\usepackage{wasysym}
\usepackage{physics}
\usepackage{dsfont}
\usepackage{hyperref}
\usepackage{tikz}
\usepackage{url}
\hypersetup{
    colorlinks=true,
    linkcolor=black,
    citecolor=blue,
    urlcolor=blue,
}
\usepackage[backend=bibtex,style=ieee]{biblatex} 
\usepackage[export]{adjustbox}
\usepackage{caption}
\usepackage{subcaption}
\newtheorem{theorem}{Theorem}[section]

\usepackage[font=small,labelfont=bf]{caption}
\usepackage{multirow}
\usepackage{multicol}
\usepackage{array}
\usepackage{rotating}
\usepackage{tabularx}
\usepackage[labelfont=bf]{caption}
\usepackage{ragged2e}
\usepackage[toc,page]{appendix}

\title{Mahler Measuring the Genetic Code of Amoebae}
\author{Siqi Chen\textsuperscript{1,2}\footnote{chensiqi1024@outlook.com (corresponding)}, Yang-Hui He\textsuperscript{2,3,4,5}\footnote{hey@maths.ox.ac.uk}, Edward Hirst\textsuperscript{3,2}\footnote{edward.hirst@city.ac.uk}, \\ Andrew Nestor\textsuperscript{6,2}\footnote{andrewnestor123@gmail.com}, Ali Zahabi\textsuperscript{7,2}\footnote{zahabi.ali@gmail.com}}
\date{\today\\ 
\textsuperscript{1}\textit{\small Balliol College, University of Oxford, OX1 3BJ, UK}\\
\textsuperscript{2}\textit{\small London Institute for Mathematical Sciences, Royal Institution, London W1S 4BS, UK}\\
\textsuperscript{3}\textit{\small Department of Mathematics, City, University of London, EC1V 0HB, UK}\\
\textsuperscript{4}\textit{\small Merton College, University of Oxford, OX1 4JD, UK}\\
\textsuperscript{5}\textit{\small School of Physics, NanKai University, Tianjin, 300071, P.R. China}\\
\textsuperscript{6}\textit{\small Hertford College, University of Oxford, OX1 3BW, UK}\\
\textsuperscript{7}\textit{\small Institut de Math\'ematiques de Bourgogne, Universit\'e Bourgogne Franche-Comt\'e, France}}

\begin{document}

\maketitle

\begin{abstract}
Amoebae from tropical geometry and the Mahler measure from number theory play important roles in quiver gauge theories and dimer models. Their dependencies on the coefficients of the Newton polynomial closely resemble each other, and they are connected via the Ronkin function. Genetic symbolic regression methods are employed to extract the numerical relationships between the 2d and 3d amoebae components and the Mahler measure. We find that the volume of the bounded complement of a d-dimensional amoeba is related to the gas phase contribution to the Mahler measure by a degree-d polynomial, with d = 2 and 3. These methods are then further extended to numerical analyses of the non-reflexive Mahler measure. Furthermore, machine learning methods are used to directly learn the topology of 3d amoebae, with strong performance. 
Additionally, analytic expressions for boundaries of certain amoebae are given.
\end{abstract}

\small
\textit{Report Number}: LIMS-2022-024
\normalsize

\newpage
\tableofcontents

\section{Introduction}

The amoebae of affine algebraic varieties are interesting objects at the intersection of tropical geometry \cite{purbhoo2006nullstellensatz,maclagan2015introduction,bogaardintroduction,Zahabi:2020hwu} in mathematics and dimer models in physics \cite{Kenyon:2003,Hanany:2005ve,Franco:2005rj,Feng:2005gw}. Amoebae are constructed out of logarithmic projections of complex varieties described by toric diagrams. These toric diagrams are lattice polytopes whose dimensions can be associated to complex coordinates and vertices to monomial terms in the varieties defining equation, via the Newton polynomial \cite{Franco:2005sm,Butti:2005vn,Bao:2020sqg}. How the topology and geometry of the surface changes under the amoeba projection makes them particularly interesting objects of study.

The Mahler measure, which was first introduced by Kurt Mahler in 1962 \cite{Mahler1962OnSI}, can be interpreted as the limit height function and the free energy in these dimer models \cite{Zahabi:2018znq,Bao:2021gxf}. Further to this in crystal melting models \cite{Okounkov:2003sp,Ooguri:2008yb,Yamazaki:2011wy,Bao:2022oyn}, the Mahler measure and the amoeba are closely related by the Ronkin function which is the limit shape of the molten crystal; with relation to quiver Yangians \cite{Bao:2022fpk,Bao:2022jhy}. In particular, the Mahler measure is the Ronkin function at the origin and the gradient of the Ronkin function is constant over each components of the amoeba complement. In \cite{Bao:2021fjd}, a number of observations were made which explicated how the dynamical aspects of the gauge theory are encoded in the Mahler measure by defining a concept called the Mahler flow. So far, the appearance of the Mahler measure in physics has only been studied in the context of 2-dimensional reflexive polygons, which have only a single free parameter in the Mahler flow. Thus, there is lots of room to dive deeper into the properties and relations related to the Mahler measure in broader context, such as in the case of non-reflexive polytopes.

In recent years, there has been an increasing effort in applying data science techniques to mathematical sciences based on the observation that mathematical data often take the form of labelled or unlabelled tensors that naturally resemble the data structure required in machine learning (ML). In mathematical physics, this was initiated in the exploration of string landscape \cite{He:2017aed,Krefl:2017yox,Carifio:2017bov,Ruehle:2017mzq} and extended to a broader range of topics in \cite{Gal:2020dyc,Larfors:2021pbb,Krippendorf:2021lee,Gao:2021xbs,He:2020eva,Chen:2020dxg,Chen:2020jjw,Bao:2020nbi,Bao:2021auj,Bao:2021ofk,Berman:2021mcw,Arias-Tamargo:2022qgb,He:2022fxp,Berglund:2022gvm}. Interested readers can refer to \cite{He:2018jtw,Bao:2022rup,He:2022cpz} for comprehensive reviews on this application.

Specifically, \cite{Bao:2021olg} integrated these two directions by applying ML techniques to study the amoebae and tropical geometry, taking advantage of the classification and image-processing power of ML. In this paper, we extend the discussion in \cite{Bao:2021olg} to 3-dimensional amoebae, consider the Mahler flow proposed in \cite{Bao:2021fjd} in greater details in the context of non-reflexive polytopes, and then apply standard ML techniques to make precise the qualitative relation between the Mahler measure and the bounded amoeba complements observed and conjectured in \cite{Bao:2021fjd}, as well as considering the implications in theories built from non-reflexive polytopes as discussed in \cite{Bao:2020kji}. Since computing exact properties of the amoeba has been a challenge with analytic results mostly concerning its approximations and special limits, for example in \cite{purbhoo2006nullstellensatz,KenyonHarnack}, our numerical results obtained from ML could provide insights in its understanding in more general scenarios. 

The paper is organised as follows. Section §\ref{sec:2} reviews some preliminaries about amoebae and the Mahler measure and their relations in dimer models which motivate this paper. The following Sections §\ref{sec:3} and §\ref{sec:4} discuss some interesting physical properties related to amoebae and the Mahler measure respectively. More specifically, in Section §\ref{sec:3}, we apply feed-forward neural networks and convolutional neural networks to ML the second Betti number of the 3-dimensional amoebae associated with reflexive polytopes, based on the discussion of 2-dimensional amoebae in \cite{Bao:2021olg}. Section §\ref{sec:4} extends the discussion of the Mahler flow of reflexive polytopes in \cite{Bao:2021fjd} to the case of non-reflexive polytopes, where there are more than one amoeba holes, leading to interesting dynamics. In Section §\ref{sec:5}, we consider the more physically relevant quantities, namely the relations between the Mahler measure and the area (volume) of the bounded complementary region of the amoebae, implementing a genetic algorithm for symbolic regression. In doing so, we also obtain analytic expressions for the boundary of some amoebae. Finally, Section §\ref{sec:6} discusses the main results in this paper and possible future directions.

\section{Preliminaries}\label{sec:2}

\subsection{The Amoeba}

The amoeba, \(\mathcal{A}_V\subset \mathbb{R}^r\), of an algebraic hypersurface, \(V_{\mathbb{C}}(f) \subset \mathbb{C}^r\), is defined as the image of the logarithmic map,
\begin{equation}
    \mathcal{A}_{V} \equiv \mathrm{Log}(V_{\mathbb{C}}(f)),
\end{equation}
where
\begin{equation}
    \mathrm{Log}(z_1,\ldots,z_n)\equiv(\mathrm{log} |z_1|,\ldots,\mathrm{log} |z_n|).
\end{equation}
Since the hypersurface \(V_{\mathbb{C}}(f) = \{z\in \mathbb{C}^n: f(z)=0\}\) is the zero locus of the function \(f\), the corresponding amoeba may also be denoted as \(\mathcal{A}_f\).

The function of interest is the Newton polynomial defined with respect to a Newton polytope which is a convex lattice polytope, also known as a toric diagram. In the case of $n$ complex dimensions, the Newton polynomial is of the form \(P(\textbf{z})=\sum_{\mathbf{p}} c_{\mathbf{p}}\textbf{z}^{\mathbf{p}}\), summing over the polytope vertices $\textbf{p}$ each with coordinates $p_i$ in the $i$-th lattice dimension. In particular, in three complex dimensional \((r=3)\) cases, Newton polynomial is of the form \(P(u,v,w)=\sum_{\mathbf{p}} c_{(p_1,p_2,p_3)}u^{p_1}v^{p_2}w^{p_3}\); and now denoting $(z_1,z_2,z_3) \mapsto (u,v,w)$ to emphasise the restriction to $r=3$.

An amoeba can be approximated using \textbf{lopsidedness} which is defined as follows. A list of positive numbers \(\{c_1,\ldots,c_n\}\) is \textbf{lopsided} if one of the numbers is greater than the sum of the rest of numbers. If \(\{c_1,\ldots,c_n\}\) is not lopsided, there exists a list of phases \(\{\phi_i\}\) such that \(\sum_i \phi_i c_n =0\), via the triangle inequality \cite{bogaardintroduction}. Thus, the lopsided amoeba, \(\mathcal{LA}_f\), is defined by
\begin{equation}
    \mathcal{LA}_f \equiv \{\textbf{a}\in \mathbb{R}^r|\hspace{2pt}f\{\textbf{a}\}\hspace{2pt} \mathrm{is}\hspace{2pt} \mathrm{not}\hspace{2pt} \mathrm{lopsided}\},
\end{equation}
so that \(\mathcal{LA}_f\supseteq \mathcal{A}_f\).

Let \(n\) be a positive integer, \(\textbf{x}\in \mathbb{R}^r\), and \(f(\textbf{x})\) a (Newton) polynomial, define \(\Tilde{f}_n\) to be 
\begin{equation}
    \Tilde{f}_n(\textbf{x}):= \prod_{k_1=0}^{n-1}\cdots\prod_{k_r=0}^{n-1}f(e^{2\pi i k_1/n}x_1,\dots,e^{2\pi i k_r/n}x_r),
\end{equation}
which is a cyclic resultant defined as
\begin{equation}
    \Tilde{f}_n=\mathrm{res}_{u_r}(\mathrm{res}_{u_{r-1}}(\dots\mathrm{res}_{u_1}(f(u_1x_1,\dots,u_rx_r),u^n_1-1)\dots,u^n_{r-1}-1),u_r^n-1)
\end{equation}
where \(\mathrm{res}_{u}(f,g)\) is the resultant of \(f,g\) with respect to the variable \(u\).

\begin{theorem}
The lopsided amoeba \(\mathcal{LA}_{\Tilde{f}_n}\) converges uniformly to \(\mathcal{A}_f\) as \(n \rightarrow \infty\), where \(\Tilde{f}_n\) is the cyclic resultant of \(f\) \,. 
\end{theorem}
The Newton polytope of \(\Tilde{f}_n\) is \(n^r\Delta(f)\), as a dilation of the original polytope \cite{purbhoo2006nullstellensatz}.

\begin{figure}[!ht]
  \centering
  \includegraphics[width=.3\linewidth]{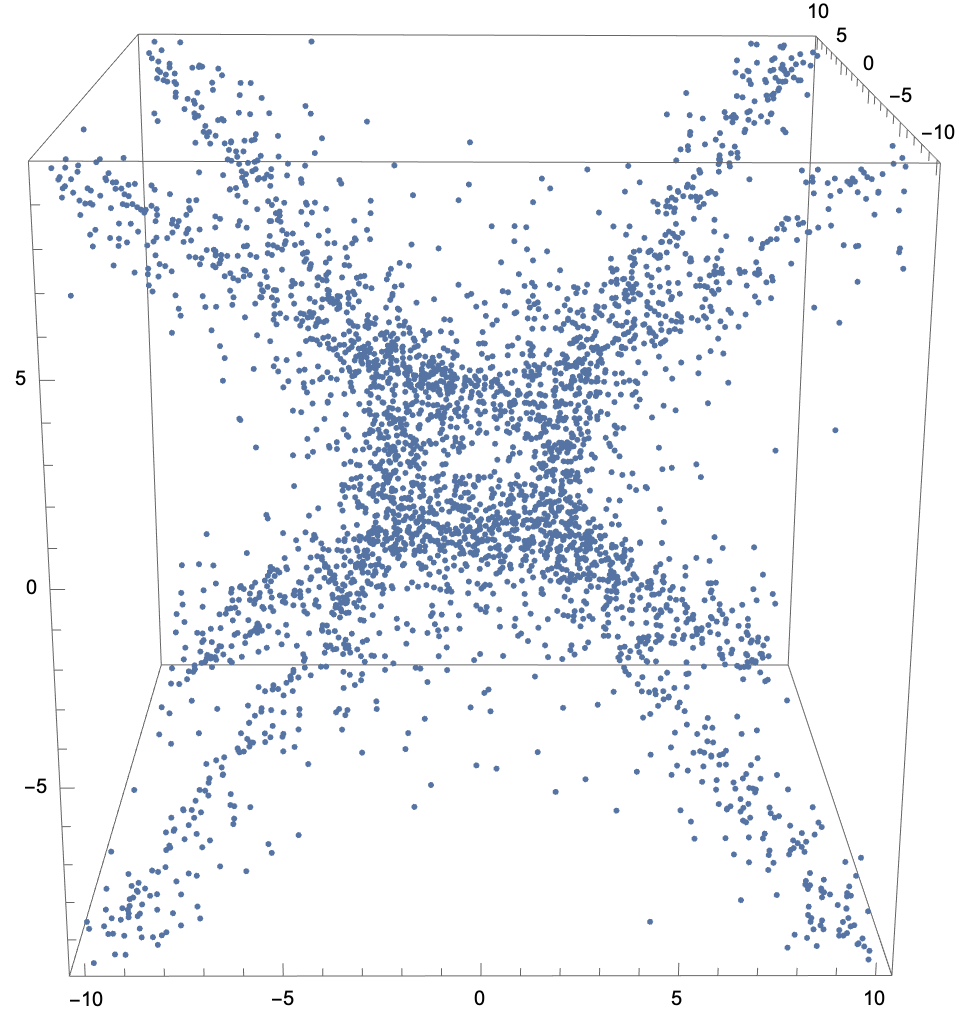}
  \caption{Amoeba of the hypersurface \(P(z_1,z_2,z_3)=z_1+z_1^{-1}+z_2+z_2^{-1}+z_3+z_3^{-1}+10=0\)\,.}
  \label{ap1}
\end{figure}

An example of a 3-dimensional amoeba is given in Figure \ref{ap1}, generated through Monte Carlo sampling of points on the Riemann surface.

\begin{figure}[!ht]
   \begin{minipage}{0.48\textwidth}
     \centering
     \includegraphics[width=.5\linewidth]{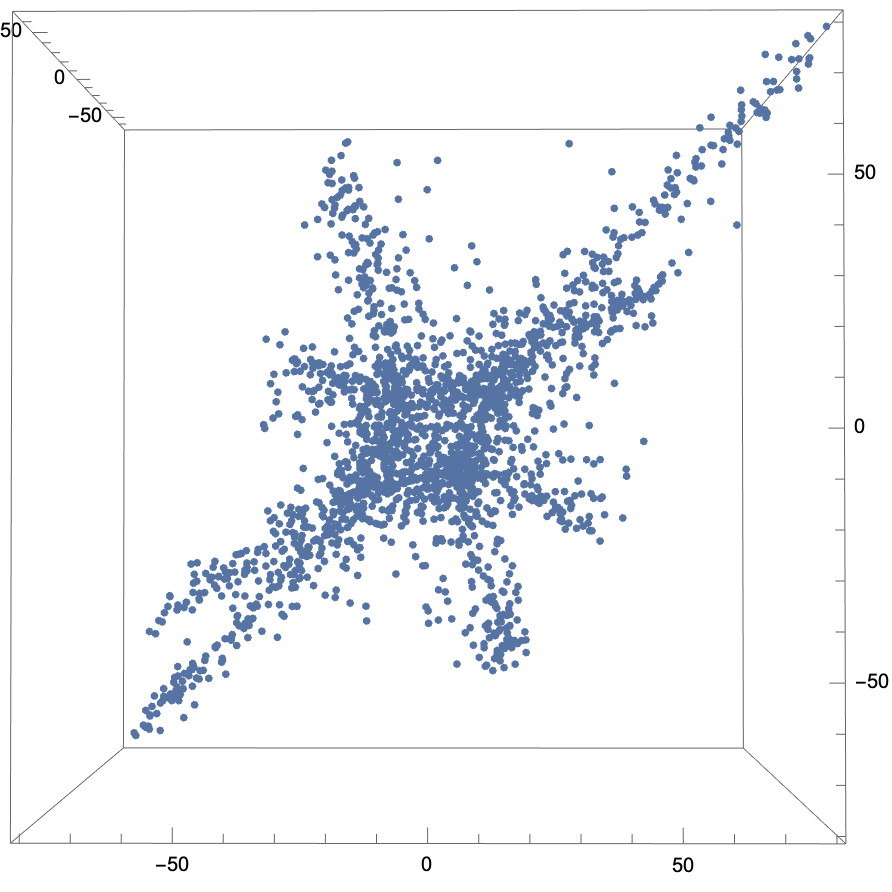}
   \end{minipage}\hfill
   \begin{minipage}{0.48\textwidth}
     \centering
     \includegraphics[width=.5\linewidth]{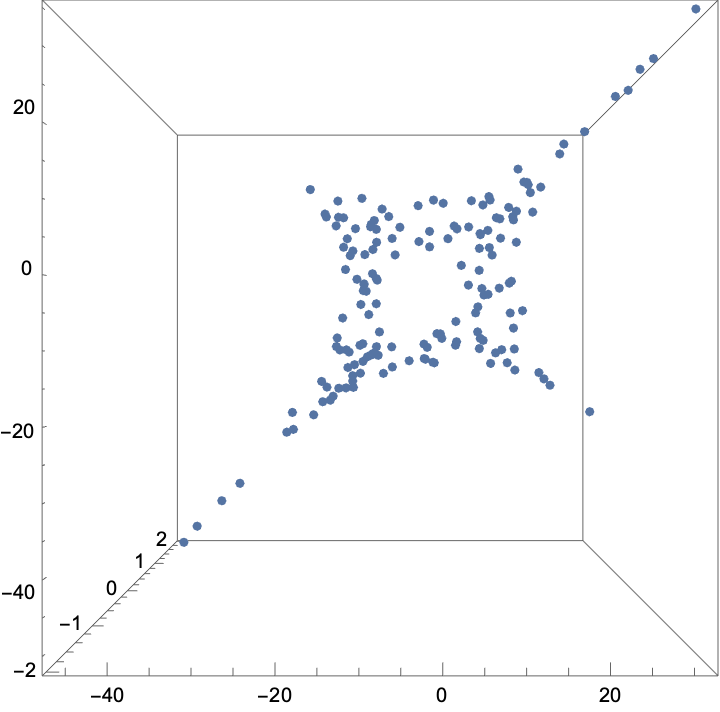}
   \end{minipage}
   \caption{The amoeba and its cross-section in \textbf{Figure 1} after transformation while preserving its topology.}
   \label{awithtransformation}
\end{figure}

Additionally, to improve the visualisation of the amoeba image, a $GL(3,\mathbb{Z})$ transformation can be performed, such that the Monte Carlo generated points occur as a more even sample across the full amoeba \cite{Bao:2021olg}. These transformations although changing the geometry preserve the amoebae topology, in particular the number of cavities (3-dimensional holes). The transformation matrix used here is
$M=\begin{pmatrix}
    5 & 1 & 2  \\
    1 & 2 & 5  \\
    1 & 5 & 1
  \end{pmatrix}$, 
  producing the amoeba shown in Figure \ref{awithtransformation}. The complementary components of an amoeba may be bounded or unbounded. In Figure \ref{awithtransformation}, there is a single bounded 3-dimensional hole (cavity), which is emphasised through plotting a cross-section of this amoeba. The number of such cavities depends on the choice of coefficients of the Newton polynomial, and is bounded above by the number of internal points in the respective toric diagram. 
The variation of the Newton polynomial coefficients considered changes the Riemann surface geometry but preserves the topology and existence of holes, however as the coefficients change the amoeba projection of this surface changes and coefficient values where the topology of the respective amoeba changes is the focus of interest in this study. 
It is worth noting here also there will be coefficient choices that make the Riemann surface singular, and change its topology, but we leave consideration of the respective amoeba transitions at these Riemann surface topological transitions to future work.

A lattice polytope \(\Delta_n\) is reflexive if its dual polytope is also a lattice polytope in \(\mathbb{Z}^n\). A necessary but not sufficient condition for reflexivity is for the polytope to have a single interior point, and this unique interior point is taken to be the origin.

Each lattice polytope can be associated with a compact toric variety with complex dimension equal to the polytope lattice dimension. For a reflexive polytopes, the corresponding compact toric variety is a Gorenstein toric Fano variety. Separately a non-compact toric Calabi-Yau \((n+1)\)-fold can also be created from the polytope by embedding it in $\mathbb{Z}^{n+1}$, setting $p_{n+1}=1 \ \forall \textbf{p} \in \Delta_n$, and using the respective fan; effectively constructing the non-compact $CY_4$ as the affine cone over the comapct Fano variety.

The toric \(CY_4\) singularities (from the non-compact construction with 3d lattice polyhedra) can be probed by \(D1-\)branes to give rise to the classical mesonic moduli space of the 2d \(\mathcal{N}=(0,2)\) gauge theory. These theories are encoded by the periodic quiver diagrams which specify their matter content involving two types of matter fields and gauge symmetry \cite{Franco:2015tya}. The graph dual to the periodic quivers on \(T^3\) represents brane configurations of NS5-brane and D4-branes. The complex surface defined by the zero locus of the Newton polynomial of the toric \(CY_4\) is the surface wrapped by the NS5-brane, which can be studied using the (co)amoeba/algae projection \cite{Feng:2005gw}.

\subsection{Mahler measure}

The Mahler measure was first introduced in algebraic number theory in \cite{Mahler1962OnSI}, and it is defined as such\footnote{The Mahler measure is often referred to the exponential quantity, \(\exp (m(P))\), in the literature.}. Given a non-zero Laurent polynomial in \(n\) complex variables, \(P(z_1,\ldots,z_n)\in \mathbb{C}[z^{\pm 1}_1,\ldots,z^{\pm 1}_n]\), the Mahler measure \(m(P)\) is given by
\begin{equation}
\label{def:mahler}
   m(P)=\frac{1}{(2\pi i)^n}\int_{|z_1|=1}\dots\int_{|z_n|=1}\log|P(z_1,\dots,z_n)|\frac{\textup{d}z_1}{z_1}\dots\frac{\textup{d}z_n}{z_n}.
\end{equation}

In this paper, we focus on two- and three-variable Laurent polynomials. For simplicity, consider the two-variable Laurent polynomials of the form 
\begin{equation}\label{kNP}
P(z,w)=k-p(z,w) \ ,
\end{equation}
where $p(z,w)$ does not have a constant term, and $|k|>\max\limits_{|z|=|w|=1}|p(z,w)|$. Therefore, Mahler measure \eqref{def:mahler} becomes
\begin{equation}
        m(P)=
        \text{Re}\left(\frac{1}{(2\pi i)^2}\int_{|z|=|w|=1}\log(k-p(z,w))\frac{\textup{d}z}{z}\frac{\textup{d}w}{w}\right).
\end{equation}

The series expansion of $\log(k-p(z,w))$ converges uniformly on the support of the integration path and leads to
\begin{equation}
\label{Mahler exp}
    m(P)=\log k+\int_k^{\infty}(u_0(t)-1)\frac{\text{d}t}{t},
\end{equation}
where
\begin{equation}
    u_0(k)=\frac{1}{(2\pi i)^2}\int_{|z|=|w|=1}\frac{1}{1-k^{-1}p(z,w)}\frac{\textup{d}z}{z}\frac{\textup{d}w}{w}.
\end{equation}
The Mahler measure \eqref{Mahler exp} leads to the \textit{Mahler flow equation}: 
\begin{equation}
    \frac{\text{d}m(P)}{\textup{d}\log k}=k\frac{\text{d}m(P)}{\text{d}k}=u_0(k),
\end{equation}
and using the Mahler flow equation, the monotonic behaviour of the Mahler measure is observed in \cite{Bao:2021fjd}:
The Mahler measure $m(P)$ strictly increases when $|k|$ increases from $\max\limits_{|z|=|w|=1}|p(z,w)|$ to $\infty$. In Section §\ref{sec:4}, we study a refinement of this result for non-reflexive polynomials.

\subsection{Ronkin function and dimer model}
In this section, we introduce the Ronkin function in order to elaborate on the relations between the Mahler measure and the amoebae. Further numerical study of such relations is explained in Section §\ref{sec:5}.

The $n$-dimensional Ronkin function is defined as
\begin{equation}
 \label{Ronkin}
    R(x_1, x_2, ..., x_n)\coloneqq \frac{1}{(2\pi i)^n}\int_{|z_1|=1}\dots\int_{|z_n|=1}\log|P(e^{x_1}z_1,\dots,e^{x_n}z_n)|\frac{\textup{d}z_1}{z_1}\dots\frac{\textup{d}z_n}{z_n}.
\end{equation}
2-dimensional Ronkin function plays a crucial role in Physics. In the context of dimer model and crystal melting model, Ronkin function is the limit shape of these models, in the scaling limit. The amoebae and the Mahler measure are the phase structure and the free energy of the model, respectively \cite{Kenyon:2003}.

\begin{figure}[!ht]
   \begin{minipage}{0.48\textwidth}
     \centering
     \includegraphics[width=.8\linewidth]{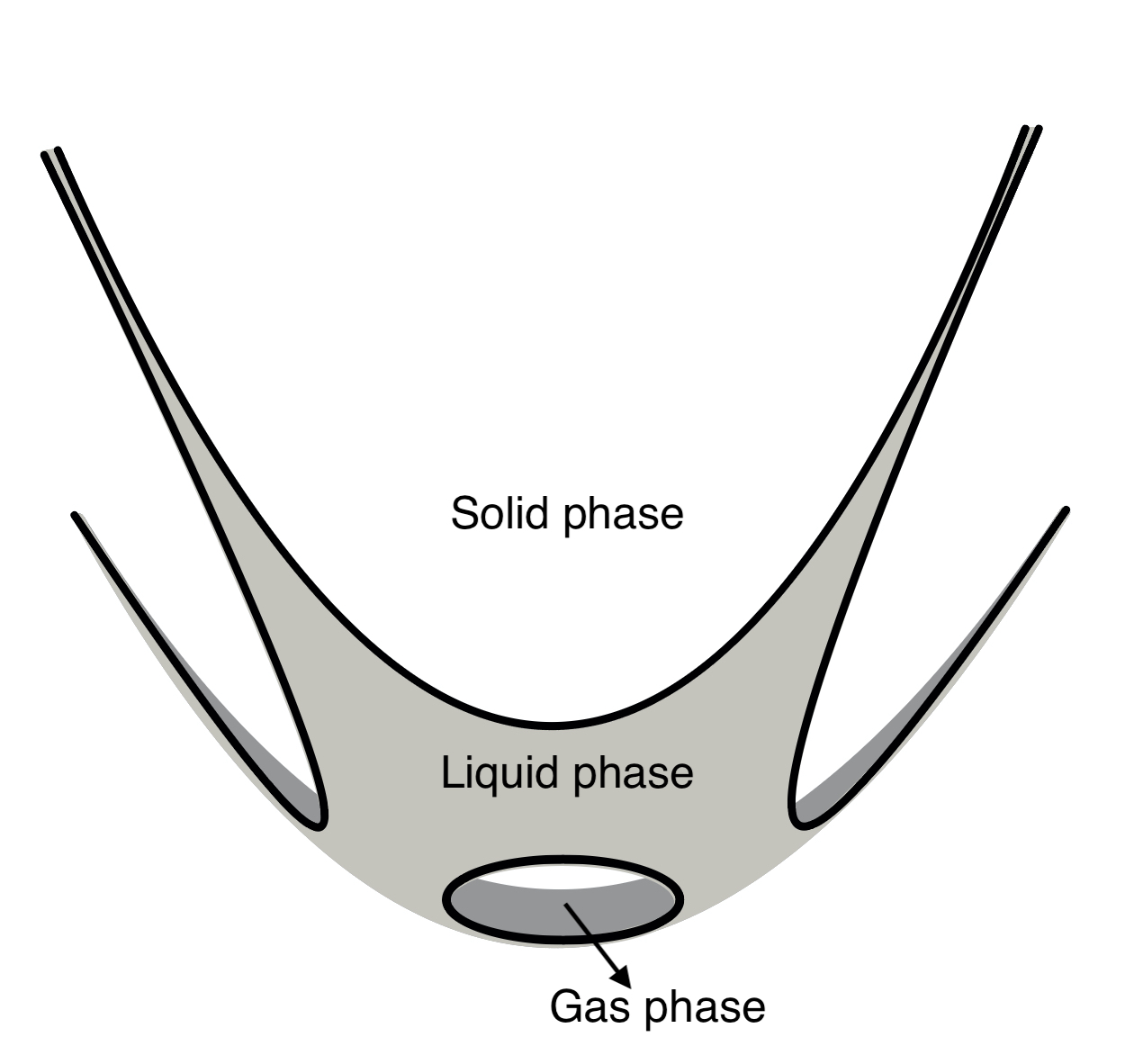}
   \end{minipage}\hfill
   \begin{minipage}{0.48\textwidth}
     \centering
     \includegraphics[width=.8\linewidth]{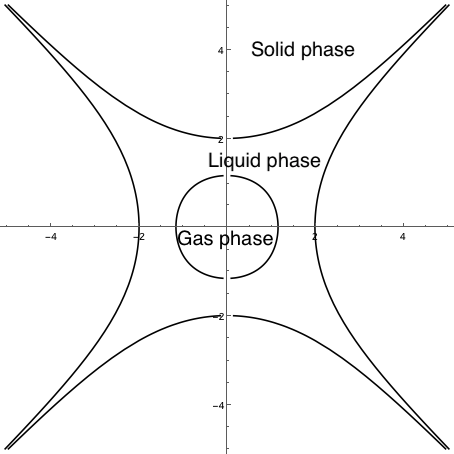}
   \end{minipage}
   \caption{The Ronkin function (left) and the amoeba of \(\mathbb{F}_0\) (right). Figures are adapted from \cite{Bao:2021fjd}.}
   \label{ronkin and amoeba}
\end{figure}

Following \cite{Kenyon:2003,kenyonlec,kenyon2005dimer}, the dimer model and the crystal melting model have the statistical interpretation as the solid, liquid, and gas phases with respect to the height fluctuations. This is illustrated in Figure \ref{ronkin and amoeba}. Specifically, the solid phase is where the height fluctuations are bounded almost certainly; liquid phase is where the covariance in the height function is unbounded as the distance between two distant points goes to infinity; and the gas phase is where the covariance of the average height difference is bounded but itself is unbounded (detailed discussion can be found in \cite{kenyon2005dimer}). In crystal melting, the opening of the amoeba hole (oval) corresponds to the gas phase.
Thus, the critical value of \(k\) at which the amoeba hole appears characterises the phase transition from the liquid phase to the gas phase. This motivates an associated definition of different phase contributions to the Mahler measure, proposed in \cite{Bao:2021fjd}. In particular, the liquid and gas phase contributions to the Mahler measure, \(m_{l,g}\), are defined as
\begin{equation}
\begin{split}
        m_l(P)&=\begin{cases}
            m(P) \, &\mathrm{for} \, k\leq k_c \\
            m(P(k_c)) \, &\mathrm{for}\, k >k_c
        \end{cases},\\
        m_g(P)&=m(P)-m(P(k_c))\qquad \text{for}\qquad k\geq k_c.
\end{split}
\end{equation}

Moreover, Ronkin function provides a unified framework for the study of amoebae and the Mahler measure. Indeed, from the definition \eqref{Ronkin}, it is clear that the Mahler measure is the Ronkin function evaluated at the origin. On the other hand, one can observe that the boundaries of the Ronkin function coincide with the boundaries of the amoebae. In fact, the boundaries of the amoebae are the limit shape of the 2-dimensional crystal model, living on the facets of the 3-dimensional model, where the boundaries of the Ronkin function are located \cite{Zahabi:2020hwu}.

Moreover, the Ronkin function is strictly convex over the amoeba and is linear in each complement region of the amoeba \cite{Bao:2021fjd,FORSBERG200045,mikhalkin2004amoebas}, so the gradient of the Ronkin function is linear for each amoeba complement. This is important for the derivation of the expressions for the boundary of the amoeba, which is elaborated in Section §\ref{sec:area_bhc}.

A particularly interesting boundary of the amoebae is the boundary of the bounded complement of the amoebae. At some critical point $k=k_c$, the bounded complement of the amoebae is degenerated to a point, which can always be transformed to the origin. Beyond the critical point $k>k_c$, the complement starts to grow in size along with the growth of the Mahler measure. This is referred to Mahler flow introduced in \cite{Bao:2021fjd}. Thus it is natural to expect that the Mahler measure is related to the bounded complement of the amoebae and perhaps its area. From the physics perspective, the bounded complement of the amoebae is the gas phase of the dimer model and its entropy is related to the Mahler measure of the gas phase $m_g$, \cite{Bao:2021fjd}. The relations between the area of the complement and $m_g$ are numerically studied in concrete examples in Section §\ref{sec:5}. However, the precise analytic relation between the area of the gas phase and the Mahler measure is not yet understood and we hope to further study this problem in future.

\section{Machine learning 3d amoebae Betti numbers from coefficients}\label{sec:3}
In this section, a variety of example complex 3d Riemann surfaces are considered, for each surface a set of polynomial coefficient vectors are generated for the respective Newton polynomial; each coefficient set giving a geometrically different surface. Each of these surfaces will have a different amoeba projection with potentially different topology under the projection.

The aim of this investigation is to establish how well ML architectures can learn to predict the second Betti number, $b_2$, dictating the number of 3-dimensional cavities, from the polynomial coefficients alone.
For each of the example surfaces, across the set of generated amoebae the $b_2$ values are calculated using the topological data analysis technique of persistent homology on Monte Carlo sampled point clouds of the amoeba.
These values are used as the outputs to be learnt from the coefficient vector inputs.

\subsection{Estimating Betti numbers with persistent homology}

The $k$-th homology group \(\mathbf{H}_k(X)\) of a topological space \(X\) is a key concept in algebraic topology. It is defined as the quotient group of the cycle group \(\mathbf{Z}_k\) by the boundary group \(\mathbf{B}_k\),
\begin{equation}
    \mathbf{H}_k \equiv \mathbf{Z}_k/\mathbf{B}_k,
\end{equation}
where
\begin{equation}
    \mathbf{Z}_k \equiv \mathrm{Ker}(\partial_k), \hspace{4pt} \mathbf{B}_k \equiv \mathrm{Im}(\partial_{k+1}),
\end{equation}
under the boundary operator $\partial_k$, which in the simplicial complex context maps $k$-simplices to their boundaries made up of $(k-1)$-simplices. Thus, the dimension of the \(k\)-th homology group \(\mathbf{H}_k(X)\), i.e., the \(k\)-th Betti number \(b_k\), counts the number of \(k\)-dimensional holes in \(X\) (the number of cycles that are not boundaries of some simplicial complexes). The largest homology group one can consider is bounded by the dimension of $X$, such that in the case of 3-dimensional amoebae, the first homology group of interest is $\mathbf{H}_2(X)$ with dimension \(b_2\), as the boundary of a 3-dimensional cavity is of dimension two.

Since the amoeba can be easily sampled to obtain the point cloud data for the space, its topological invariants can be obtained directly using a filtration starting from these points, via topological data analysis. This filtration of complexes is created by first considering the sampled points in $\mathbb{R}^3$, and a respective simplicial complex of as many points. Then imagining a 3d ball of radius $\delta$ centred on each point, the value of $\delta$ is continuously increased from 0 to $\infty$ and at each $\delta$ value where there is a new intersection of the balls the respective simplicial complex is updated to produce the next complex in the filtration. When $(k+1)$ balls intersect a $k$-simplex is drawn between their respective points in the simplicial complex (up to $k=3$ for these 3d data clouds). 

The \((p,q)\)-persistent \(k\)-homology \(\mathbf{H}^{p,q}_k\) hence describes the birth $(p)$ and death $(q)$ of \(k\)-cycles created and subsequently filled as the complex changes through the filtration. There are many available algorithms and software tools for computing persistent homology, and we adopted the \texttt{python} package \texttt{ripser} due to its relative efficiency \cite{ctralie2018ripser}.

\subsection{ML architecture}

As in \cite{Bao:2021olg}, we compared feed-forward neural networks and convolutional neural networks, coded in \texttt{Mathematica} \cite{Mathematica}, to ML the number of cavities present in the amoebae from the coefficients. The architectures are the same as in \cite{Bao:2021olg}:

\textbf{MLP}: one hidden layer of 100 perceptrons and ReLU activation function.

\textbf{CNN}: four 1d convolutional layers, each followed by a Leaky ReLU layer and a 1d MaxPooling layer.

For all neural networks, we used learning rates of \(0.001\) and Adam optimizer. We also used a 5-fold cross validation to compute the standard errors.
The input data are the coefficients of a particular Newton polynomial and the output is the second Betti number of the corresponding amoeba,
\begin{equation}
    \{c_1,\ldots,c_n\}\rightarrow b_2\,.
\end{equation}

\subsection{Example: \texorpdfstring{\(\mathbb{P}^1\times\mathbb{P}^1\times\mathbb{P}^1\)}{P1P1P1}} 

\begin{figure}[!ht]
   \begin{minipage}{0.48\textwidth}
     \centering
     \includegraphics[width=.6\linewidth]{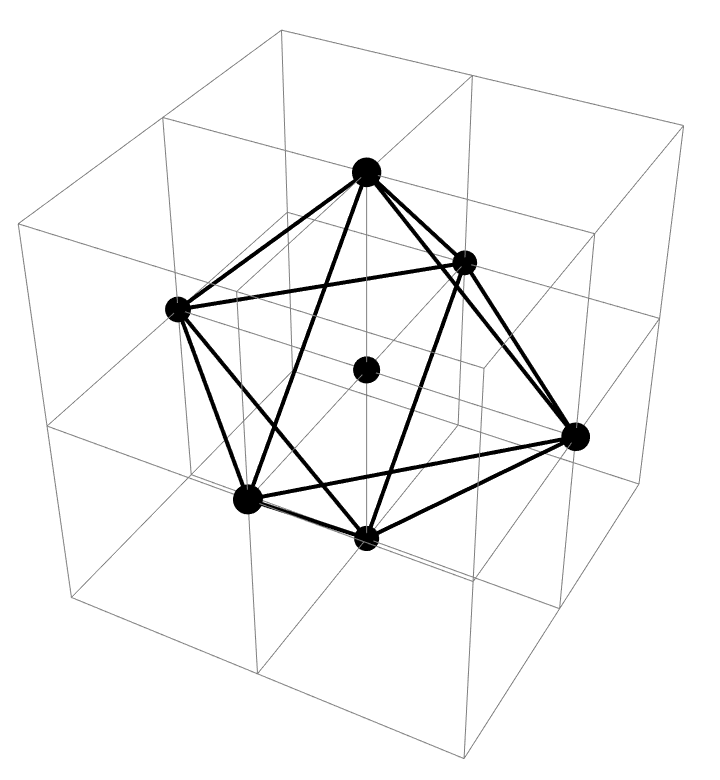}
     \caption{Toric diagram for \(\mathbb{P}^1\times\mathbb{P}^1\times\mathbb{P}^1\)\,.}
     \label{p1p1p1}
   \end{minipage}\hfill
   \begin{minipage}{0.48\textwidth}
     \centering
     \includegraphics[width=.6\linewidth]{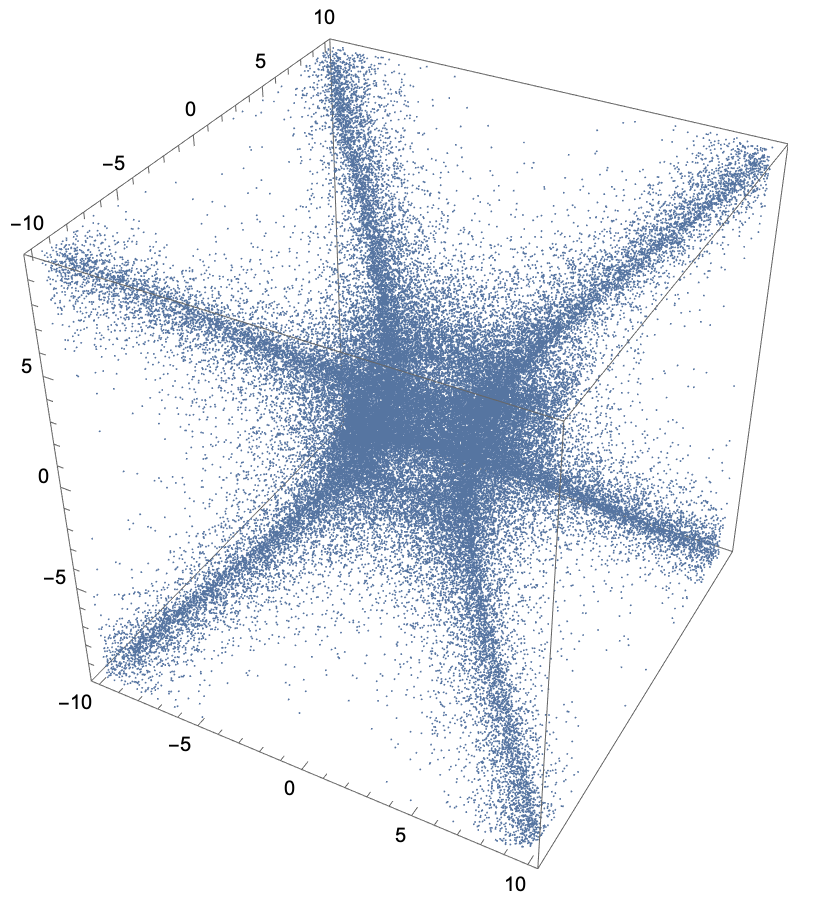}
     \caption{An example of the corresponding \(\mathbb{P}^1\times\mathbb{P}^1\times\mathbb{P}^1\) amoeba from Monte Carlo sampling.}
     \label{p1p1p1amb}
   \end{minipage}
\end{figure}

Consider the example surface of \(P(z_1,z_2,z_3)=c_1z_1+c_2z_1^{-1}+c_3z_2+c_4z_2^{-1}+c_5z_3+c_6z_3^{-1}+c_7=0\), where the corresponding toric diagram is shown in Figure \ref{p1p1p1}, which is analogous to the toric diagram of \(\mathbb{F}_0\) with an extra $\mathbb{P}^1$ fibration. An example of the associated amoeba is given in Figure \ref{p1p1p1amb} from Monte Carlo sampling. Since its toric diagram has only one interior point, the maximum number of 3-dimensional cavities is one, i.e., \(b_2 = 0\) or \(1\), such that this is a binary classification.

\subsubsection{Learning persistent homology \texorpdfstring{$b_2$}{b2}}

A balanced dataset of 7200 random samples was generated of real coefficients with \(c_{(1,0,0)}, c_{(-1,0,0)}, c_{(0,1,0)}, c_{(0,-1,0)}, c_{(0,0,1)}, c_{(0,0,-1)} \in [-5,5]\) and \(c_{(0,0,0)} \in [-20,20]\). For each set of coefficients, we used \(\mathcal{LA}_{\Tilde{f}_1}\) to approximate \(\mathcal{A}_f\), and sampled \(\mathcal{LA}_{\Tilde{f}_1}\) with around \(700\) points to allow feasible computation. A matrix transformation is performed on the amoeba such that its boundary is clearer while preserving the value of \(b_2\). Then, the point cloud data is passed into the \texttt{ripser} package to obtain the persistent pairs of \(\mathbf{H}_2\).

\begin{figure}[!ht]
  \centering
  \includegraphics[width=.5\linewidth]{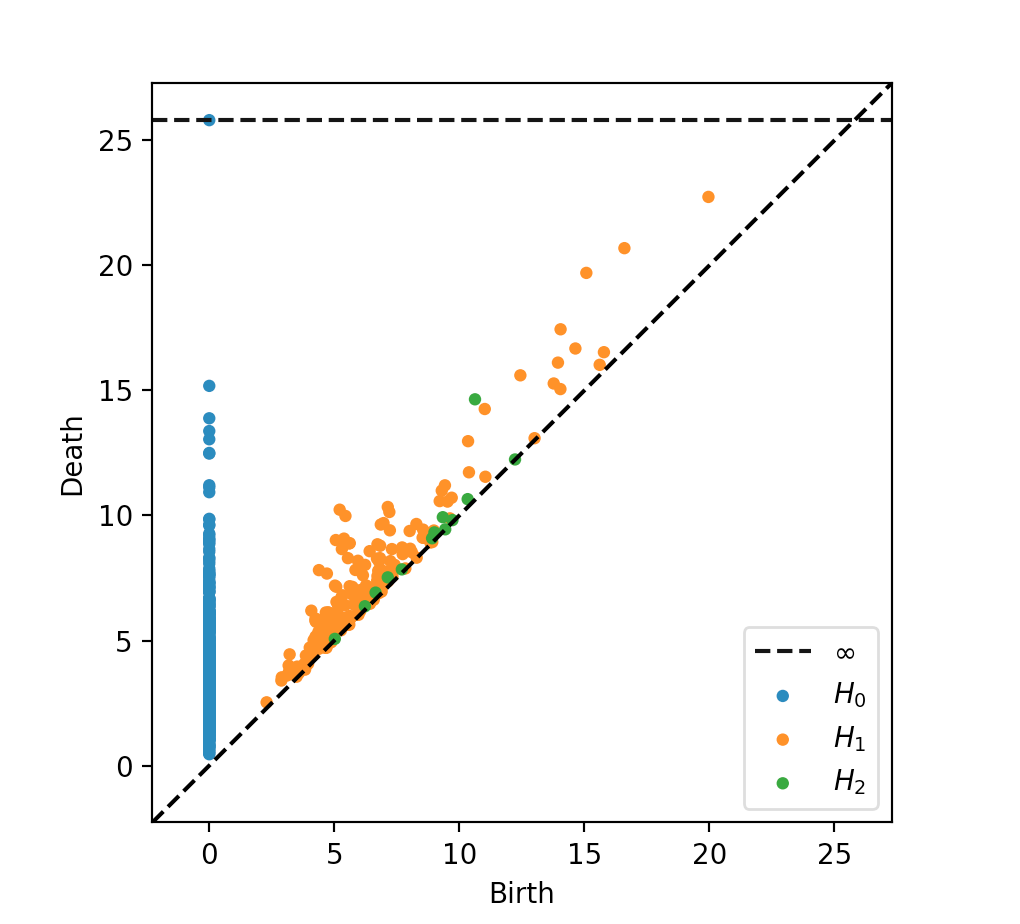}
  \caption{Example of a persistent diagram showing the homology groups \(H_0, H_1, H_2\) for the point cloud data of the amoeba in Figure \ref{ap1}. The $H_2$ point (10.62652588, 14.6450882) (represented by the green point with coordinates (10.62652588, 14.6450882)) suggests the existence of a 2-dimensional cavity, i.e., \(b_2=1\).}
  \label{p1p1p1 ph example}
\end{figure}

After obtaining all the persistent pairs, a selection is required, as when the birth ($p$) and death ($q$) times are close to each other this may be the result of point sampling not being dense enough. Thus, persistent pairs \((p,q)\) with \(q-p \leq 1.45\) are discarded as noise. 
This value was selected as a heuristic optimum for the dataset considered, and negligible classification improvements were seen with values $\sim 1$ and for larger datasets ($\sim 10 \times$).
Since there is at most one cavity, the value of \(b_2\) is determined by
\begin{equation}
    b_2 = \begin{cases} 0 \hspace{10pt} & \mathrm{No}\hspace{4pt} \mathrm{persistent}\hspace{4pt} \mathrm{pairs}\hspace{4pt} \mathrm{with}\hspace{4pt} q-p >1.45\,;\\
    1 \hspace{10pt} & \mathrm{Otherwise}\,.
    \end{cases}
\end{equation}
The identification of the $b_2$ Betti number from the topological data analysis is exemplified in Figure \ref{p1p1p1 ph example}. Within this, the main source of error comes from the number of sampling points and the selection of the persistent pairs.

These $b_2$ values extracted from the persistent homology where used as the data labels for each amoeba.
The subsequent ML hence performed the binary classification task of learning the $b_2$ value from the input vector of amoeba coefficients.
Two NN architectures were used, and classification performance was measured with accuracy as the proportion of correctly predicted $b_2$ values.
Across the 5-fold cross validation runs for both architectures, the performance measures of accuracy (ACC) and Matthews Correlation Coefficient (MCC) were:
\begin{align}
    &\text{MLP:} \quad \text{ACC: } 0.771 \pm 0.014\,, \quad \text{MCC: } 0.543 \pm 0.029\,, \\
    &\text{CNN:} \quad \text{ACC: } 0.776 \pm 0.014 \,, \quad \text{MCC: } 0.550 \pm 0.031\,.
\end{align}
Note also that performance could be marginally improved by increasing the number of sampling points at a cost of longer computation time for the persistent homology.

\subsubsection{Learning analytic lopsidedness \texorpdfstring{\(b_2\)}{b2}}

This example is simple enough that the condition for the number of cavities ($b_2$) can be derived in a similar way as for the example of \(\mathbb{F}_0\) in \(2d\), using lopsidedness. The condition obtained is
\begin{equation}
     b_2 = \begin{cases} 0 \hspace{10pt} & |c_7|\leq 2|c_1c_2|^{1/2}+2|c_3c_4|^{1/2}+2|c_5c_6|^{1/2}\,;\\
    1 \hspace{10pt} & \mathrm{Otherwise}\,.
    \end{cases}
\end{equation}

Now performing the ML using the analytic condition from the lopsided amoeba approximation to generate the $b_2$ output values for each input amoeba coefficient vector, the results improved.
A balanced dataset of 5400 random samples was used to achieve learning measures for each architecture
\begin{align}
    &\text{MLP:} \quad \text{ACC: } 0.937 \pm 0.008\,, \quad \text{MCC: } 0.874 \pm 0.016\,, \\
    &\text{CNN:} \quad \text{ACC: } 0.894 \pm 0.014 \,, \quad \text{MCC: } 0.789 \pm 0.026\,.
\end{align}

The mismatch between two datasets is mostly due to the sampling points not being dense enough such that the separation of the points becomes comparable with the size of the cavity. Plotting the corresponding amoeba shows that it is difficult to tell the number of 3-dimensional cavities by eye in such cases.

\subsubsection{MDS projection}

\begin{figure}[!ht]
   \begin{minipage}{0.48\textwidth}
     \centering
     \includegraphics[width=.7\linewidth]{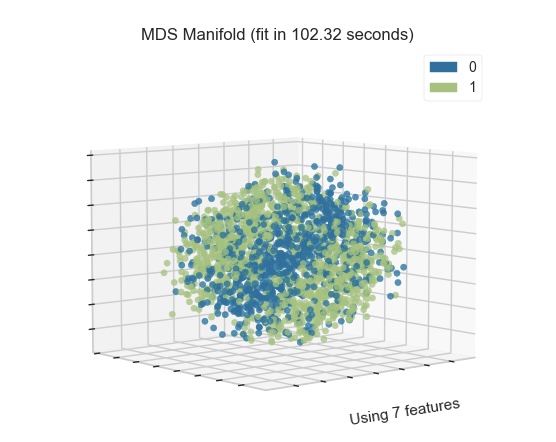}
   \end{minipage}\hfill
   \begin{minipage}{0.48\textwidth}
     \centering
     \includegraphics[width=.8\linewidth]{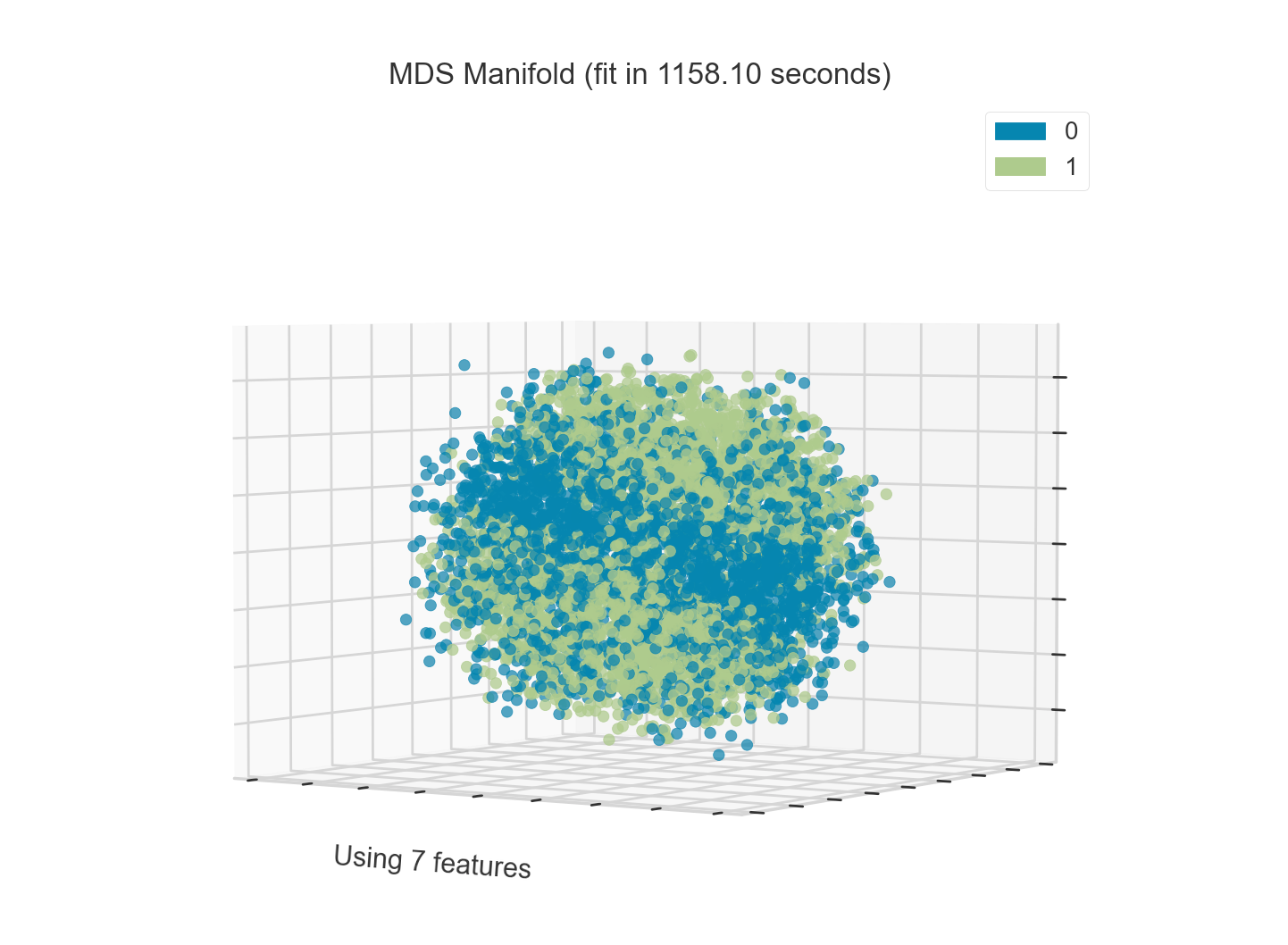}
   \end{minipage}
    \caption{MDS manifold projection on dataset obtained using persistent homology (left) and analytic condition (right).}
    \label{mds}
\end{figure}

Using the \texttt{yellowbrick} package \cite{bengfort_yellowbrick_2019}, multi-dimensional scaling (MDS) projections (Figure \ref{mds}) on the dataset obtained via persistent homology and the dataset obtained via analytic condition show similar separations. 
This MDS method performs non-linear dimensionality reduction of the $\mathbb{R}^7$ space of coefficient vectors into $\mathbb{R}^3$ for effective visualisation, and amoeba (as coefficient vector points) are coloured according to their computed $b_2$ value (via persistent homology, or analytically).

The difference in these plots may be attributed to poor sampling over the amoeba leading to false results for the persistent homology, or conversely may be due to the error caused by approximating the true amoeba by its lopsided counterpart for the analytic condition derivation.
Both these features highlight the subtlety in determining amoebae topology.

\subsection{Summary of the ML results}
Across the three 3d examples that we considered (details are given in Appendix \ref{appendix1}), the ML architectures perform similarly learning the $b_2$ Betti numbers computed from either persistent homology or lopsidedness. For ease of comparison the ML results are repeated for all 3 examples in Table \ref{tab:ML summary}, and the MDS projections computed for each in Table \ref{tab:mds summary}.

\begin{table}[!ht]
    \centering
    \addtolength{\leftskip} {-2cm}
    \addtolength{\rightskip}{-2cm}
    \begin{tabular}{|>{\centering\arraybackslash}m{2em}|>{\centering\arraybackslash}m{2em}||>{\centering\arraybackslash}m{5em}|>{\centering\arraybackslash}m{5em}|>{\centering\arraybackslash}m{5em}|>{\centering\arraybackslash}m{5em}|>{\centering\arraybackslash}m{5em}|>{\centering\arraybackslash}m{5em}|}
    \hline
    \multicolumn{2}{|c||}{\multirow{2}{*}{Surface}} & \multicolumn{2}{c|}{\(\mathbb{P}^1\times \mathbb{P}^1\times\mathbb{P}^1\)}&\multicolumn{2}{c|}{\(\mathbb{P}^3\)}&\multicolumn{2}{c|}{ \(\mathbb{P}^2\times\mathbb{P}^1\) } \\
    \cline{3-8}
    \multicolumn{2}{|c||}{}& PH & Analytic & PH & Analytic & PH & Analytic\\
    \hline
    \multirow{2}{4em}{ACC} & MLP & \(0.771 \pm0.014\) & \(0.937\pm0.008\) & \(0.840\pm 0.015\) &\(0.939\pm0.009\) & \(0.830\pm0.016\) & \(0.947\pm0.007\)\\
    \cline{2-8}
    & CNN & \(0.776\pm 0.014\) & \(0.894\pm0.014\) & \(0.727\pm0.031\) & \(0.910\pm0.010\) & \(0.825\pm0.027\) & \(0.920\pm0.011\) \\
    \hline
    \multirow{2}{4em}{MCC} & MLP & \(0.543\pm0.029\) & \(0.874\pm0.016\) & \(0.699\pm0.022\) & \(0.876\pm0.017\) & \(0.652\pm0.035\) & \(0.893\pm0.014\) \\
    \cline{2-8}
    & CNN & \(0.550\pm0.031\) & \(0.789\pm0.026\) & \(0.457\pm0.073\) & \(0.819\pm 0.019\) & \(0.630\pm 0.063\) & \(0.841\pm 0.023\)\\
    \hline
    \end{tabular}
    \caption{Summary of the ML results, learning the homology of amoebae constructed from the stated Riemann surfaces with varying coefficients. Learning was performed by MLP and CNN architectures, predicting the $b_2$ values computed using persistent homology (PH) or lopsidedness (Analytic). Performance was measured with accuracy (ACC) and MCC over the 5-fold cross validation.}
    \label{tab:ML summary}
\end{table}

\begin{table}[!ht]
    \centering
    \addtolength{\leftskip} {-2cm}
    \addtolength{\rightskip}{-2cm}
    \begin{tabular}{|>{\centering\arraybackslash}m{6em}||>{\centering\arraybackslash}m{9em}|>{\centering\arraybackslash}m{9em}|>{\centering\arraybackslash}m{9em}|}
    \hline
    \(\mathrm{Surface}\) &  \(\mathbb{P}^1\times\mathbb{P}^1\times\mathbb{P}^1\)&
     \(\mathbb{P}^3\)&\(\mathbb{P}^2\times\mathbb{P}^1\)\\
     \hline
    MDS Projection (PH) & \raisebox{-.5\height}{ \includegraphics[height=30mm]{Fig/MDS.jpeg}} &\raisebox{-.5\height}{ \includegraphics[height=30mm]{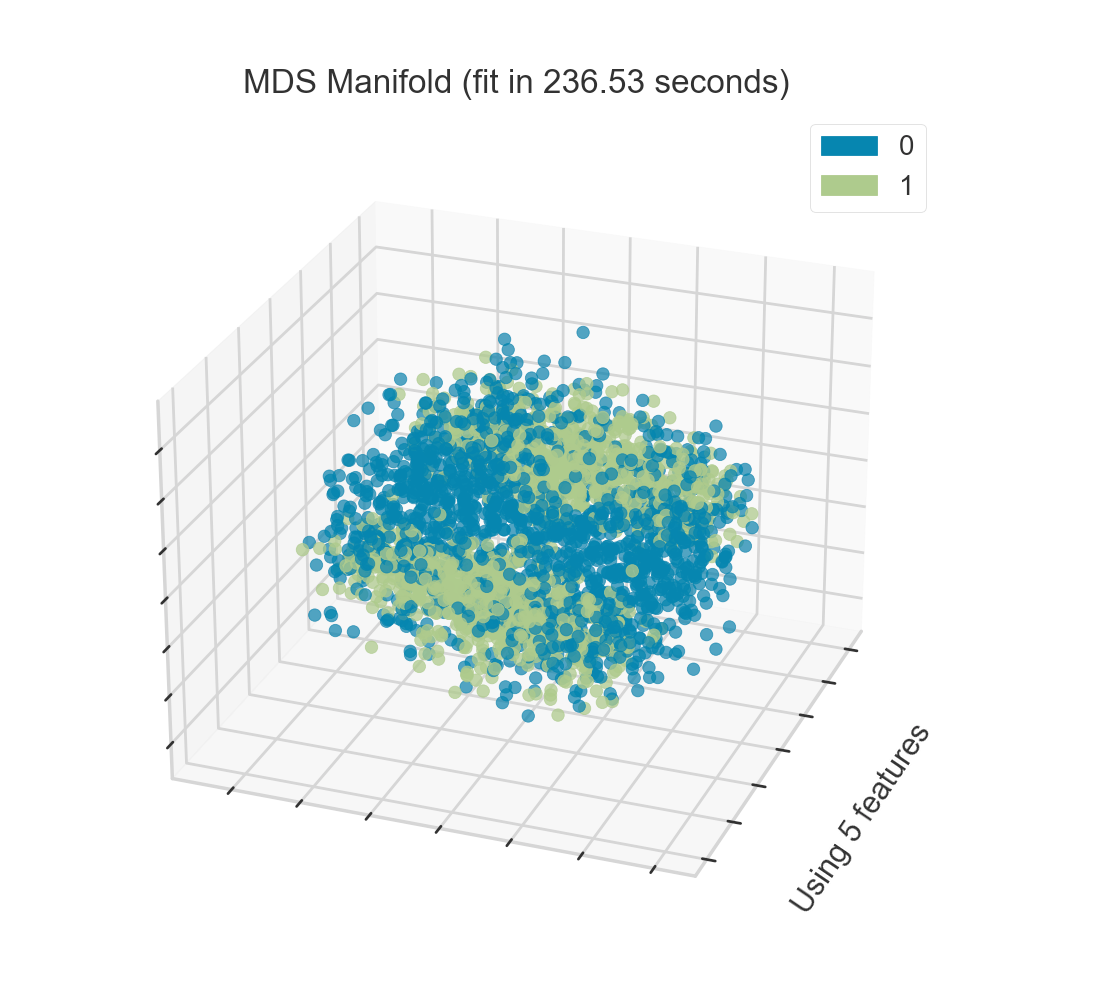}} &\raisebox{-.5\height}{ \includegraphics[height=30mm]{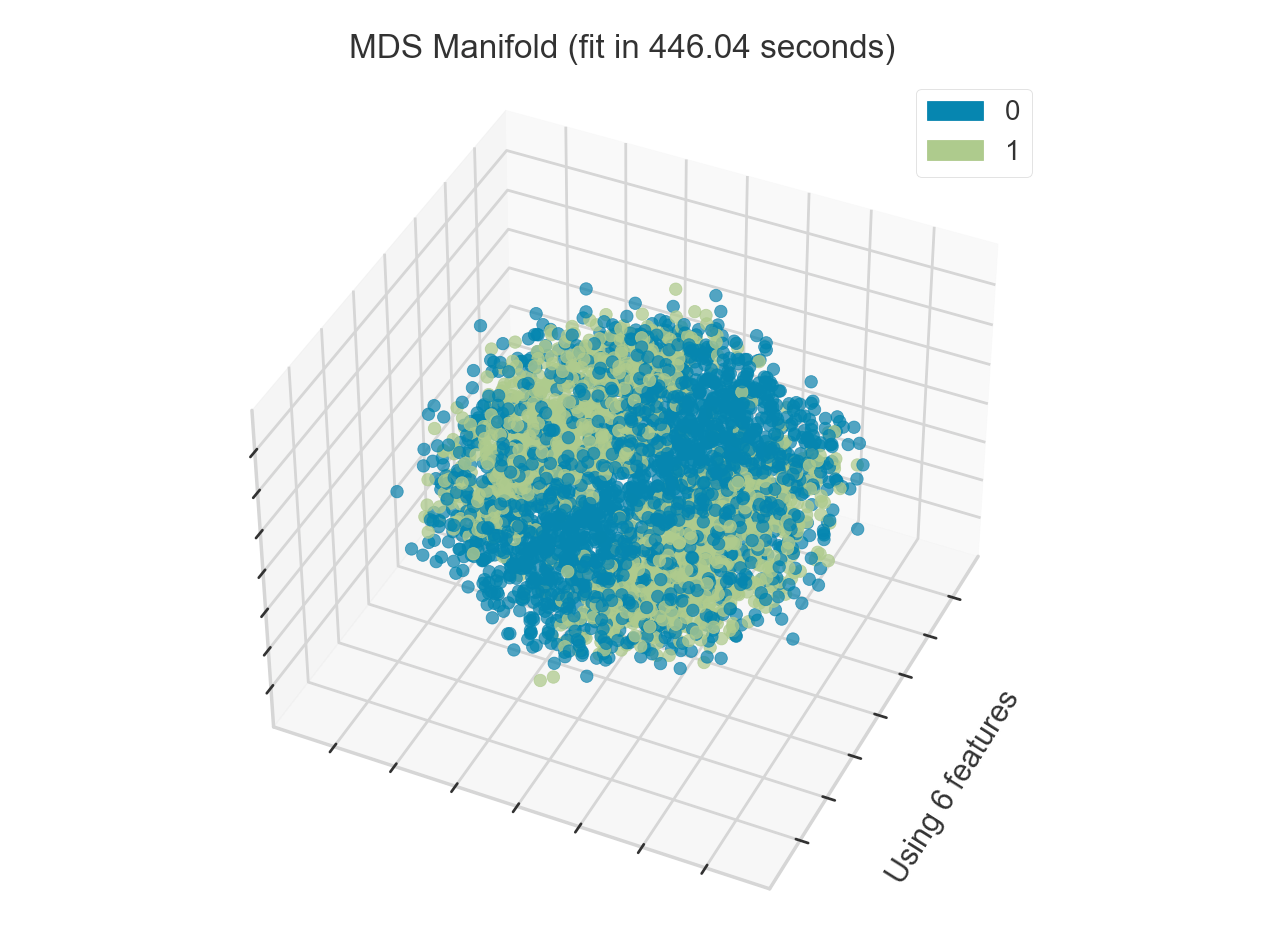}}\\
    \hline
    MDS Projection (Analytic) & \raisebox{-.5\height}{ \includegraphics[height=30mm]{Fig/Mdsa.png}}&
    \raisebox{-.5\height}{ \includegraphics[height=30mm]{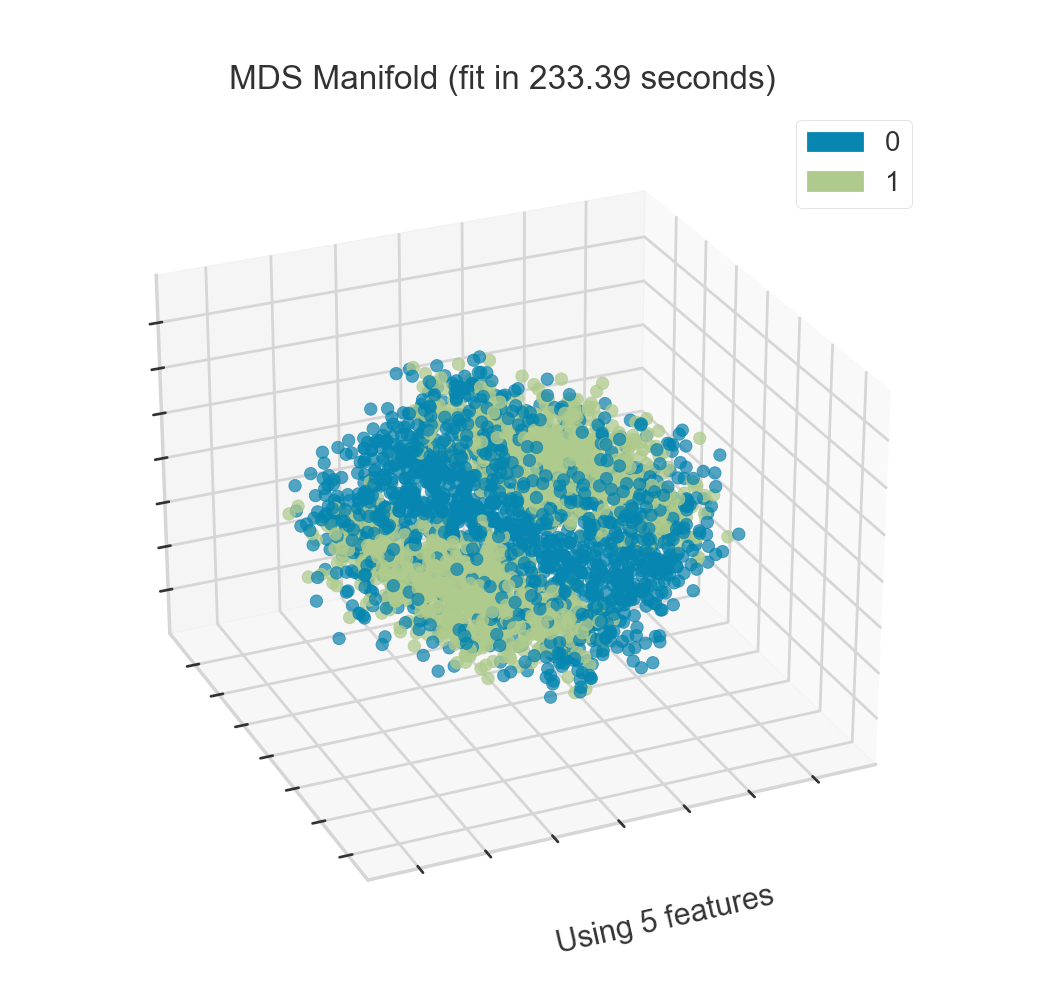}}& \raisebox{-.5\height}{ \includegraphics[height=30mm]{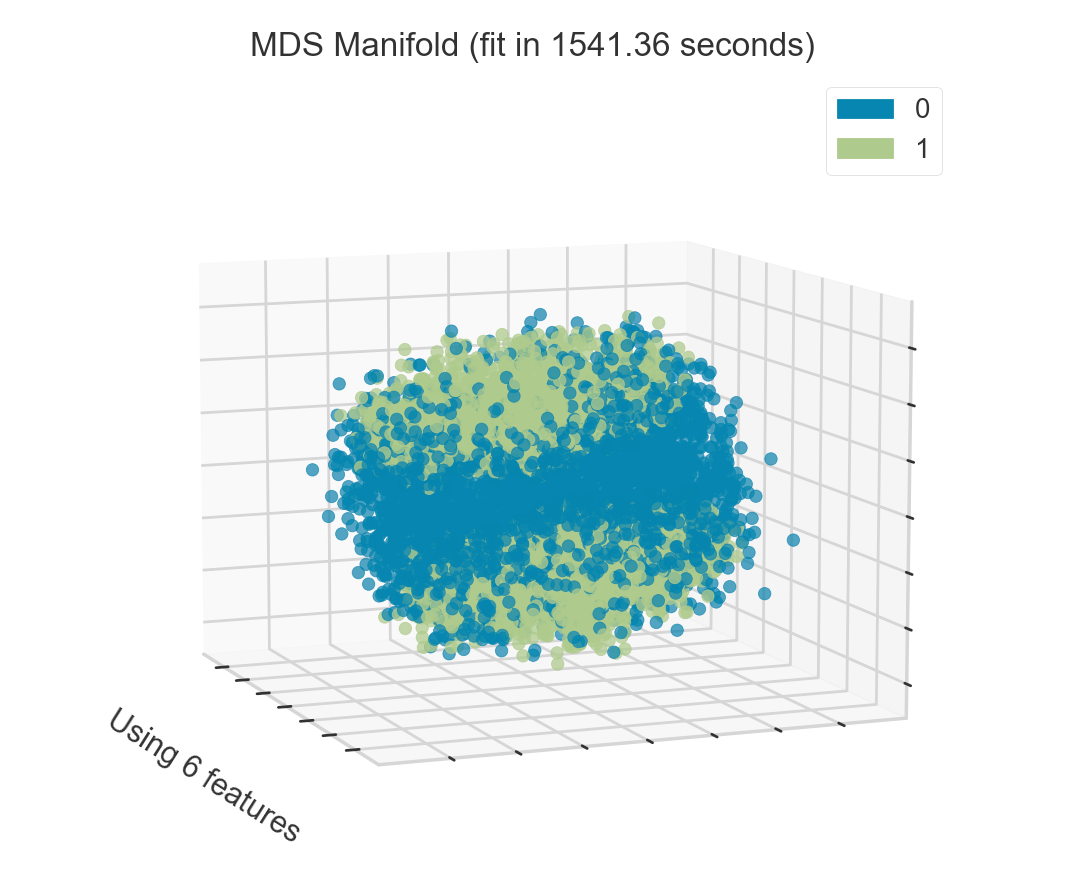}}\\
    \hline
    \end{tabular}
    \caption{MDS projections for each of the three Riemann surface examples considered, colouring according to the $b_2$ values $\{0,1\}$ computed via persistent homology (PH) or lopsidedness (Analytic) respectively.}
    \label{tab:mds summary}
\end{table}

\section{Non-Reflexive Mahler measure and Mahler flow}\label{sec:4}

As we mention in Section \S\ref{sec:2}, a reflexive polytope $\Delta$ on $\mathbb{Z}^n$ is one whose dual polytope
\begin{equation}
    \Delta^{\circ}=\{\textbf{v}\in \mathbb{Z}^n|\textbf{v}\cdot\textbf{u}\leq -1,\forall \textbf{u} \in \Delta\}
\end{equation}
is also reflexive. For $n=2$, we can show that $\Delta$ is reflexive iff the polytope has one interior point. In this section, we focus on polytopes with two interior points, which are therefore non-reflexive.

We can deal with the Mahler measure of non-reflexive polytopes in a similar way to the reflexive case introduced in Section \S\ref{sec:2}. We first consider polynomials of the form $P(z,w)=k_1-p(z,w)$, where all coefficients of $p(z,w)$ are positive. For polytopes with two interior points, we can write this as $P(z,w)=k_1-k_2z^nw^m-p'(z,w)$, where $p'(z,w)=p(z,w)-k_2z^nw^m$, and the position of the second interior point is $(n,m)$. For cases where $k_2>\textrm{max}|k_1-p'(z,w)|$, we can calculate the Mahler measure $m(P)$ using Cauchy's residue theorem. We factor out $\log (k_2z^nw^m)$ and are left with:  
\begin{equation}
    m(P)=\textrm{Re}\left(\log k_2+\frac{1}{(2\pi i)^2}\int_{|w|,|z|=1}\log \left( 1-\frac{1}{k_2z^nw^m}(k_1-p'(z,w)) \right)\frac{dz}{z}\frac{dw}{w}\right),
    \end{equation}
where the $\log k_2$ term contributes to the residue, and therefore to the Mahler measure. The $\log (z^nw^m)$ term also contributes to the residue, but since it is purely imaginary, it does not contribute to the measure. To get the full value of the Mahler measure, we expand the $\log (1-(k_2(z^{n}w^{m})^{-1}(k_1-p'(z,w))))$ in powers of the second argument. A full example of this can be seen in Appendix \ref{appendix3}.\par
Similar to Eq.\eqref{Mahler exp}, we can also write the above equation as:
\begin{equation}\label{mexpansionequation}
    m(P)=\log k_2 + \int^{\infty}_{k_2}(u_2(t)-1)\frac{dt}{t},
    \end{equation}
where
\begin{equation}
    u_2(k_2) = \frac{1}{(2\pi i)^2} \int_{|w|,|z|=1}\frac{1}{1-(k_2z^{n}w^{m})^{-1}(k_1-p'(z,w))}\frac{dz}{z}\frac{dw}{w}.
    \end{equation} 
As in the single variable case, $u_2(k_2)$ is the period of a holomorphic 1-form $\omega_Y$ on the curve $Y$ defined by $1-(k_2(z^{n}w^{m})^{-1}(k_1-p'(z,w)))$, and it therefore satisfies the Picard-Fuchs
equation \cite{Villegas1999,}:
\begin{equation}
    A(k_2)\frac{d^2u_2(k_2)}{dk_2^2}+B(k_2)\frac{du_2(k_2)}{dk_2}+C(k_2)u_2(k_2) =0.
    \end{equation} 
Combined with the similar equation for computing the Mahler measure when the polynomial is lopsided in favour of $k_1$ (i.e. $k_1 > \max |p(z,w)|$), we now have a means of calculating the measure for the whole $k_1k_2$-plane except for a strip of width $\sqrt{2}\textrm{max}|p'(z,w)|$ centered along $k_1=k_2$. Within these two disconnected parts of the $k_1,k_2$-plane, the Mahler measure behaves as we expect.\par
We can redefine the Mahler flow, first introduced in \cite{Bao:2021fjd}, using two equations, one for each disconnected section:
\begin{equation}
\begin{split}
\frac{\partial m(P)}{\partial \log k_1} &= u_0(k_1)\\
\frac{\partial m(P)}{\partial \log k_2} &= u_2(k_2).
\end{split}
\end{equation}

As $u_0$ and $u_2$ both represent periods, they are always positive. Therefore, the Mahler measure is always increasing as we move along each flow. When travelling perpendicular to the respective flows, however, this is not necessarily true, as we will see in the next subsection.\par
As in the reflexive case, many polytopes also have lattice points lying on the edges. We often like to vary the coefficients of these edge points as well as the interior points. If the coefficient of this point is such that the polynomial becomes lopsided in its favour, we deal with it analogously to how we did above. In general, for a polynomial of the form $P(\textbf{z})=\sum_{\textbf{n}}c_{\textbf{n}}\textbf{z}^{\textbf{n}}$, where $\textbf{z}=(z_1,\dots ,z_i)$ and $\textbf{n}=(n_1,\dots ,n_i)$ are lattice points, as any $c_{\textbf{n}}$ tends to $\infty$, the Mahler measure $m(P)$ tends to $\max_{\textbf{n}}\log c_{\textbf{n}}$.

In the limit of large $(k_1,k_2)$, the Mahler measure tends to $\max(\log k_1,\log k_2)$, and we therefore get an infinite measure at the tropical limit at infinity.

\subsection{Example: \texorpdfstring{\(\mathbb{C}^3/\mathbb{Z}_5\)}{c3z5}}

For a concrete example, we will analyse the surface of $\mathbb{C}^3/\mathbb{Z}_5$ toric Calabi-Yau threefold whose associated toric diagram is pictured below. As mentioned above, for cases where $k_1>\max |p(z,w)|$ or $k_2>\max|k_1-p'(z,w)|$, we can expand the Newton polynomial, taking only the constant term. In this section, we have primarily used \texttt{Mathematica} \cite{Mathematica} for computations, and the infinite sum is truncated by taking the first 200 terms for a reasonable approximation.
\begin{center}
\tikzset{every picture/.style={line width=0.75pt}} 
\begin{tikzpicture}[x=0.75pt,y=0.75pt,yscale=-1,xscale=1]

\draw [color={rgb, 255:red, 155; green, 155; blue, 155 }  ,draw opacity=1 ]   (510,30) -- (510,130) ;
\draw [color={rgb, 255:red, 155; green, 155; blue, 155 }  ,draw opacity=1 ]   (536,30) -- (536,130) ;
\draw [color={rgb, 255:red, 155; green, 155; blue, 155 }  ,draw opacity=1 ]   (562,30) -- (562,130) ;
\draw [color={rgb, 255:red, 155; green, 155; blue, 155 }  ,draw opacity=1 ]   (588,30) -- (588,130) ;
\draw [color={rgb, 255:red, 155; green, 155; blue, 155 }  ,draw opacity=1 ]   (614,30) -- (614,130) ;
\draw [color={rgb, 255:red, 155; green, 155; blue, 155 }  ,draw opacity=1 ]   (640,30) -- (640,130) ;
\draw [color={rgb, 255:red, 155; green, 155; blue, 155 }  ,draw opacity=1 ]   (510,30) -- (640,30) ;
\draw [color={rgb, 255:red, 155; green, 155; blue, 155 }  ,draw opacity=1 ]   (510,55) -- (640,55) ;
\draw [color={rgb, 255:red, 155; green, 155; blue, 155 }  ,draw opacity=1 ]   (510,80) -- (640,80) ;
\draw [color={rgb, 255:red, 155; green, 155; blue, 155 }  ,draw opacity=1 ]   (510,105) -- (640,105) ;
\draw [color={rgb, 255:red, 155; green, 155; blue, 155 }  ,draw opacity=1 ]   (510,130) -- (640,130) ;
\draw  [color={rgb, 255:red, 0; green, 0; blue, 0 }  ,draw opacity=1 ][fill={rgb, 255:red, 0; green, 0; blue, 0 }  ,fill opacity=1 ] (559.4,80) .. controls (559.4,78.62) and (560.56,77.5) .. (562,77.5) .. controls (563.44,77.5) and (564.6,78.62) .. (564.6,80) .. controls (564.6,81.38) and (563.44,82.5) .. (562,82.5) .. controls (560.56,82.5) and (559.4,81.38) .. (559.4,80) -- cycle ;
\draw  [color={rgb, 255:red, 0; green, 0; blue, 0 }  ,draw opacity=1 ][fill={rgb, 255:red, 0; green, 0; blue, 0 }  ,fill opacity=1 ] (585.4,80) .. controls (585.4,78.62) and (586.56,77.5) .. (588,77.5) .. controls (589.44,77.5) and (590.6,78.62) .. (590.6,80) .. controls (590.6,81.38) and (589.44,82.5) .. (588,82.5) .. controls (586.56,82.5) and (585.4,81.38) .. (585.4,80) -- cycle ;
\draw  [color={rgb, 255:red, 0; green, 0; blue, 0 }  ,draw opacity=1 ][fill={rgb, 255:red, 0; green, 0; blue, 0 }  ,fill opacity=1 ] (585.4,55) .. controls (585.4,53.62) and (586.56,52.5) .. (588,52.5) .. controls (589.44,52.5) and (590.6,53.62) .. (590.6,55) .. controls (590.6,56.38) and (589.44,57.5) .. (588,57.5) .. controls (586.56,57.5) and (585.4,56.38) .. (585.4,55) -- cycle ;
\draw  [color={rgb, 255:red, 0; green, 0; blue, 0 }  ,draw opacity=1 ][fill={rgb, 255:red, 0; green, 0; blue, 0 }  ,fill opacity=1 ] (611.4,105) .. controls (611.4,103.62) and (612.56,102.5) .. (614,102.5) .. controls (615.44,102.5) and (616.6,103.62) .. (616.6,105) .. controls (616.6,106.38) and (615.44,107.5) .. (614,107.5) .. controls (612.56,107.5) and (611.4,106.38) .. (611.4,105) -- cycle ;
\draw  [color={rgb, 255:red, 0; green, 0; blue, 0 }  ,draw opacity=1 ][fill={rgb, 255:red, 0; green, 0; blue, 0 }  ,fill opacity=1 ] (533.4,80) .. controls (533.4,78.62) and (534.56,77.5) .. (536,77.5) .. controls (537.44,77.5) and (538.6,78.62) .. (538.6,80) .. controls (538.6,81.38) and (537.44,82.5) .. (536,82.5) .. controls (534.56,82.5) and (533.4,81.38) .. (533.4,80) -- cycle ;
\draw [line width=1.5]    (588,55) -- (614,105) ;
\draw [line width=1.5]    (588,55) -- (536,80) ;
\draw [line width=1.5]    (614,105) -- (536,80) ;

\draw (588.93,80.4) node [anchor=north west][inner sep=0.75pt]  [font=\scriptsize]  {$r$};
\draw (562.93,79.57) node [anchor=north west][inner sep=0.75pt]  [font=\scriptsize]  {$s$};
\draw (614.87,109.48) node [anchor=north west][inner sep=0.75pt]  [font=\scriptsize]  {$p_{1}$};
\draw (521.27,81.15) node [anchor=north west][inner sep=0.75pt]  [font=\scriptsize]  {$p_{2}$};
\draw (595.8,55.32) node [anchor=north west][inner sep=0.75pt]  [font=\scriptsize]  {$p_{3}$};

\end{tikzpicture}
    \end{center}
    
There are two expansions of the Mahler measure, depending on the values of $k_1$ and $k_2$. In both cases we take the origin to be the left interior point (labelled $s$ in the above figure). This corresponds to a Newton polynomial given by $P(z,w)= k_1-k_2z-zw-z^2w^{-1}-z^{-1}$. First, we expand for cases where the Newton polynomial is lopsided in favor of $k_1$. We get a Mahler measure given by:
\begin{equation}\label{eqc3z5}
    m_1\left(P_s(z,w)\right) = \log k_1 - \sum_{n=1}^{\infty}\sum_{i=0}^{n}\binom{n}{i}\binom{n-i}{\frac{n-i}{2}}\binom{i}{\frac{5i-3n}{4}}\frac{k_2^{\frac{5i-3n}{4}}}{k_1^nn}.
    \end{equation} 

Similarly, when the polynomial is lopsided in favor of $k_2$, we get an expression given by:
\begin{equation}
    m_2\left(P_s(z,w)\right) = \log k_2 - \sum_{n=1}^{\infty}\sum_{i=0}^{n}\binom{n}{i}\binom{i}{\frac{i}{2}}\binom{n-i}{\frac{3i-2n}{2}}\frac{k_1^{\frac{4n-5i}{2}}(-1)^{\frac{5i-2n}{2}}}{k_2^nn}.
    \end{equation}

In both cases, we have constraints on allowed combinations of $n$ and $i$, such that for every binomial coefficient $\binom{n}{r}$ all coefficients are positive integers, and $n\geq r$ (if not, the contributing summand is zero). This greatly reduces the number of terms we need to calculate, reducing the computing time. 

\begin{figure}[!ht]
\begin{minipage}{0.55\textwidth}
    \centering
     \includegraphics[width=.8\linewidth]{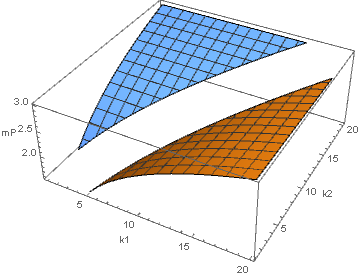}
     \end{minipage}\hfill
     \begin{minipage}{0.45\textwidth}
     \includegraphics[width=.7\linewidth]{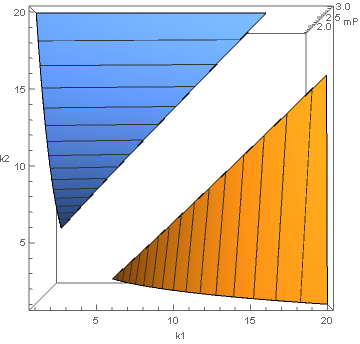}
     \end{minipage}
     \caption{The Mahler measure of the $\mathbb{C}^3/\mathbb{Z}_5$ polynomial. As expected, we see two disconnected components.}
    \label{c3z5 mahler expansion}
     \end{figure}

From Figure \ref{c3z5 mahler expansion}, we see that in each component, the Mahler measure increases monotonically along the respective Mahler flows. As we increase the value of $k_1$ and/or $k_2$, the plot tends to $\max (\log k_1,\log k_2)$. At large values of $k_1$, the plot therefore looks like $\log k_1$ as we move parallel to the $k_1$-axis and likewise for $k_2$. We can explicitly check this monotonic increase by using the definition of the Mahler flow, differentiating the above equations:

\begin{equation}\label{eq.31}
    \frac{\partial m_1(P)}{\partial \log k_1} = 1 + \sum_{n=1}^{\infty}\sum_{i=0}^{n}\binom{n}{i}\binom{n-i}{\frac{n-i}{2}}\binom{i}{\frac{5i-3n}{4}}\frac{k_2^{\frac{5i-3n}{4}}}{k_1^n}.
\end{equation}

As $k_1$ and $k_2$ are always positive, the right hand side of Eq.\eqref{eq.31} is clearly always positive and the Mahler measure always increases. This is not necessarily true while travelling along the perpendicular direction (on the same component of the surface). In this case, we obtain

\begin{equation}\label{eq.32}
 \frac{\partial m_1(P)}{\partial \log k_2} = -\sum_{n=1}^{\infty}\sum_{i=0}^{n}\binom{n}{i}\binom{n-i}{\frac{n-i}{2}}\binom{i}{\frac{5i-3n}{4}}\frac{5i-3n}{4}\frac{k_2^{\frac{5i-3n}{4}}}{k_1^{n}}.
\end{equation}

One of the conditions for the third binomial coefficient in Eq.\eqref{eq.32} to be defined is that $(5i-3n)\geq 0$. Since all other terms are also necessarily positive, this derivative is negative. As we travel along a path of constant $k_1$ within the $k_1$ component, the Mahler measure is therefore always decreasing. We can see this behaviour in the orange surface in Figure \ref{c3z5 mahler expansion}.

Although we observe the same behaviour for the $k_2$ section of the plot in Figure \ref{c3z5 mahler expansion}, it is not as immediately obvious from the derivatives. First examining the behaviour along the Mahler flow as defined above, we get:

\begin{equation}\label{eq.33}
    \frac{\partial m_2(P)}{\partial \log k_2} = 1 + \sum_{n=1}^{\infty}\sum_{i=0}^{n}\binom{n}{i}\binom{i}{\frac{i}{2}}\binom{n-i}{\frac{3i-2n}{2}}\frac{k_1^{\frac{4n-5i}{2}}(-1)^{\frac{5i-2n}{2}}}{k_2^n}.
\end{equation}

\begin{figure}[!ht]
    \centering
     \includegraphics[width=.7\linewidth]{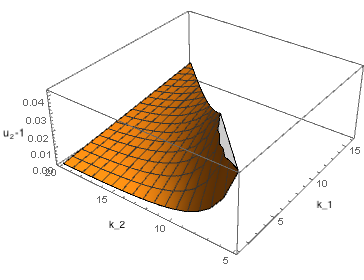}
      \caption{Derivative of the second term in the $k_2$ section of the $\mathbb{C}^3/\mathbb{Z}_5$ expansion.}
    \label{c3z5 decreasing derivative}
     \end{figure}

The factor of $-1$ means that we will have some negative terms in the expansion in Eq.\eqref{eq.33}. In order to have a monotonically increasing Mahler measure, the second term on the right hand side of Eq.\eqref{eq.33} must be greater than $-1$ for all values of $(k_1,k_2)$ within the blue region in Figure \ref{c3z5 mahler expansion}, i.e., for all values of $(k_1,k_2)$ which satisfy $k_1<k_2-4$. Specifically, we plot this second term for these values of $(k_1,k_2)$ as in Figure \ref{c3z5 decreasing derivative}. We see a decrease in the size of the term as we move along $k_1$, but it never goes below zero. For the values mentioned above, the sum over $n$ will always converge. This corresponds to the value of each consecutive term decreasing. As we move along $k_2$, we decrease the size of each term, causing the sum to converge to a smaller number. As $k_2\rightarrow\infty$, this term tends to zero, and the derivative tends to 1, as expected. Moving along $k_1$ also decreases the Mahler measure, though the gradient is much less than its equivalent in the $k_1$ section. This is again expected, as all terms in the $k_1$ section are negative, while the sign of the terms in the $k_2$ section alternate. This gradient is given by:

\begin{equation}
     \frac{\partial m_2(P)}{\partial \log k_1} = -\sum_{n=1}^{\infty}\sum_{i=0}^{n}\binom{n}{i}\binom{i}{\frac{i}{2}}\binom{n-i}{\frac{3i-2n}{2}}\frac{4n-5i}{2}\frac{k_1^{\frac{4n-5i}{2}}(-1)^{\frac{5i-2n}{2}}}{k_2^nn}.
    \end{equation}\par

\subsection{Numerical analysis}
\begin{figure}[ht!]
\begin{minipage}{0.45\textwidth}
    \centering
     \includegraphics[width=.9\linewidth]{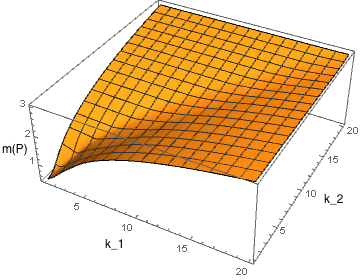}
     \end{minipage}\hfill
     \begin{minipage}{0.45\textwidth}
     \includegraphics[width=.7\linewidth]{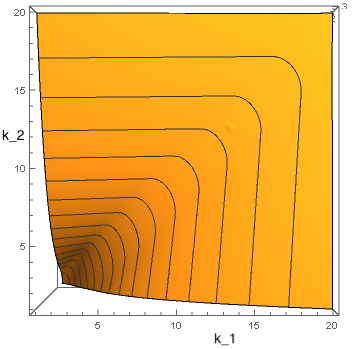}
     \end{minipage}
     \caption{The Mahler measure of the $\mathbb{C}^3/\mathbb{Z}_5$ polynomial calculated numerically.}
    \label{c3z5 mahler numeric}
     \end{figure}

Although we cannot obtain a similar expression for the Mahler measure when $|k_1-k_2|\leq \max|p'(z,w)|$ using the expansion method, we can resort to direct numerical integration to obtain values. Specifically, results from numerical integration in the case of $\mathbb{C}^3/\mathbb{Z}_5$ are plotted in Figure \ref{c3z5 mahler numeric}. Polynomials whose measure can not be calculated using the expansion method will have poles for certain values of $(z,w)$, which means we will have to integrate over singularities. Nevertheless, results are still accurate to at least 5 decimal places when tested against known exact results, such as those found in \cite{smyth_2008} and results found using the expansion method above. These singularities correspond to instances when the origin lies within the interior of the related amoeba. In general, shorter computation time is required for numerical integration for polynomials with many terms than using the expansion method above. \par

\subsection{Summary of results for non-reflexive Mahler measure}

We repeated this analysis for more non-reflexive polytopes and obtained expressions for their expansions for large $k_1$ and $k_2$, which are summarised in Table \ref{mahler_results_table} for clarity. We also plotted the Mahler measure numerically in each case. Although we only performed these expansions around interior points, similar expressions can be obtained when the polynomial is lopsided in favour of points lying on the polytope edges. As we can see from the plots in Table \ref{mahler_results_table}, the numerical and expansion methods give the same results wherever the expansion is defined. Within each section for the expansion plots, the Mahler measure increases monotonically along the Mahler flow, but may decrease or increase when moving perpendicular to it. 
This variability is particularly visible in the $k_2$ section, where the expansion series alternates its sign. In the $k_1$ section, we do not see this as there is no factor of $-1$, and all terms in the expansion are negative. This results in a decreasing measure.

  \begin{table}[ht!]
         \centering
         \begin{tabular}{|c|c|c|}
         \hline
            Toric diagram & Mahler measure (Expansion) & Mahler measure (Numerical)\\
              \hline
 \raisebox{-.5\height}{ \includegraphics[width=40mm,height=30mm]{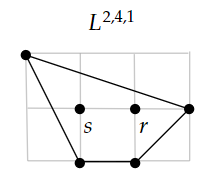}}
&  \raisebox{-.5\height}{ \includegraphics[width=30mm,height=30mm]{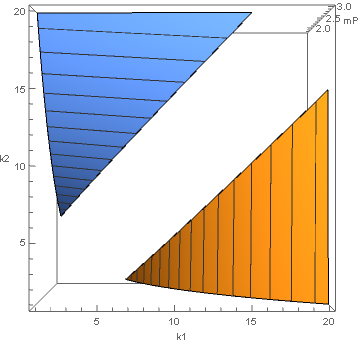}}&  \raisebox{-.5\height}{ \includegraphics[width=30mm,height=30mm]{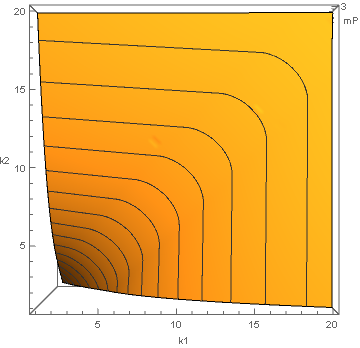}}\\
             \hline
              \multicolumn{3}{|c|}{$m_1\left(P_s(z,w)\right) = \log k_1 - \sum\limits_{n=1}^{\infty}\sum\limits_{i=0}^{n}\sum\limits_{l=0}^{n-i}\binom{n}{i}\binom{i}{\frac{i}{2}}\binom{n-i}{l}\binom{\frac{i}{2}}{\frac{5i+2l-4n}{2}}\frac{k_2^{l}}{k_1^nn}$}\\[3ex]
            \multicolumn{3}{|c|}{$m_2\left(P_s(z,w)\right) = \log k_2 - \sum\limits_{n=1}^{\infty}\sum\limits_{i=0}^{n}\sum\limits_{l=0}^{n-i}\binom{n}{i}\binom{i}{\frac{i}{2}}\binom{n-i}{l}\binom{\frac{i}{2}}{n-2l-2i}\frac{k_1^{l}(-1)^{n-2i-l}}{k_2^nn}$}\\[3ex]
            \hline
            \multicolumn{3}{c}{}\\[0.3ex]
              \hline
                Toric diagram & Mahler measure (Expansion) & Mahler measure (Numerical)\\
              \hline
             \raisebox{-.5\height}{ \includegraphics[width=40mm,height=30mm]{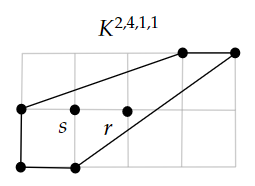}} & \raisebox{-.5\height}{ \includegraphics[width=30mm,height=30mm]{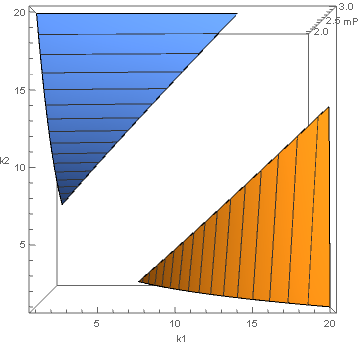}} & \raisebox{-.5\height}{ \includegraphics[width=30mm,height=30mm]{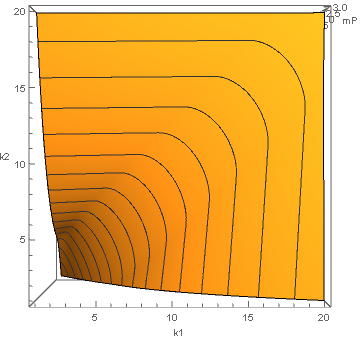}}\\
               \hline
                \multicolumn{3}{|c|}{$m_1\left(P_s(z,w)\right) = \log k_1 - \sum\limits_{n=1}^{\infty}\sum\limits_{i=0}^{n}\sum\limits_{l=0}^{n-i}\sum\limits_{h=0}^{\frac{i}{2}}\binom{n}{i}\binom{n-i}{l}\binom{i}{\frac{i}{2}}\binom{\frac{i}{2}}{n+h-2i-2l}\frac{k_2^{l}}{k_1^nn}$}\\[3ex]
            \multicolumn{3}{|c|}{$m_2\left(P_s(z,w)\right) = \log k_2 - \sum\limits_{n=1}^{\infty}\sum\limits_{i=0}^{n}\sum\limits_{l=0}^{n-i}\sum\limits_{h=0}^{\frac{i}{2}}\binom{n}{i}\binom{i}{\frac{i}{2}}\binom{n-i}{l}\binom{\frac{i}{2}}{h}\binom{\frac{i}{2}}{h+2n-l-2i}\frac{k_1^{l}(-1)^{n-l}}{k_2^nn}$}\\[3ex]
            \hline
             \multicolumn{3}{c}{}\\[0.3ex]
              \hline
                Toric diagram & Mahler measure (Expansion) & Mahler measure (Numerical)\\
              \hline
             \raisebox{-.5\height}{ \includegraphics[width=35
mm,height=30mm]{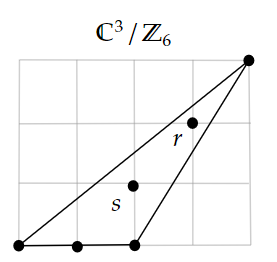}} & \raisebox{-.5\height}{ \includegraphics[width=30mm,height=30mm]{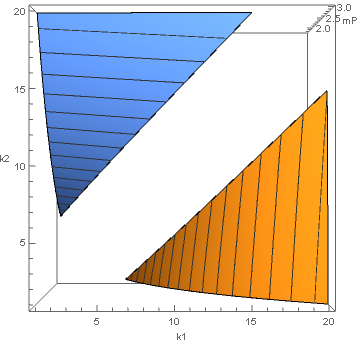}} & \raisebox{-.5\height}{ \includegraphics[width=30mm,height=30mm]{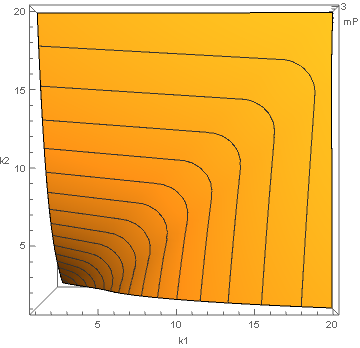}}\\
               \hline
                \multicolumn{3}{|c|}{$m_1\left(P_s(z,w)\right) = \log k_1 - \sum\limits_{n=1}^{\infty}\sum\limits_{i=0}^{n}\sum\limits_{l=0}^{i}\binom{n}{i}\binom{n-i}{2n-3i}\binom{i}{l}\binom{l}{2l-i}\frac{k_2^{2n-3i}}{k_1^nn}$}\\[3ex]
            \multicolumn{3}{|c|}{$m_2\left(P_s(z,w)\right) = \log k_2 - \sum\limits_{n=1}^{\infty}\sum\limits_{i=0}^{n}\sum\limits_{l=0}^{i}\binom{n}{i}\binom{n-i}{\frac{n-3i}{2}}\binom{i}{l}\binom{l}{i-l}\frac{k_1^{\frac{n-3i}{2}}(-1)^{\frac{n+3i}{2}}}{k_2^nn}$}\\[3ex]
            \hline
         \end{tabular}
         \caption{Summary of results for some non-reflexive polytopes. Plots generated by the expansion method and the numerical method are consistent with each other. As we travel along the Mahler flow, the Measure increases monotonically.}
         \label{mahler_results_table}
     \end{table}

\section{ML, Amoeba, and Mahler measure}\label{sec:5}

It is noticed in \cite{Bao:2021fjd} that the changes in the liquid and gas phase contributions to the Mahler measure along the Mahler flow are similar to the changes in the area of the amoeba and the area of the bounded amoeba complement. The conjecture is that given a Newton polynomial \(P(z,w)=k-p(z,w)\), the liquid phase contribution to the Mahler measure \(m_l(P)\) is solely determined by the area of the amoeba and the gas phase contribution \(m_g(P)\) is solely determined by the area of the bounded amoeba complement, i.e., its hole.

The relation between Mahler measure and amoeba is evident via the Ronkin function. The amoeba is the region where the gradient of the Ronkin function is non-linear, whereas the Mahler measure is the Ronkin function evaluated at \((0,0)\). It is possible to use ML to make this relation more precise.

\subsection{Area of the bounded amoeba complement}\label{sec:area_bhc}

Only reflexive polytopes as toric diagrams are considered such that the definition of the gas phase contributions to the Mahler measure is most obvious. Thus, we are only considering a single bounded region for the amoeba. The area of this bounded amoeba complement (the amoeba hole), \(A_h\), is obtained using both sampling and analytic solutions as a crosscheck for each other.

It is possible to sample only the bounded complement of the amoeba using lopsidedness and restricting the sampled region to the bounded region formed by its spines. This bounded region formed by its spines can be determined from Theorem 3.7 in \cite{Bao:2021fjd}.

The analytic boundary of the amoeba is derived by considering the boundary conditions where the gradient of the Ronkin function changes from being linear to being non-linear. This is when the pole in the gradient of the Ronkin function, i.e., \(P(z,w)=0\), within the integration path is independent of the phase angle of \(w=|w|e^{i\theta}\) at constant \(y=\ln|w|\), following the considerations in \cite{Maeda:2006we}.

The areas obtained from both methods agree with each other rather well, so we choose to use the analytic solutions in this section for the ease of computation.

\subsection{Symbolic regression and genetic algorithm}

Symbolic regression is a machine learning technique which allows us to determine the mathematical relationship between the independent variables and the dependent variable targets. Genetic Programming refers to the technique of automated evolution of programs, usually starting from random programs which are progressively evolved using operations analogous to naturally occurring genetic operations. The \texttt{gplearn} package is an implementation of Genetic Programming to perform symbolic regression. It first generates a population of random formulas and then each subsequent population is obtained by performing genetic operations on the fittest individuals from the preceding population. With the help of \texttt{gplearn}, we were able to obtain numerical relations between the area(volume) of the amoeba hole and the coefficients of the Newton polynomial and the numerical relations between the area(volume) of the amoeba hole and gas phase contribution to the Mahler measure.


Specifically, in this section, the genetic algorithm has the following structure in which equations are represented as trees with selected operations from \{addition, subtraction, multiplication, division, negation, square root, logarithm, inverse, absolute value\} applied to variables and constants in the range \((-10,10)\). It begins by initialising with a random population of size 5000. The raw fitness metric, the mean absolute error (MAE) in this case, of the true output values for all input values is calculated for each equation in the population to give a performance loss which is weighted by the complexity of the equation with weight 0.02. Then, the fittest \(0.4\) percent of the population are selected to evolve to successive generation of equations via the genetic operations including performing crossover with probability of 0.85, subtree mutation with probability of 0.02, leave mutation with probability 0.01, and hoist mutation with probability of 0.015. This process is iterated for 100 generations, and equations are selected early if the metric score reaches 0.001.

\subsection{2d Example: \texorpdfstring{\(\mathbb{F}_0=\mathbb{P}^1\times\mathbb{P}^1\)}{F0}}
\begin{figure}[!ht]
   \begin{minipage}{0.3\textwidth}
     \centering
     \includegraphics[width=.7\linewidth]{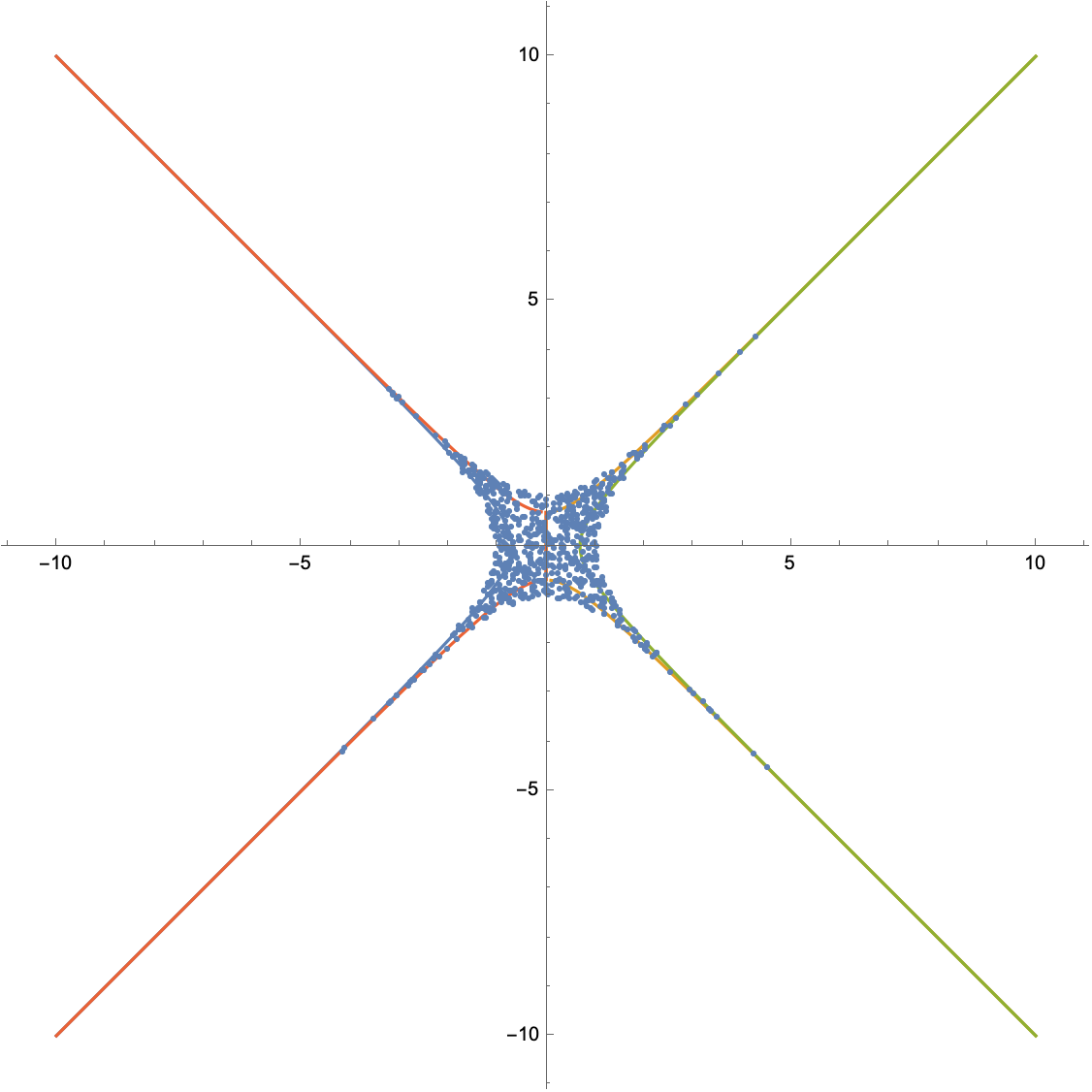}
   \end{minipage}\hfill
   \begin{minipage}{0.3\textwidth}
     \centering
     \includegraphics[width=.8\linewidth]{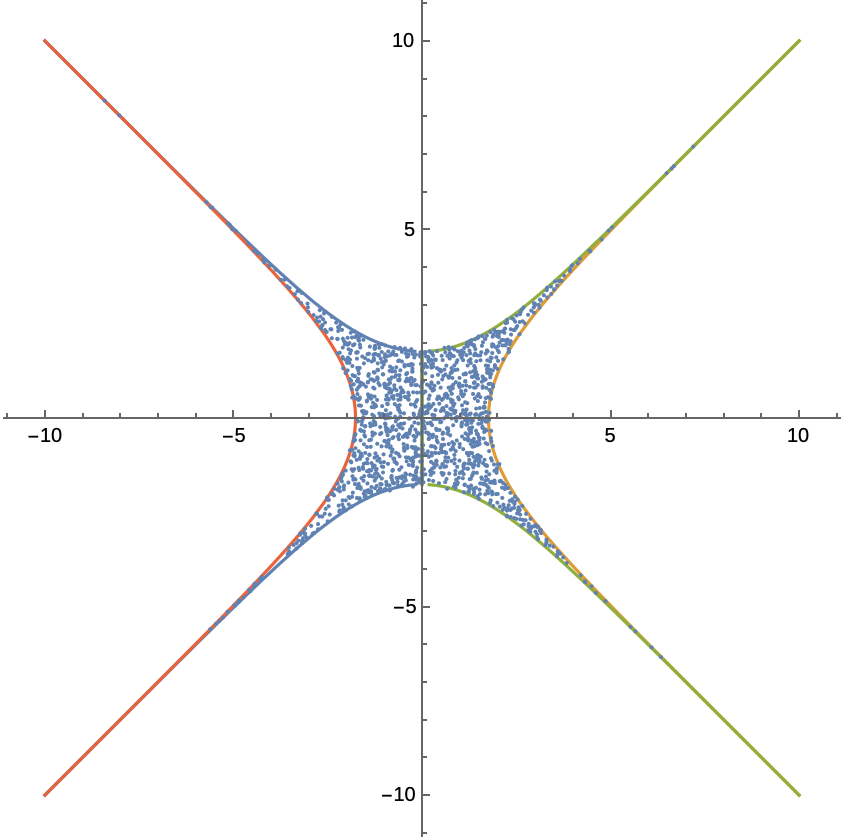}
   \end{minipage}\hfill
   \begin{minipage}{0.3\textwidth}
     \centering
     \includegraphics[width=.8\linewidth]{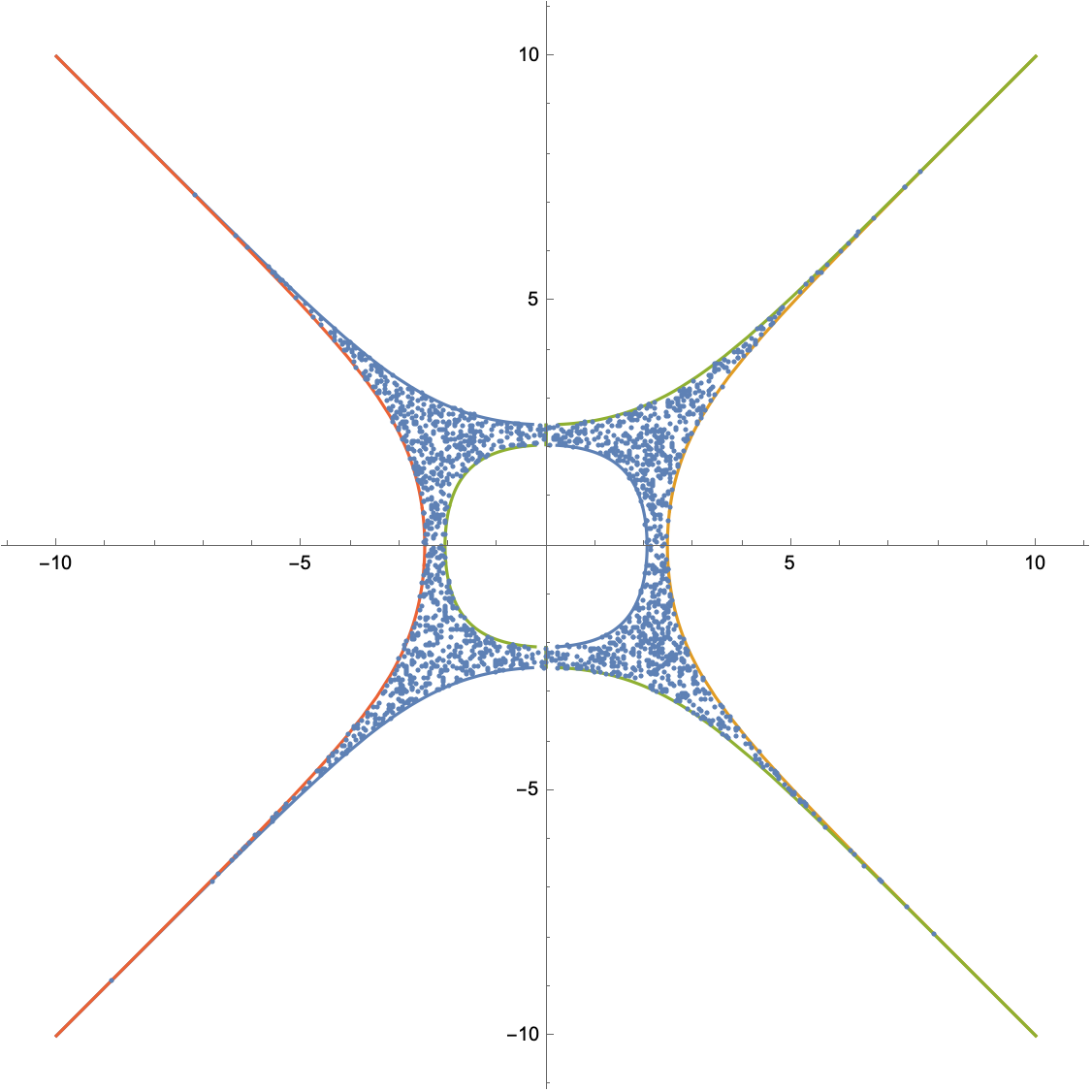}
   \end{minipage}
    \caption{The analytic boundaries of amoebae with \(k=-0.5\) (left), \(k=k_c=4\) (middle) and \(k=10\) (right).}
    \label{f0 boundary}
\end{figure}

The Newton Polynomial in this case is \(P(z,w)=k-z-z^{-1}-w-w^{-1}\). The analytic boundary of the amoeba is found to be
\begin{equation}\label{eq10}
    x=\ln \left(\left|\frac{k}{2}\pm\cosh{y}\pm
   \sqrt{\left(\frac{k}{2}\pm\cosh{y}\right)^2-1}\right|
   \right),
\end{equation}
where \(x=\ln|z|, y=\ln|w|\). The boundaries of amoebae with \(k=-0.5, 4, 10\) are plotted in Figure \ref{f0 boundary}, where $k=4$ is the critical value of $k$ at which the amoeba hole starts to appear. The boundary of the hole agrees well with the sampled boundary of the amoeba. The areas of the hole obtained by sampling and by analytic boundaries agree with each other to at least 2 decimal places depending on the density of points.

\begin{figure}[!ht]
   \begin{minipage}{0.48\textwidth}
     \centering
     \includegraphics[width=.8\linewidth]{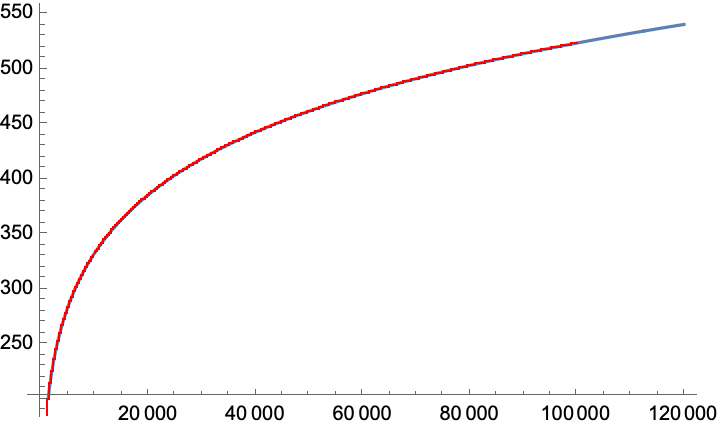}
  \caption{The relation between amoeba hole area $A_h$ and $k$. Original data points are shown red, and the relation found is plotted blue.}
  \label{f0 area vs k}
   \end{minipage}\hfill
   \begin{minipage}{0.48\textwidth}
     \centering
     \includegraphics[width=.8\linewidth]{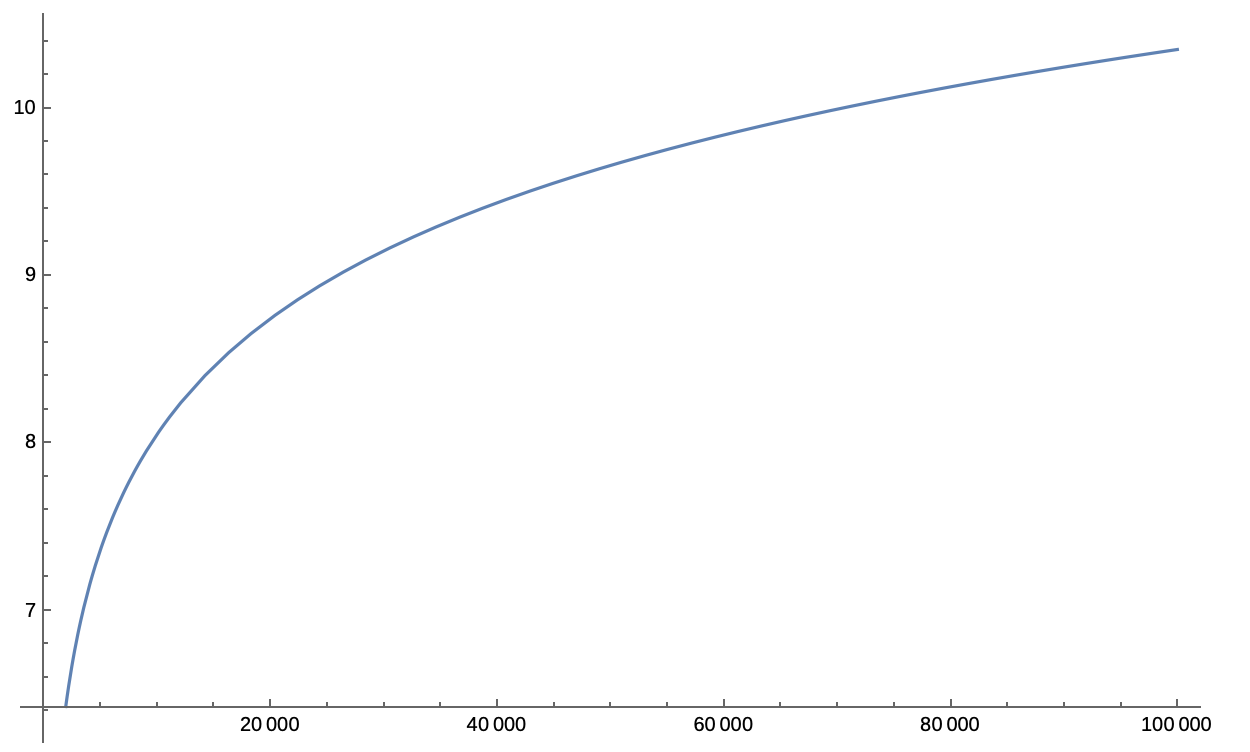}
     \caption{Plot of gas phase contribution to Mahler measure $m_g$ against $k$.}
    \label{f0mgvsk}
   \end{minipage}
\end{figure}

The relation between the amoeba hole area and the value of k is fitted with 7500 pairs of \((k,A_h)\) and is found to be
\begin{equation}\label{eq.36}
    A_h=4\ln^2 k-6.601\,,
\end{equation}
with an \(R^2\) score of 1.0000 and a mean absolute error of 0.0318. In the limit of \(k\rightarrow\infty\), the leading term scales as \(4\ln^2 k\) (as plotted in Figure \ref{f0 area vs k}). This agrees with Conjecture 3.9 in \cite{Bao:2021fjd}, and the power of $\ln k$ is given by the dimension of the amoeba.

\subsubsection{Mahler measure and hole area for \texorpdfstring{\(k\geq 4\)}{kgeq4}}

The Mahler measure for the Newton polynomial \(P(z,w)=k-z-z^{-1}-w-w^{-1}\) is
\begin{equation}
    m(P)=\ln k -2k^{-2}{}_{4}F_{3}\left(1,1,\frac{3}{2},\frac{3}{2};2,2,2;16k^{-2}\right),
\end{equation}
for \(k\geq 4\) \cite{Bao:2021fjd}. The gas phase contribution to the Mahler measure, \(m_g(P)=m(P)-m(P(k=4))\), is plotted in Figure \ref{f0mgvsk} which shows similar trend as in Figure \ref{f0 area vs k}. This suggests a possible direct relationship between \(A_h\) and \(m_g(P)\), as motivated in Section \S\ref{sec:2}.



\begin{figure}[!ht]
  \centering
  \includegraphics[width=.4\linewidth]{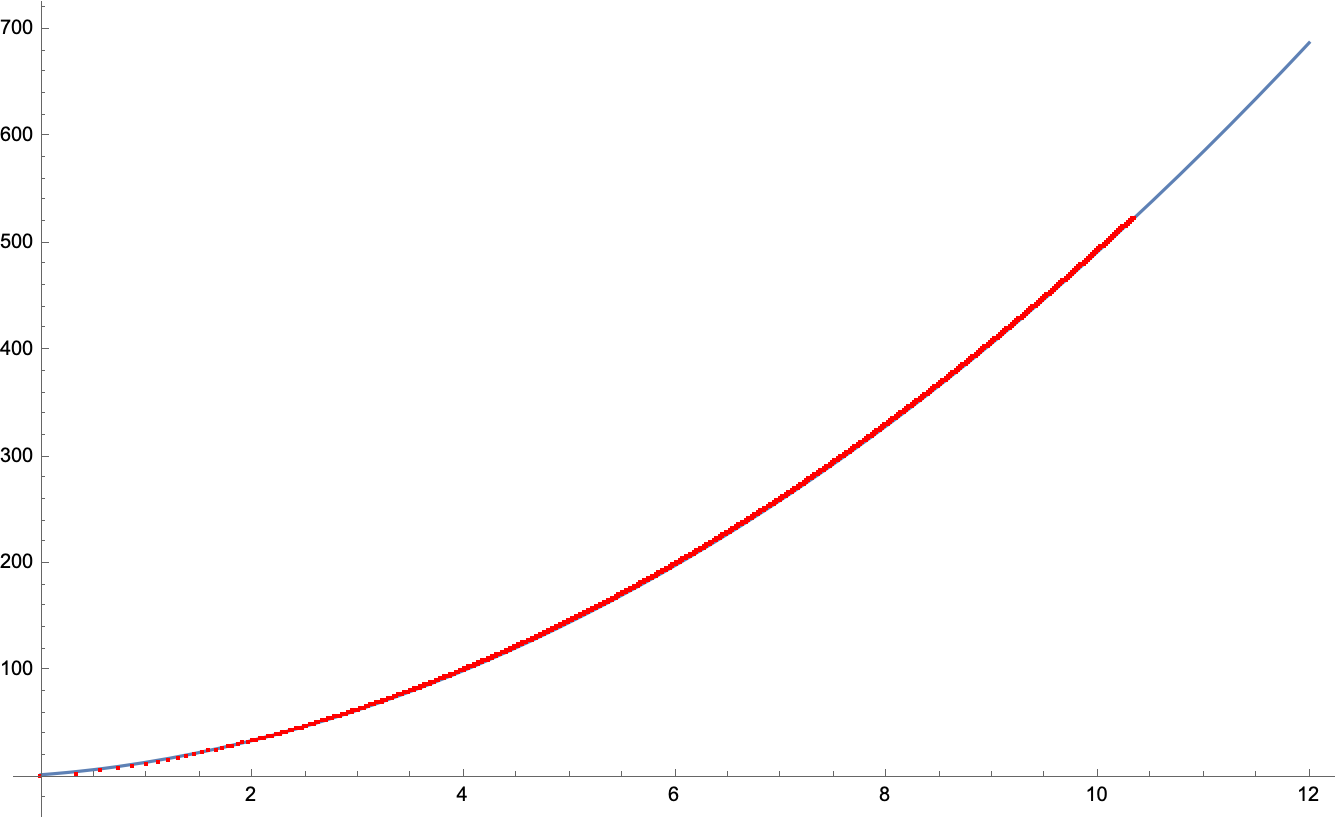}
  \caption{Data points (red) and the fitted relation (blue) between $A_h$ and $m_g$.}
  \label{mg vs a}
\end{figure}

The relation between the gas phase contribution \(m_g(P)\) and the amoeba hole area is fitted with about 50000 data pairs and found to be
\begin{equation}\label{eq.38}
    A_h=3.9804 m_g^2+9.888 m_g -1.5243 \sqrt{m_g}\,,
\end{equation}
with an \(R^2\) score of 1.0000 and a mean absolute error of 0.0249 (plotted in Figure \ref{mg vs a}). In the limit of large $k$ and hence large $m_g$, \(A_h \sim 4m_g^2\) and the leading coefficient in Eq.\eqref{eq.38} is close to the leading coefficient in Eq.\eqref{eq.36}. This is expected as the large $k$ behaviour of the Mahler measure is of $\ln k$.

\subsection{2d Example with more than one parameter: \texorpdfstring{\(Y^{2,2}\)}{Y22}}
\begin{figure}[!ht]
   \begin{minipage}{0.48\textwidth}
     \centering
     \includegraphics[width=.4\linewidth]{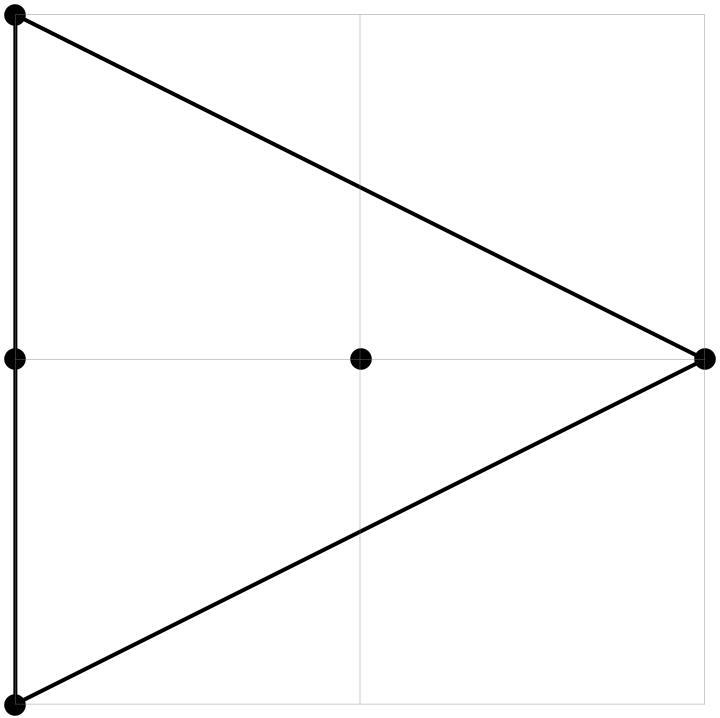}
     \caption{The toric diagram associated with \(\mathrm{Y^{2,2}}\)\,.}
     \label{y22 toric}
   \end{minipage}\hfill
   \begin{minipage}{0.48\textwidth}
     \centering
     \includegraphics[width=.4\linewidth]{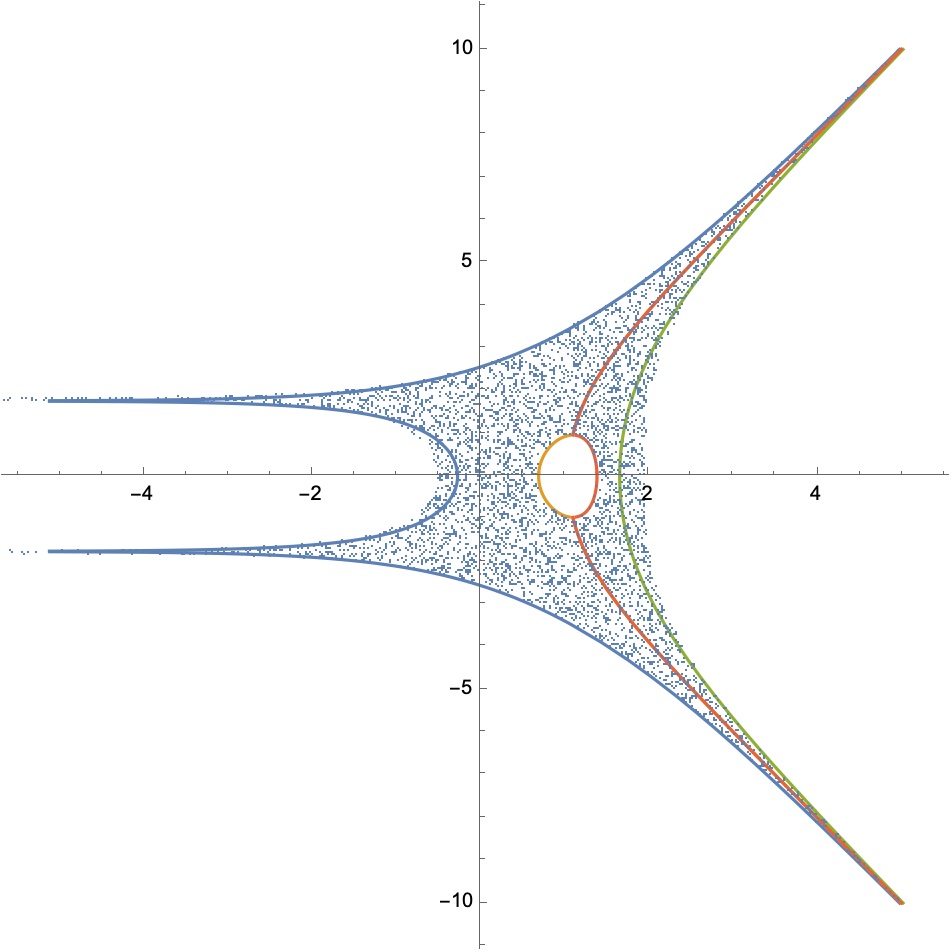}
     \caption{The boundary curves of the amoeba.}
     \label{y22 boundary}
   \end{minipage}
\end{figure}

We also considered an example which has more than one coefficient of the Newton polynomial that is not constant. Specifically, we looked at the surface \(Y^{2,2}\) with the associated Newton polynomial \(P(z,w)=z^2+bz+k+r(w+w^{-1})\) and coefficients \((b,k,r)\). Its toric diagram is given in Figure \ref{y22 toric}. The physical interpretation of the coefficients can be found in \cite{Maeda:2006we}.

The analytic boundaries of the amoeba can be determined using the same method as before, and they are given by
\begin{equation}
    x=\ln \left|-\frac{b}{2}\pm\sqrt{\frac{b^2}{4}-\left(k\pm 2r\cosh{y}\right)}\right|,
\end{equation}
where \(x=\ln|z|, y=\ln|w|\). The boundary of the amoeba hole agrees with the sampled boundary of the \(n=1\) lopsided amoeba, as shown in Figure \ref{y22 boundary}.

The relation between the coefficients \(b,k,r\) and the amoeba hole area is fitted with 40328 pairs of \((\{b,k,r\},A_h)\) where \(0<b,k,r\leq 40\) and \(A_h \neq 0\), and it is found to be
\begin{equation}
    A_h=\sqrt{\bigg(b + 2.4142\sqrt{r} - r\bigg) \bigg(\frac{b}{\sqrt{r}} - \ln k\bigg)}
\end{equation}
with an \(R^2\) score of 0.9519 and a mean absolute error of 1.0478. The area of the amoeba hole does not scale as \(\ln^2 k\) in the large \(k\) limit at constant \(b\) and \(r\), and it is most significantly affected by the value of \(b\) instead of \(k\).

Moreover, the gas phase contribution to the Mahler measure cannot be analogously defined here because the Mahler measure takes different values for coefficients that give the same hole area. An example is given in Table \ref{tab:area and mahler for b,k,r}.

\begin{table}[!ht]
    \centering
    \addtolength{\leftskip} {-2cm}
    \addtolength{\rightskip}{-2cm}
    \begin{tabular}{|>{\centering\arraybackslash}m{4em}||>{\centering\arraybackslash}m{5em}|>{\centering\arraybackslash}m{5em}|>{\centering\arraybackslash}m{5em}|>{\centering\arraybackslash}m{5em}|>{\centering\arraybackslash}m{5em}|>{\centering\arraybackslash}m{5em}|}
    \hline
    \((b,k,r)\)&\((4, 1, 1)\) & \((8, 4, 4)\)& \((12, 9, 9)\)&\((16, 16,
   16)\) & \((20, 25, 25)\) &\((24, 36, 36)\)\\
   \hline
   \(A_h\)&1.63644 & 1.63644 & 1.63644 & 1.63644 & 1.63644 & 1.63644\\
   \hline
   \(m(P)\)&1.43518 & 2.17779 & 2.7313 & 3.15535 & 3.51961 & 3.8322\\
   \hline
    \end{tabular}
    \caption{Example of numerical values of the Mahler measure for different sets of coefficients \((b,k,r)\) with the same amoeba hole area.}
    \label{tab:area and mahler for b,k,r}
\end{table}


\subsection{Summary of 2d results}
The 2d compact Fano varieties considered in this work, along with the respective Newton polynomials used and toric diagrams, are given in Table \ref{tab:2d examples}. The amoebae and Mahler measure information for each are respectively collected in Table \ref{tab:2d results} for ease of comparison; along with the symbolic regression results in Table \ref{tab:2d results fitting}. Another detailed example is given in Appendix \ref{appendix2}.

\begin{table}[!ht]
    \centering
    \addtolength{\leftskip} {-2cm}
    \addtolength{\rightskip}{-2cm}
    \begin{tabular}{|>{\centering\arraybackslash}m{4em}|>{\centering\arraybackslash}m{17em}|>{\centering\arraybackslash}m{10em}|}
    \hline
    Surface & Newton Polynomial & Toric Diagram\\
    \hline
    \(\mathbb{F}_0\)& \(P(z,w)=k-(z+z^{-1}+w+w^{-1})\)& \raisebox{-.5\height}{ \includegraphics[width=20mm,height=20mm]{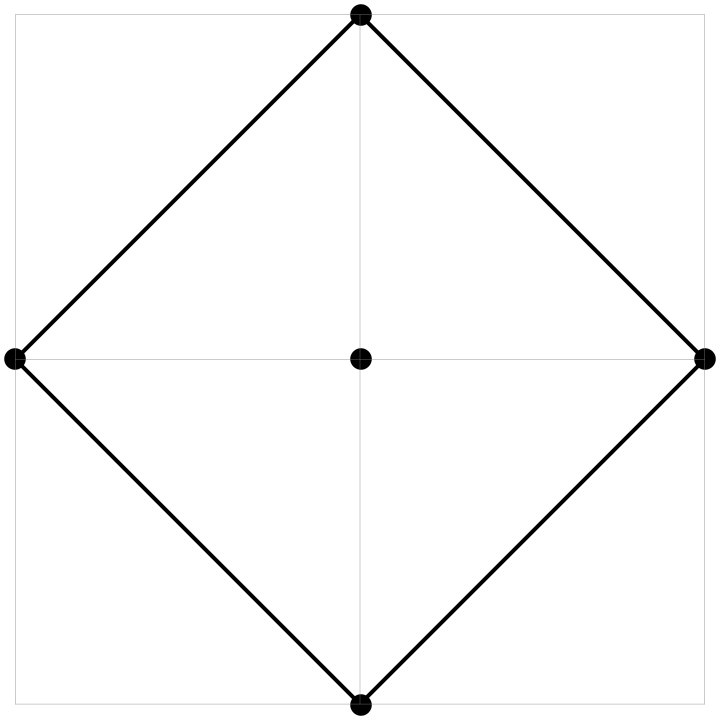}}\\
    \hline
    \(\mathbb{P}^2 (\mathrm{dP_0})\)& \(P(z,w)=k-(z+w+z^{-1}w^{-1})\)& \raisebox{-.5\height}{ \includegraphics[width=20mm,height=20mm]{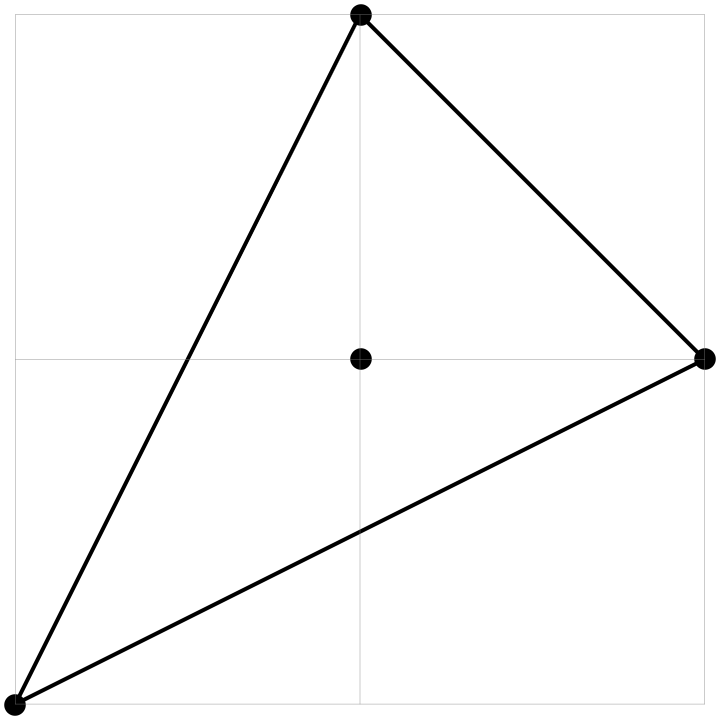}}\\
    \hline
    \(\mathrm{dP_1}\)& \(P(z,w)=k-(z+w+w^{-1}+z^{-1}w^{-1})\)& \raisebox{-.5\height}{ \includegraphics[width=20mm,height=20mm]{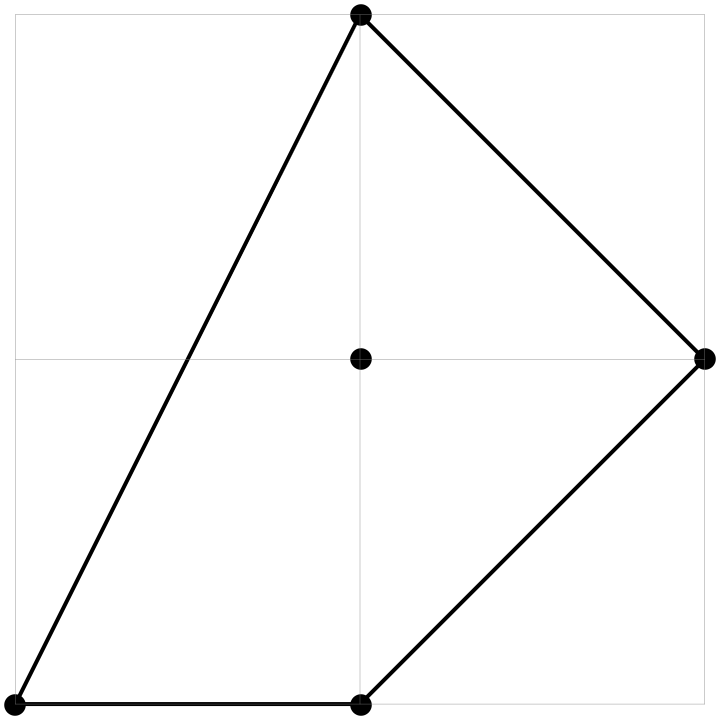}}\\
    \hline
    \(\mathrm{dP_2}\)& \(P(z,w)=k-(z+z^{-1}+w+w^{-1}+z^{-1}w^{-1})\)& \raisebox{-.5\height}{ \includegraphics[width=20mm,height=20mm]{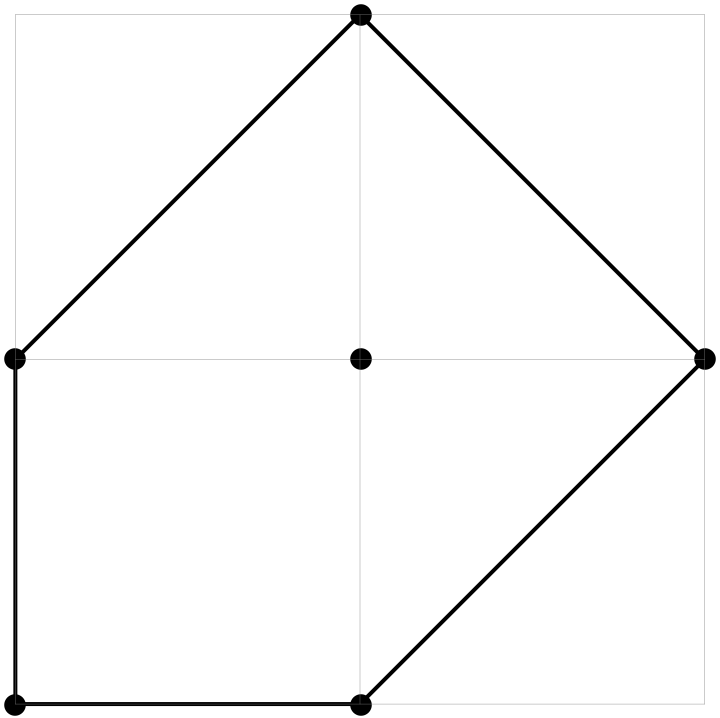}}\\
    \hline
    \(\mathrm{dP_3}\)& \(P(z,w)=k-(z+z^{-1}+w+w^{-1}+zw^{-1}+z^{-1}w)\)& \raisebox{-.5\height}{ \includegraphics[width=20mm,height=20mm]{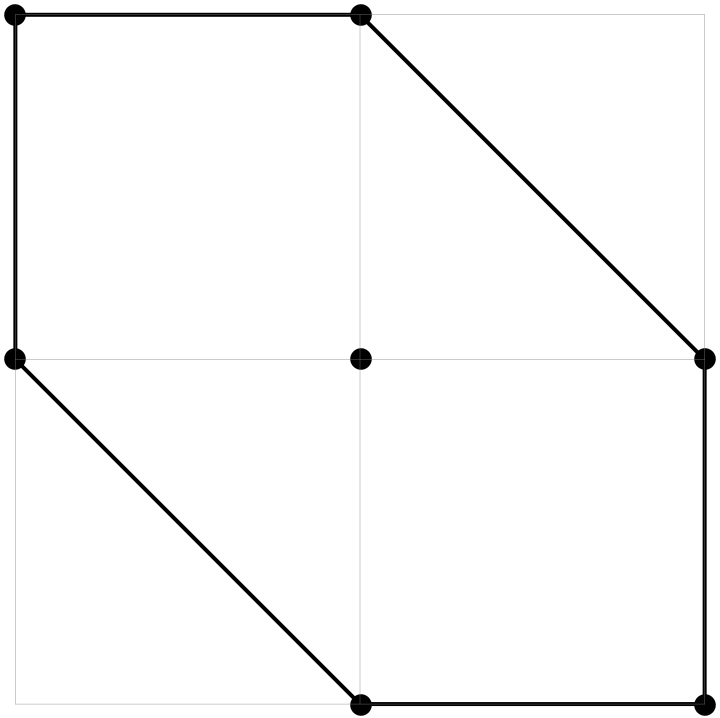}}\\
    \hline
    \end{tabular}
    \caption{Examples of toric surfaces, each with an associated specific Newton polynomial and the respective toric diagram.}
    \label{tab:2d examples}
\end{table}

\begin{sidewaystable}
    \centering
\begin{tabularx}{\textwidth}{|>{\centering\arraybackslash}m{2em}||>{\centering\arraybackslash}m{24em}|>{\centering\arraybackslash}m{10em}|>{\centering\arraybackslash}m{23em}|}
\cline{1-4}
\(\mathcal{V}\) & Amoeba Boundary & Example & Mahler measure\\
    \cline{1-4}
\(\mathbb{F}_0\) & \begin{equation*}
    x=\ln \left(\left|\frac{k}{2}\pm\cosh{y}\pm
   \sqrt{\left(\frac{k}{2}\pm\cosh{y}\right)^2-1}\right|
   \right)
\end{equation*}& \raisebox{-.5\height}{ \includegraphics[width=28mm,height=28mm]{Fig/F0boundaryk=10.png}}
   & \begin{equation*}
    m(P)=\ln k -2k^{-2}{}_{4}F_{3}\left(1,1,\frac{3}{2},\frac{3}{2};2,2,2;16k^{-2}\right)
\end{equation*}\\
\cline{1-4} 
 \(\mathbb{P}^2\) \((\mathrm{dP_0})\) & \begin{equation*}
     x= \ln \left(\left| \frac{k}{2} \pm\frac{e^y}{2}\pm\sqrt{\frac{1}{4} \left(k\pm e^y\right)^2\pm e^{-y}}\right| \right)
 \end{equation*} & \raisebox{-.5\height}{ \includegraphics[width=28mm,height=28mm]{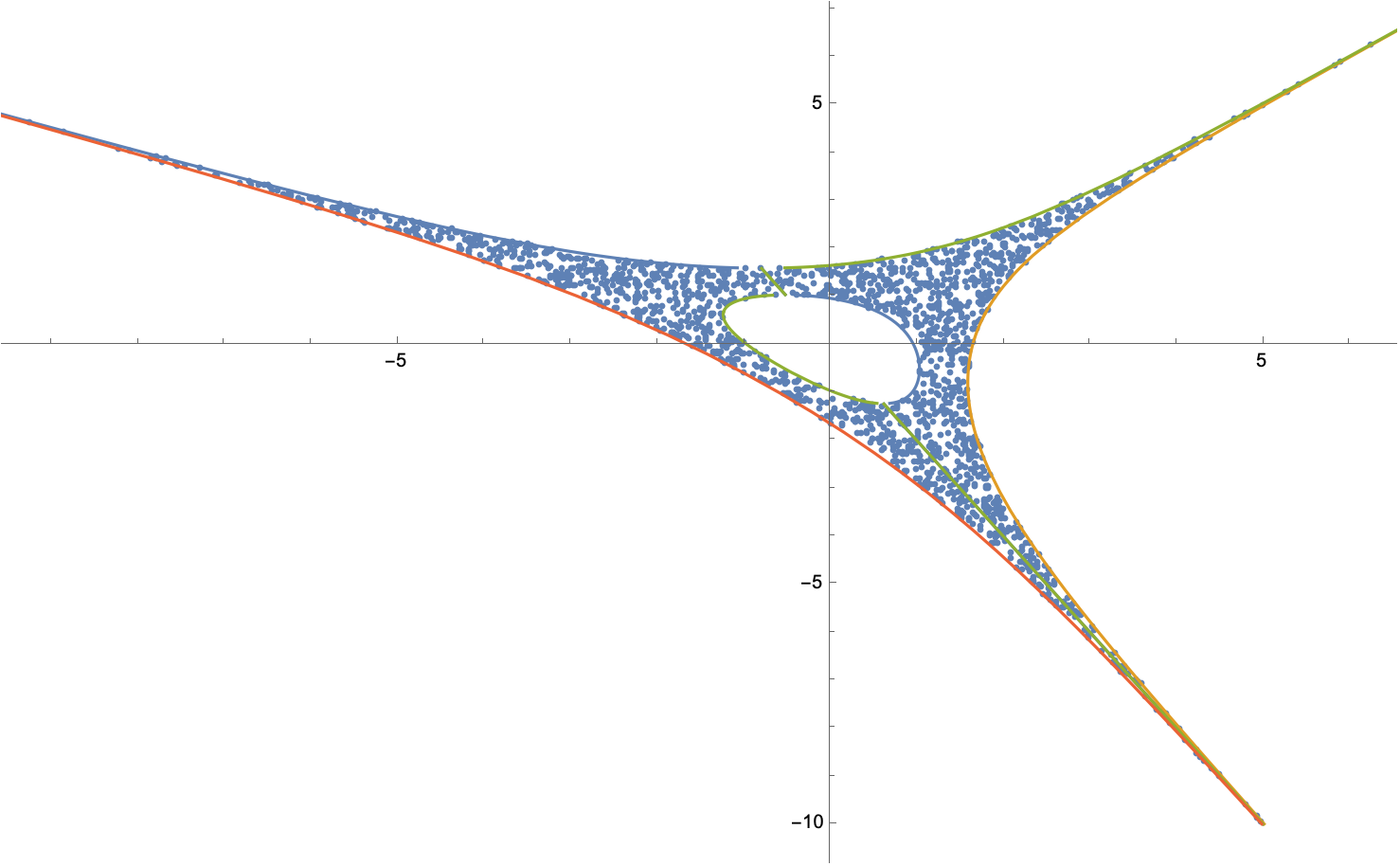}} & \begin{equation*}
     m(P)=\ln k -2k^{-3} {}_{4}F_{3}\left(1,1,\frac{4}{3},\frac{5}{3};2,2,2;27k^{-3}\right)
 \end{equation*}\\
 \cline{1-4}
 \(\mathrm{dP_1}\)&\begin{equation*}
     x=\ln\left(\left| \frac{k}{2} \pm\cosh{y}\pm\sqrt{ \left(\frac{k}{2} \pm\cosh{y}\right)^2\pm e^{-y}}\right| \right)
 \end{equation*} &\raisebox{-.5\height}{ \includegraphics[width=28mm,height=28mm]{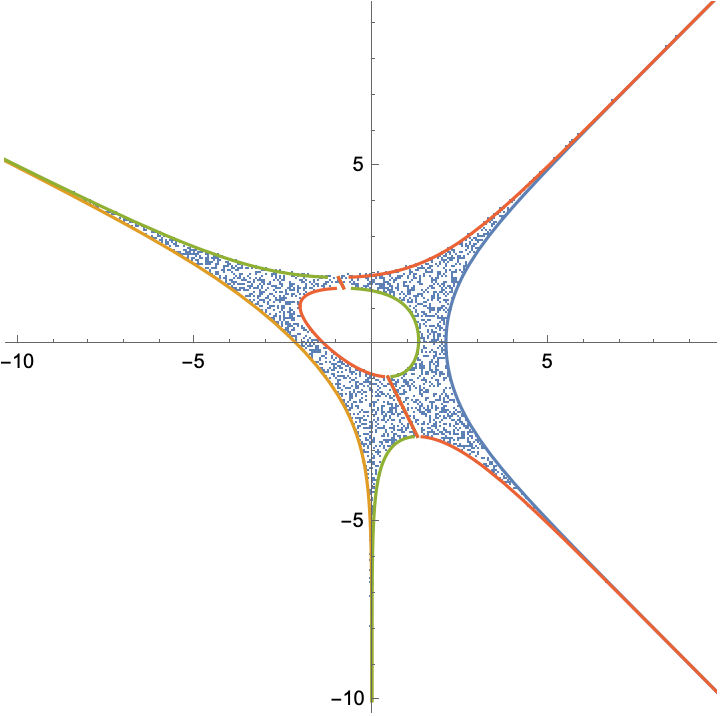}} & \begin{equation*}
     m(P)=\ln k -\sum_{n=1}^{\infty}\sum_{i=0}^{n}\frac{1}{nk^n}{n \choose i}{i\choose\frac{i}{2}}{n-i\choose\frac{2n-i}{4}},
 \end{equation*} \(2n-i \mod 4 =0\) and \(i \mod 2 =0\).\\
 \cline{1-4}
 \(\mathrm{dP_2}\)& \begin{equation*}
    x= \ln \left(\left| \frac{k}{2} \pm\cosh{y}\pm\sqrt{ \left(\frac{k}{2} \pm\cosh{y}\right)^2-(1\pm e^{-y})}\right| \right)
 \end{equation*} & \raisebox{-.5\height}{ \includegraphics[width=28mm,height=28mm]{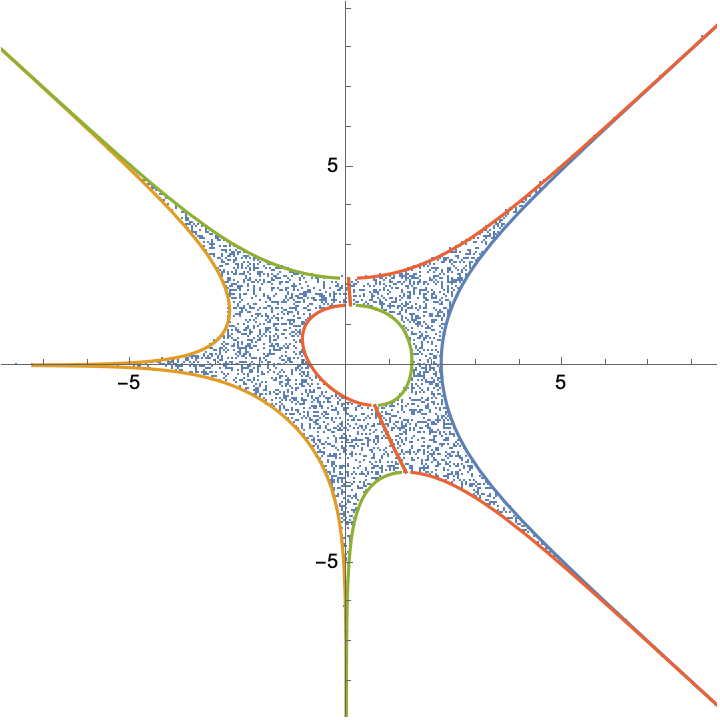}} & \begin{equation*}
    m(P)=\ln k -\sum_{n=1}^{\infty}\sum_{i=0}^{n}\sum_{j=0}^{i}\frac{1}{nk^n}{n \choose i}{n-i\choose\frac{n-i}{2}}{i\choose j}{j\choose\frac{i}{2}},
 \end{equation*} \(n-j \mod 2 =0\), \(i \mod 2=0\), and \(j \geq \frac{i}{2}\).\\
 \cline{1-4}
 \(\mathrm{dP_3}\) & \begin{equation*}\begin{aligned}
    x&= \ln \Bigg(\Big| (1\pm e^{-y})^{-1}\bigg(\frac{k}{2} \pm\cosh{y}\\&\pm\sqrt{ \left(\frac{k}{2} \pm\cosh{y}\right)^2-(2\pm 2\cosh{y})}\bigg)\Big| \Bigg)
    \end{aligned}
 \end{equation*} & \raisebox{-.5\height}{ \includegraphics[width=28mm,height=28mm]{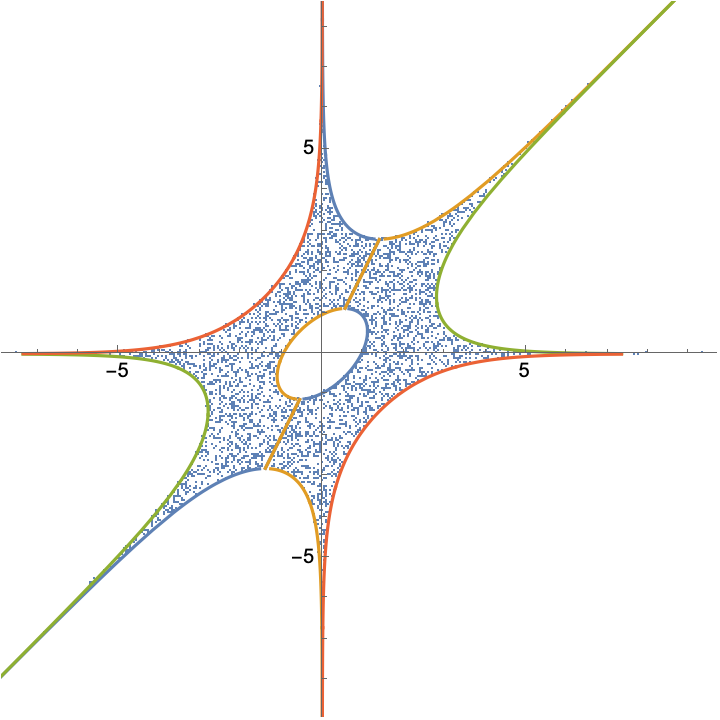}} & \begin{equation*}\begin{aligned}
     m(P)&=\ln k -\sum_{n=1}^{\infty}\sum_{i=0}^{n}\sum_{l=0}^{i}\sum_{j=0}^{n-i}\frac{1}{nk^n}\\&{n \choose i}{i\choose l}{l\choose\frac{l+n-i-2j}{2}}{i-l\choose\frac{2i+2j-l-n}{2}}{n-i\choose j},
     \end{aligned}
 \end{equation*} \(l+n \mod 2=0\), \(i\mod 2=0\), and \(n-i-2j\leq l \leq n-2j\).\\
 \cline{1-4}
\end{tabularx}
\caption{Summary of 2d results for surfaces \(\mathcal{V}\).}
    \label{tab:2d results}
\end{sidewaystable}

\begin{sidewaystable}
    \centering
\begin{tabularx}{\textwidth}{|>{\centering\arraybackslash}m{2em}||>{\centering\arraybackslash}m{19em}|>{\centering\arraybackslash}m{10em}|>{\centering\arraybackslash}m{18em}|>{\centering\arraybackslash}m{10em}|}
    \cline{1-5}
\(\mathcal{V}\) &  Fitted \(A_h(k)\) & Plot \(A_h(k)\) &  Fitted \(A_h(m_g)\) & Plot  \(A_h(m_g)\) \\
    \cline{1-5}
    \(\mathbb{F}_0\) & \begin{equation*}
    A_h=4\ln^2 k-6.601
\end{equation*}
with an \(R^2\) score of 1.0000 and a mean absolute error of 0.0318.  & \raisebox{-.5\height}{ \includegraphics[width=\linewidth]{Fig/kaf0.png}} &\begin{equation*}
    A_h=3.9804 m_g^2+9.888 m_g -1.5243 \sqrt{m_g}
\end{equation*}
with an \(R^2\) score of 1.0000 and a mean absolute error of 0.0249.
& \raisebox{-.5\height}{ \includegraphics[width=\linewidth]{Fig/mgaf0.png}}\\
 \cline{1-5}
  \(\mathbb{P}^2\) \((\mathrm{dP_0})\)&  \begin{equation*}\begin{aligned}
   &A_h=2 \ln^2 k+\ln (k \ln k)-5.49\\&+\ln k^2 \ln ((k-\ln k^2 ) (\ln (k \ln k)-5.49))
    \end{aligned}
\end{equation*}
with an \(R^2\) score of 0.9998 and a mean absolute error of 1.0983. & \raisebox{-.5\height}{ \includegraphics[width=\linewidth]{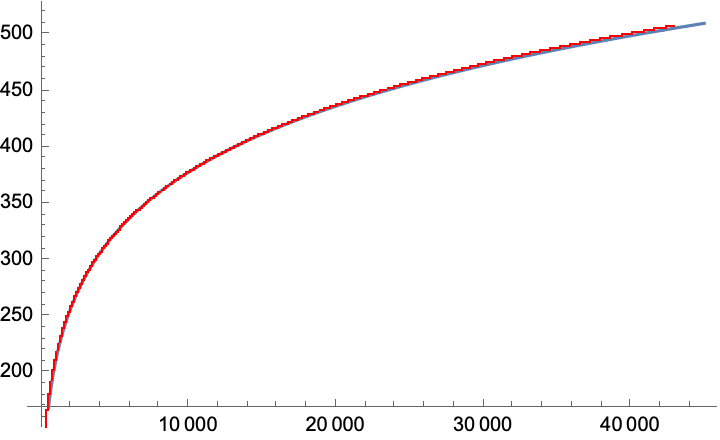}} & \begin{equation*}
    A_h = 4.5904 m_g^2+7.2486m_g+5.0980
\end{equation*}
with an \(R^2\) score of 1.0000 and a mean absolute error of 0.1340.
 & \raisebox{-.5\height}{ \includegraphics[width=\linewidth]{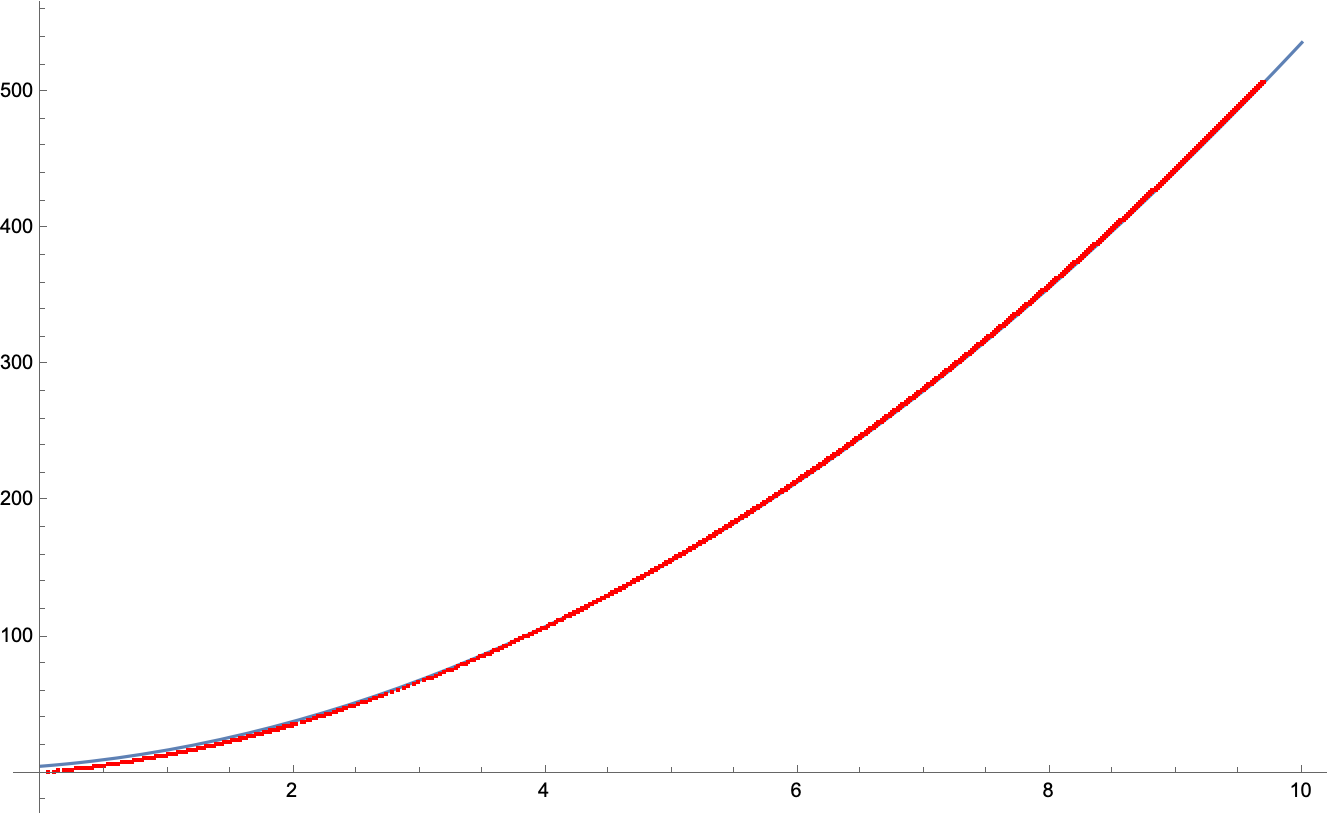}}\\
 \cline{1-5} 
 \(\mathrm{dP_1}\) & \begin{equation*}\begin{aligned}
     A_h&=3.891 \ln(k-9.457)\\&\times\ln(k-3.891\ln(x-9.457)) 
    \end{aligned}
\end{equation*}
with an \(R^2\) score of 0.9994 and a mean absolute error of 0.8515. & \raisebox{-.5\height}{ \includegraphics[width=\linewidth]{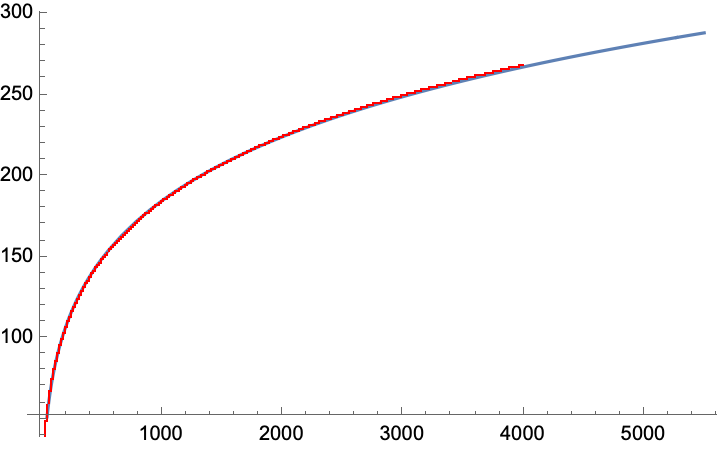}} &\begin{equation*}\begin{split}
    A_h= 4.146m_g^2+8.291m_g+\ln m_g+2.958
    \end{split}
\end{equation*}
with an \(R^2\) score of 1.0000 and a mean absolute error of 0.1514.
 & \raisebox{-.5\height}{ \includegraphics[width=\linewidth]{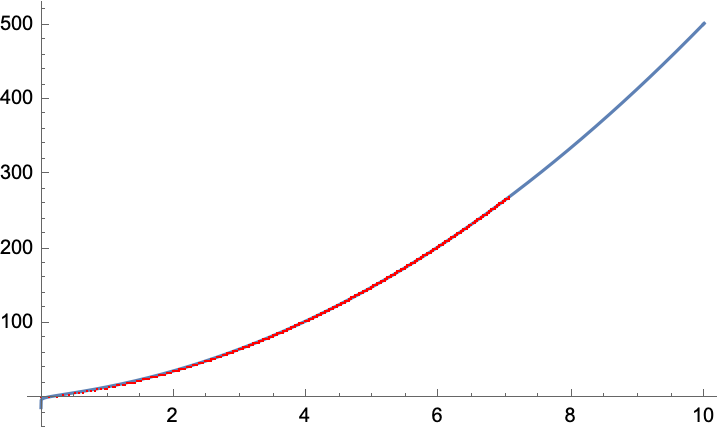}}\\
 \cline{1-5}
 \(\mathrm{dP_2}\) & \begin{equation*}\begin{split}
    A_h=3.891\ln k \ln (0.246k+\frac{k}{\ln (k\ln k)})
    \end{split}
\end{equation*}
with an \(R^2\) score of 0.9998 and a mean absolute error of 0.4549. & \raisebox{-.5\height}{ \includegraphics[width=\linewidth]{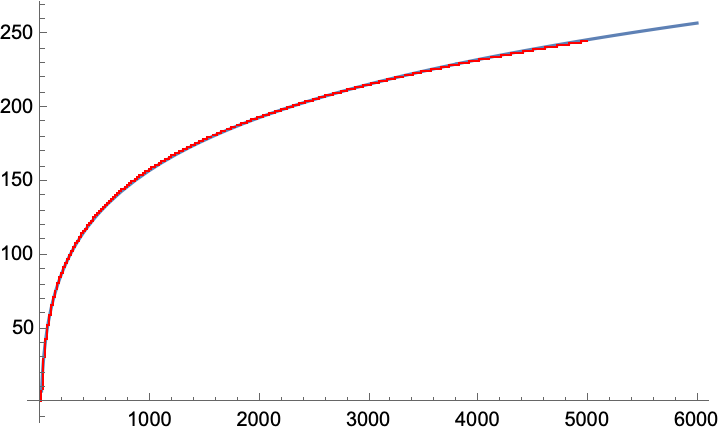}} &\begin{equation*}\begin{aligned}
    A_h&= 5.471m_g^2\sqrt{1-0.194\ln m_g}\\&+3.407m_g+7.294
    \end{aligned}
\end{equation*}
with an \(R^2\) score of 1.0000 and a mean absolute error of 0.1442.
 & \raisebox{-.5\height}{ \includegraphics[width=\linewidth]{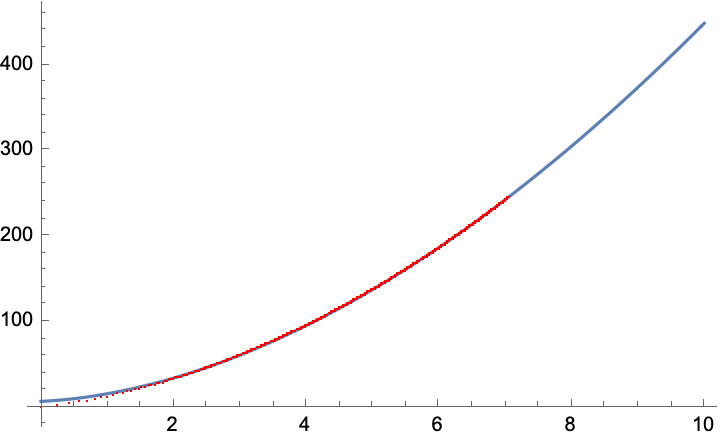}}\\
 \cline{1-5}
 \(\mathrm{dP_3}\) & \begin{equation*}\begin{aligned}
    &A_h=2.783(\ln k +1)\\&\times(\ln k -\frac{1}{\ln(2.510(0.025k+1)^{1/4})})-5.809
    \end{aligned}
\end{equation*}
with an \(R^2\) score of 1.000 and a mean absolute error of 0.2014. & \raisebox{-.5\height}{ \includegraphics[width=\linewidth]{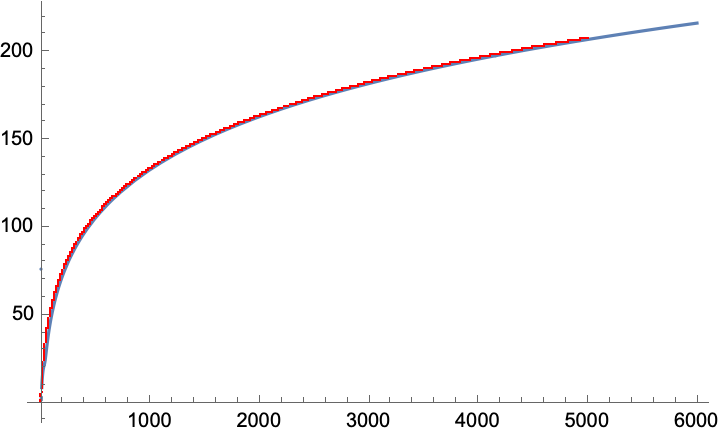}} &\begin{equation*}\begin{split}
    A_h= 3m_g^2+9.623m_g-1.573
    \end{split}
\end{equation*}
with an \(R^2\) score of 1.0000 and a mean absolute error of 0.0618.
 & \raisebox{-.5\height}{ \includegraphics[width=\linewidth]{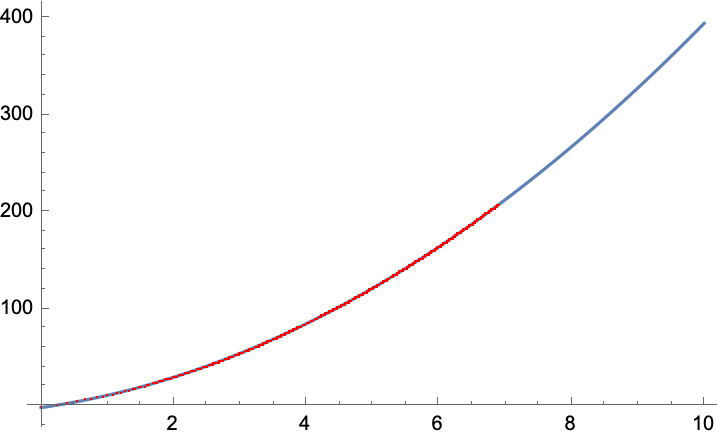}}\\
 \cline{1-5}
\end{tabularx}
\caption{Summary of 2d results for surfaces \(\mathcal{V}\).}
    \label{tab:2d results fitting}
\end{sidewaystable}

\subsection{3d Example: \texorpdfstring{\(\mathbb{P}^1\times\mathbb{P}^1\times\mathbb{P}^1\)}{P1P1P1}}

\begin{figure}[!ht]
   \begin{minipage}{0.48\textwidth}
     \centering
     \includegraphics[width=.6\linewidth]{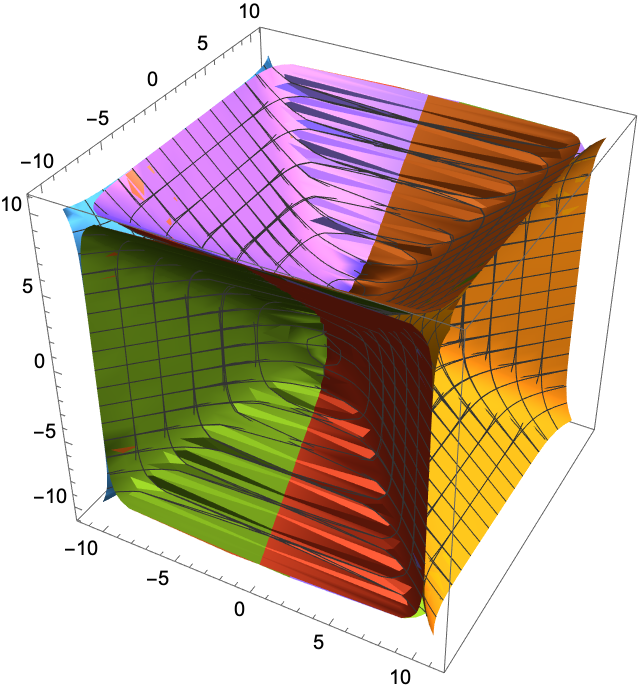}
   \end{minipage}\hfill
   \begin{minipage}{0.48\textwidth}
     \centering
     \includegraphics[width=.6\linewidth]{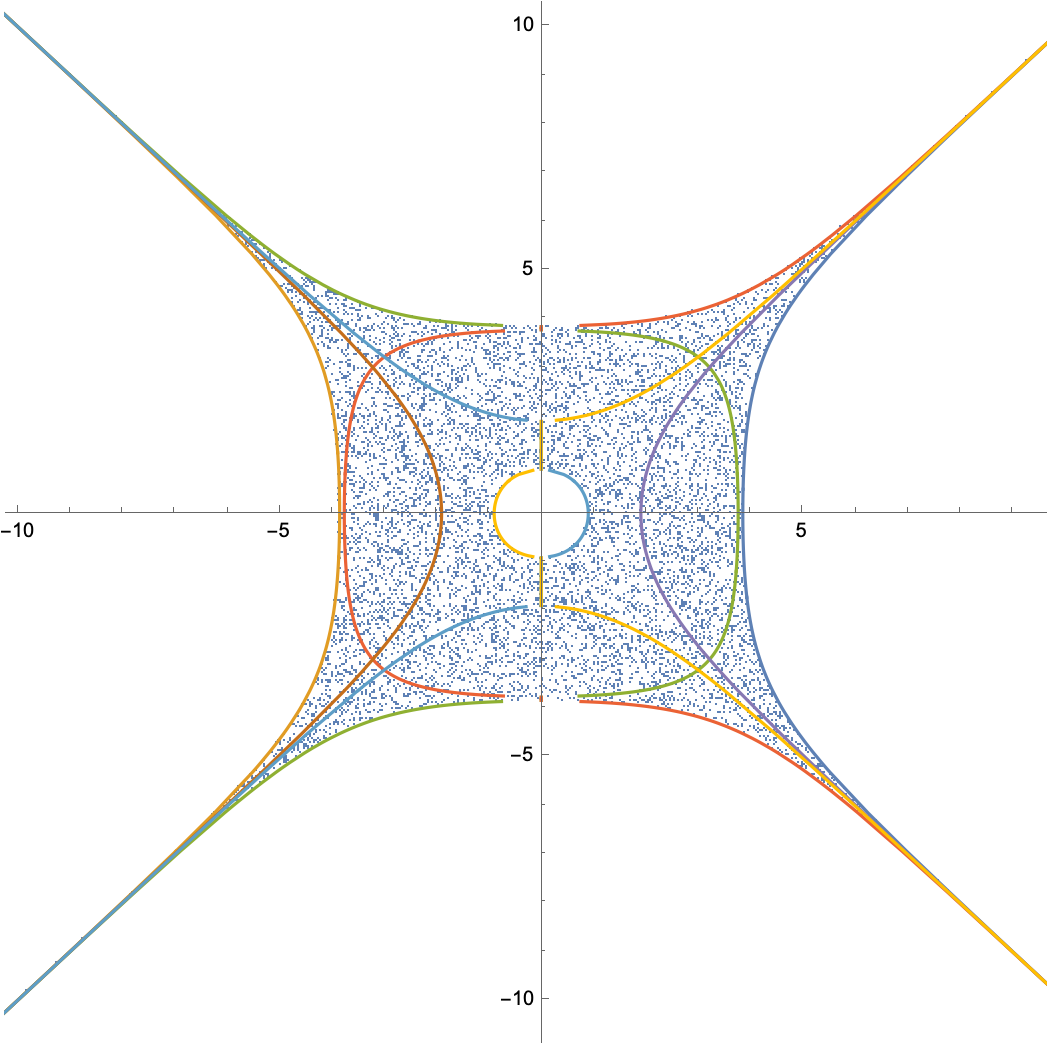}
   \end{minipage}
   \caption{Boundary surfaces of the amoeba corresponding to \(P(u,v,w)=k-u-u^{-1}-v-v^{-1}-w-w^{-1}\) and its cross-section at $z=0$.}
    \label{p1p1p1boundary}
\end{figure}

Similar methods are applied to the 3-dimensional example of \(\mathbb{P}^1\times\mathbb{P}^1\times\mathbb{P}^1\). The analytic expressions for the boundary surfaces of the associated amoeba are given by (Figure \ref{p1p1p1boundary})
\begin{equation}
    x=\ln \left(\left|\frac{| k| }{2}\pm\cosh{y}\pm\cosh{z}\pm\sqrt{\left(\frac{| k| }{2}\pm\cosh{y}\pm\cosh{z}\right)^2-1}\right| \right),
\end{equation}
where \(x=\ln|u|, y=\ln|v|, z=\ln|w|\) and the \(\pm\) sign in front of two \(\cosh \)'s are the same for both \(y\) and \(z\).

In particular, the boundary surfaces of the amoeba hole are formed by
\begin{equation}
    x=\ln \left(\left| \frac{| k| }{2}-\cosh{y}-\cosh{z}\pm\sqrt{\left(\frac{| k| }{2}--\cosh{y}-\cosh{z}\right)^2-1}\right| \right).
\end{equation}

\begin{figure}[!ht]
  \centering
  \includegraphics[width=.4\linewidth]{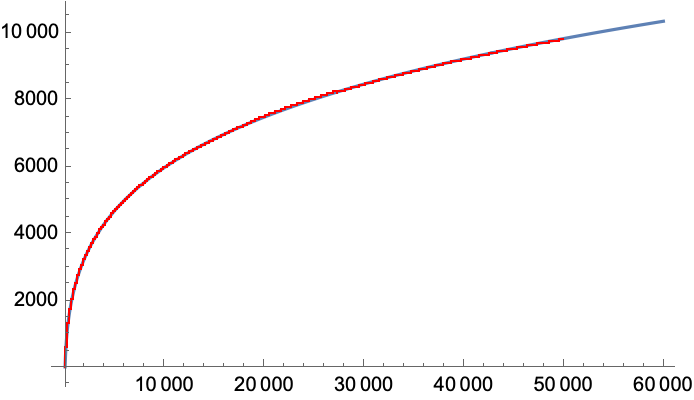}
  \caption{Data points (red) and the fitted relation (blue) for $V_h$ against $k$.}
  \label{k vs v}
\end{figure}

The numerical relation between the volume of the bounded complement of the amoeba, $V_h$, and the value of $k$ is found by fitting 20000 pairs of \((k,V_h)\) values. It is found to be
\begin{equation}
    V_h(k)=\ln (k+\ln (0.4854 k)) \left(8.374 \ln \left(0.4854 k-\ln ^2(0.4854 k)\right)+\ln \left(\frac{7.19}{\ln k}\right)\right) \ln
   (k+\ln k)\,,
\end{equation}
with an \(R^2\) score of 1.0000 and a mean absolute error of 3.2689 (plotted in Figure \ref{k vs v}). In the limit of \(k\rightarrow\infty\), the leading term scales as \(\ln^3 k\), which again agrees with Conjecture 3.9 in \cite{Bao:2021fjd}.

\begin{figure}[!ht]
  \centering
  \includegraphics[width=.4\linewidth]{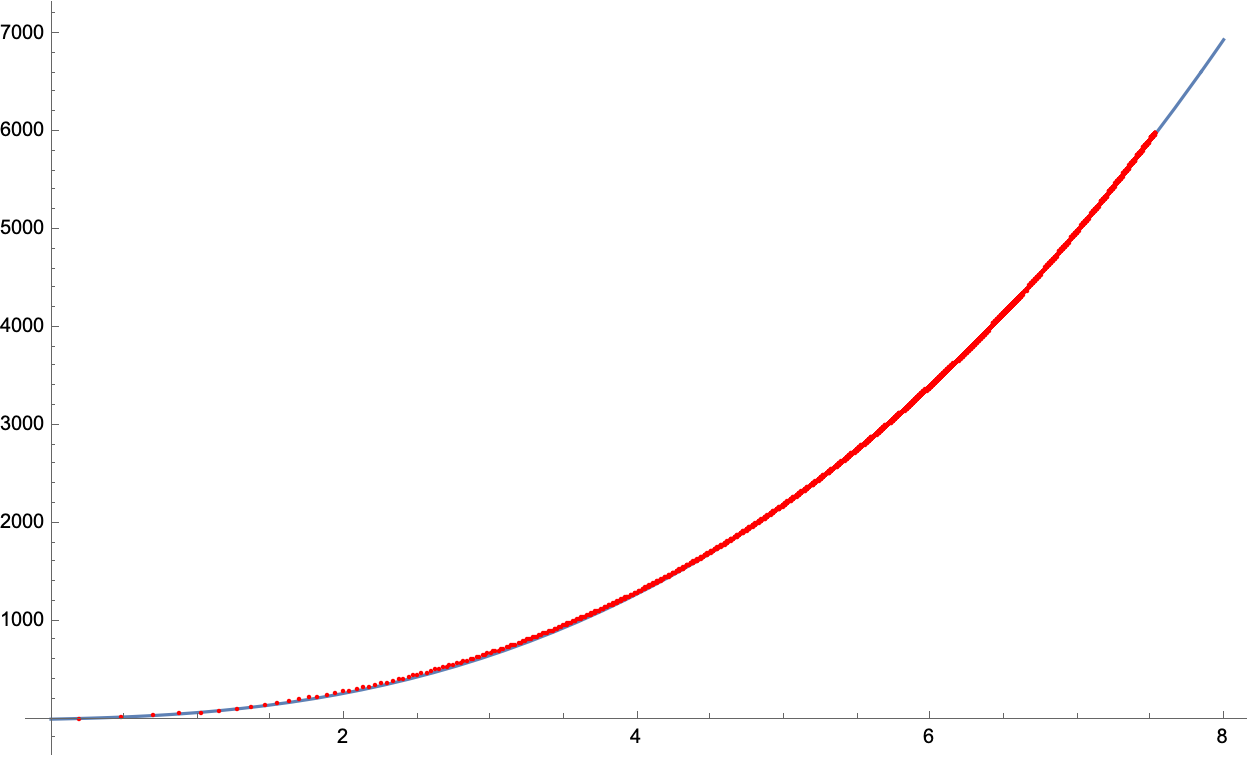}
  \caption{Data points (red) and the fitted relation (blue) between gas phase Mahler measure $m_g$ and amoeba hole volume $V_h$.}
  \label{v vs mg k=6}
\end{figure}

Using Taylor expansion and Cauchy residue theorem, Mahler measure for \(P(u,v,w)=k-u-u^{-1}-v-v^{-1}-w-w^{-1}\) as a function of \(k\) for \(k>6\) is found to be
\begin{equation}\label{eq:27}
    m(P)=\ln k - \sum_{n=1}^\infty\frac{1}{2nk^{2n}}{2n\choose n}\sum_{l=0}^{n}{2l\choose l}{n\choose l}^2.
\end{equation}
If there exists an associated 3-dimensional dimer model that allows a similar interpretation of the Mahler measure in different phases, the gas phase contribution to the Mahler measure \(m_g(P)\) may be analogously defined as \(m_g(P)=m(P)-m(P(k_c))\). For the expansion method used in obtaining Eq.\eqref{eq:27} to be valid, \(k\) must be greater than \(\max_{|u|,|v|,|w|=1} |u+u^{-1}+v+v^{-1}+w+w^{-1}|=6\). Thus, the critical value of \(k\) is \(k_c=6\), which is also the value at which the bounded amoeba complement begins to form. The relation between the volume of the amoeba hole and the analogously defined \(m_g\) is fitted with 5000 pairs of values, and is found to be
\begin{equation}
   V_h=3.835 \left| m_g \left(7.968-\sqrt{m_g} (-9.743 m_g+\ln (m_g-9.664)+\ln (0.170 m_g))\right)\right|\,,
\end{equation}
with an \(R^2\) score of 1.0000 and a mean absolute error of 6.0229 (plotted in Figure \ref{v vs mg k=6}). In the large $k$ limit, $V_h$ is found to scale as $m_g^{5/2}$ which deviates from the expected power of $3$. This may be due to the erratic nature of genetic algorithm, so we explicitly tested this conjectured relation again with specific ansatz which will be elaborated in the following subsection.

\subsection{Summary of 3d results}
In Tables \ref{tab:3d examples}, \ref{tab:3d results}, and \ref{tab:3d results fitting}, we summarise our results of ML the relationship between the coefficient \(k\), the volume of bounded complementary region of amoeba, and the Mahler measure for clarity.

\begin{table}[!ht]
    \centering
    \addtolength{\leftskip} {-2cm}
    \addtolength{\rightskip}{-2cm}
    \begin{tabular}{|>{\centering\arraybackslash}m{6em}|>{\centering\arraybackslash}m{17em}|>{\centering\arraybackslash}m{6em}|>{\centering\arraybackslash}m{6em}|>{\centering\arraybackslash}m{6em}|}
    \hline
    Surface & Newton Polynomial & Toric Diagram & Boundary & Cross section at z=0\\
    \hline
    \(\mathbb{P}^1\times\mathbb{P}^1\times\mathbb{P}^1\)& \(P(z,u,w)=k-(z+z^{-1}+u+u^{-1}+w+w^{-1})\)& \raisebox{-.5\height}{ \includegraphics[width=20mm,height=20mm]{Fig/p1p1p1.png}}& \raisebox{-.5\height}{ \includegraphics[width=20mm,height=20mm]{Fig/p1p1p1boundary.png}}& \raisebox{-.5\height}{ \includegraphics[width=20mm,height=20mm]{Fig/p1p1p1boundarycrosssection.png}}\\
    \hline
    \(\mathbb{P}^3\)& \(P(z,u,w)=k-(z+u+w+z^{-1}u^{-1}w^{-1})\)& \raisebox{-.5\height}{ \includegraphics[width=20mm,height=20mm]{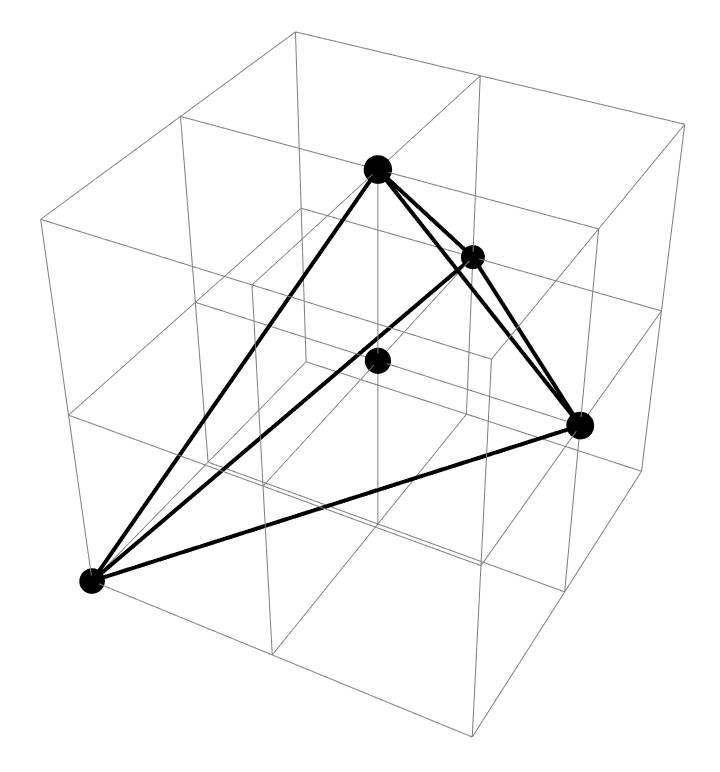}}& \raisebox{-.5\height}{ \includegraphics[width=20mm,height=20mm]{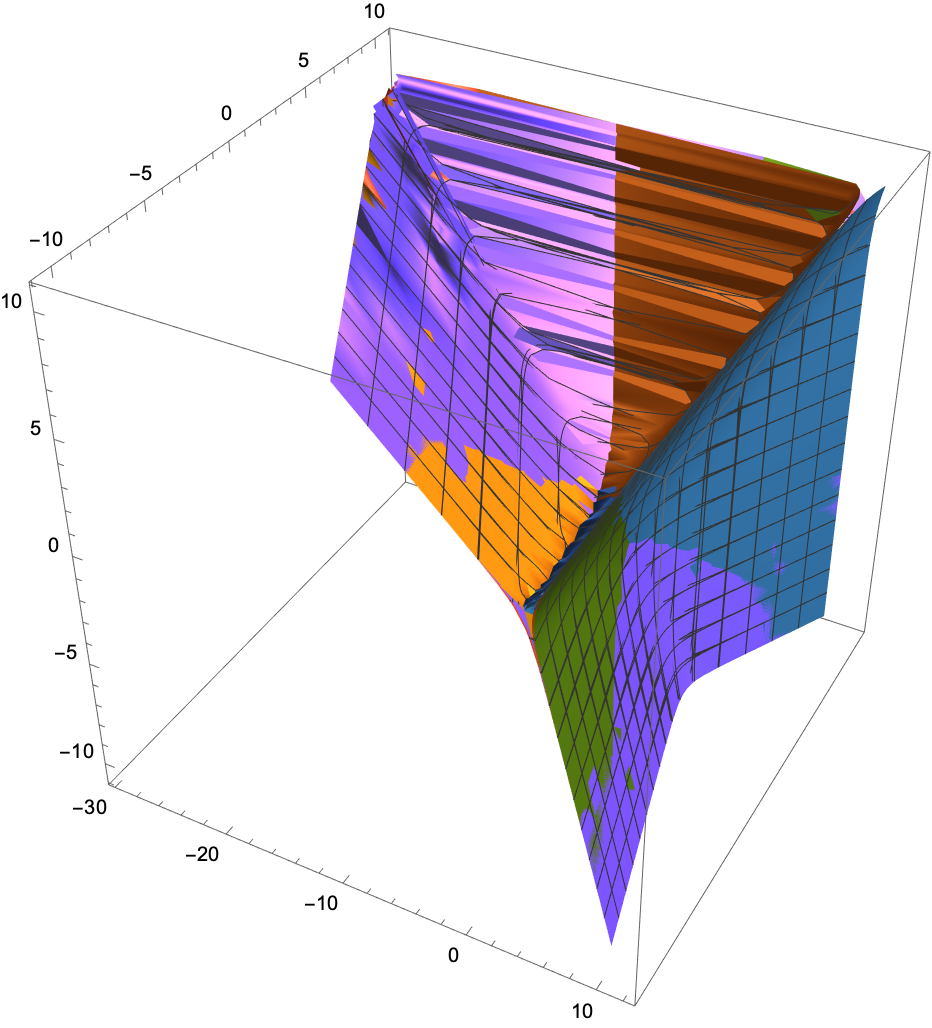}}& \raisebox{-.5\height}{ \includegraphics[width=20mm,height=20mm]{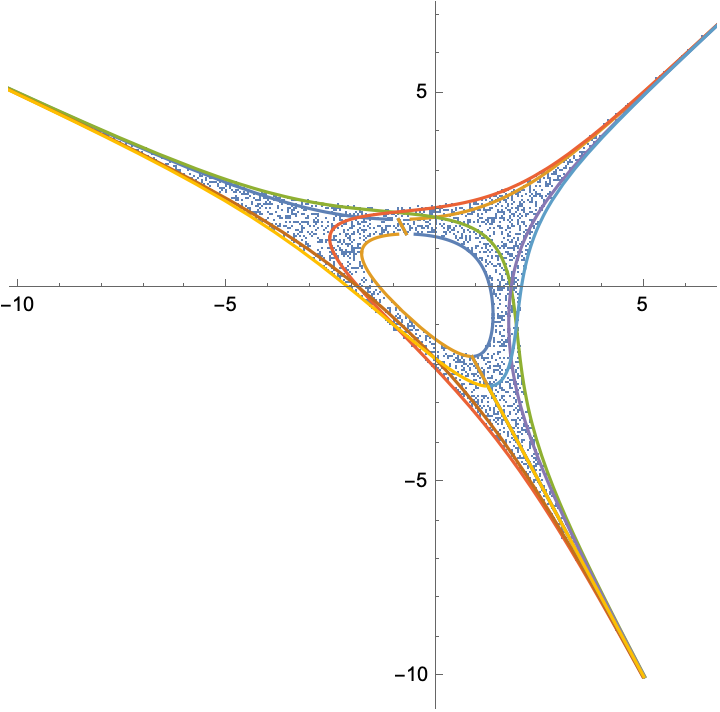}}\\
    \hline
    \(\mathbb{P}^2\times\mathbb{P}^1\)& \(P(z,u,w)=k-(z+u+w+z^{-1}+u^{-1}w^{-1})\)& \raisebox{-.5\height}{ \includegraphics[width=20mm,height=20mm]{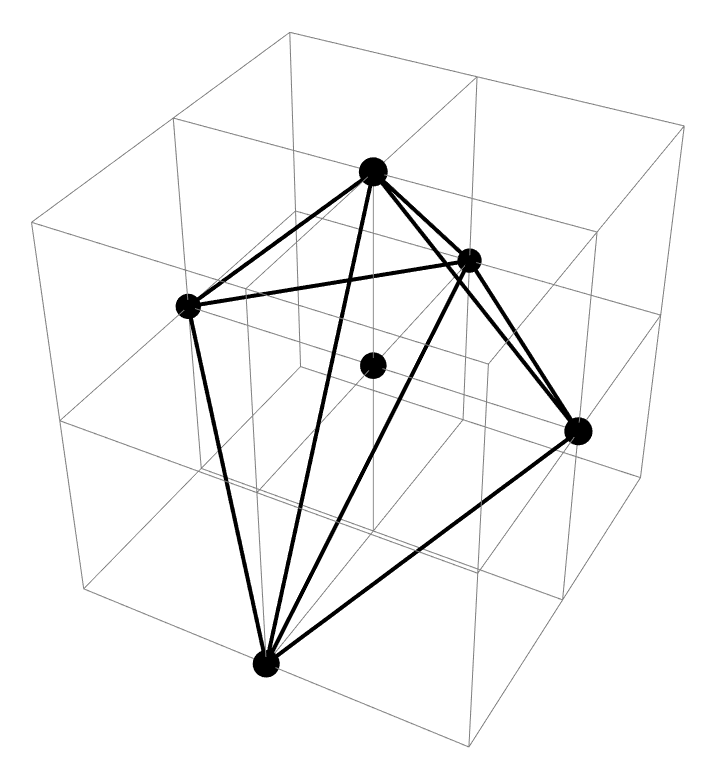}}& \raisebox{-.5\height}{ \includegraphics[width=20mm,height=20mm]{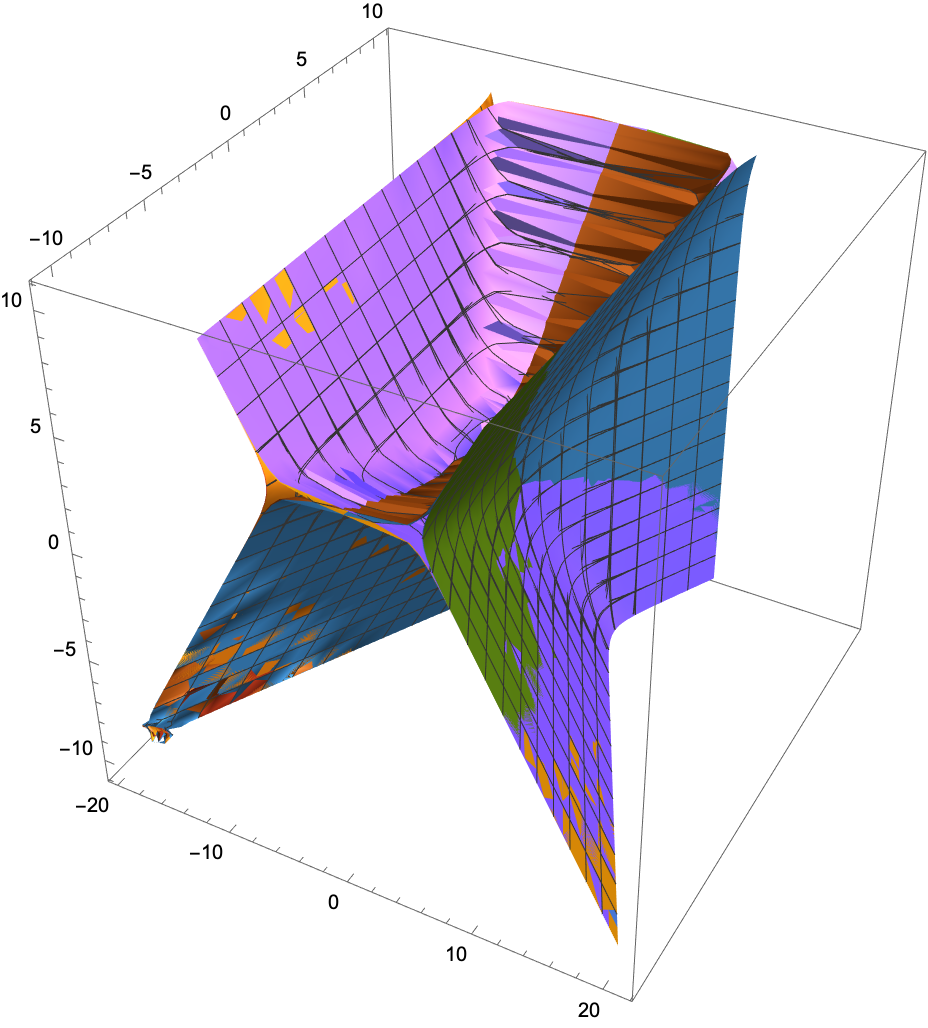}}& \raisebox{-.5\height}{ \includegraphics[width=20mm,height=20mm]{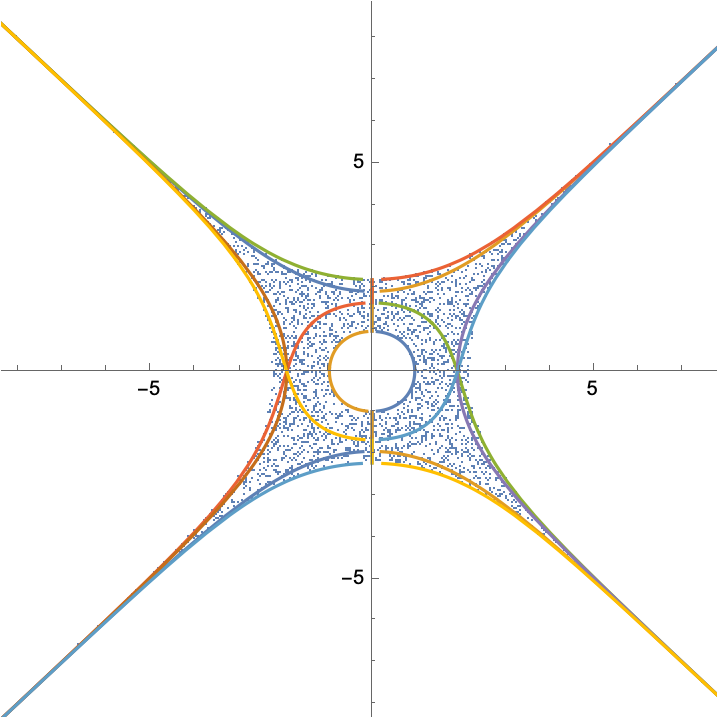}}\\
    \hline
    \end{tabular}
    \caption{Examples of 3d Fano varieties and their associated Newton polynomials, toric diagrams, plots of the boundary, and cross-sections of their amoebae.}
    \label{tab:3d examples}
\end{table}

\begin{sidewaystable}
    \centering
\begin{tabularx}{\textwidth}{|>{\centering\arraybackslash}m{6em}||>{\centering\arraybackslash}m{33em}|>{\centering\arraybackslash}m{20em}|}
    \cline{1-3}
  Surface & Amoeba Boundary  & Mahler measure\\
    \cline{1-3}
\(\mathbb{P}^1\times\mathbb{P}^1\times\mathbb{P}^1\) & \begin{equation*}
    x=\ln \left(\left|\frac{| k| }{2}\pm\cosh{y}\pm\cosh{z}\pm\sqrt{\left(\frac{| k| }{2}\pm\cosh{y}\pm\cosh{z}\right)^2-1}\right| \right)
\end{equation*}
   & \begin{equation*}
    m(P)=\ln k - \sum_{n=1}^\infty\sum_{l=0}^{n}\frac{1}{2nk^{2n}}{2n\choose n}{2l\choose l}{n\choose l}^2
\end{equation*}\\
\cline{1-3} 
 \(\mathbb{P}^3\)  & \begin{equation*}
     x= \ln \left(\left| \frac{|k| -(\pm e^y\pm e^z)}{2}\pm\sqrt{\left(\frac{|k| -(\pm e^y\pm e^z)}{2}\right)^2-(\pm e^{-y})(\pm e^{-z})}\right| \right)
 \end{equation*} & \begin{equation*}
     m(P)=\ln k -\sum_{n=1}^\infty\frac{1}{4nk^{4n}}{4n\choose 2n}{2n\choose n}^2
 \end{equation*}\\
 \cline{1-3}
 \(\mathbb{P}^2\times\mathbb{P}^1\)&\begin{equation*}\begin{aligned}
     x&=\ln\Bigg(\bigg| \frac{|k|-\left(\pm e^y\pm e^z+(\pm e^{-y})(\pm e^{-z})\right)}{2} \\&\pm\sqrt{ \left(\frac{|k|-\left(\pm e^y\pm e^z+(\pm e^{-y})(\pm e^{-z})\right)}{2}\right)^2-1}\bigg| \Bigg)\end{aligned}
 \end{equation*} & \begin{equation*}
     m(P)=\ln k -\sum_{n=1}^{\infty}\sum_{i\geq\frac{n}{2}}^n\frac{{n \choose i}{i\choose 2n-3i}{n-i\choose 2n-3i}{4i-2n \choose 2i-n}}{nk^n}
 \end{equation*} \\
 \cline{1-3}
\end{tabularx}
\caption{Summary of 3d results.}
    \label{tab:3d results}
\end{sidewaystable}

\begin{sidewaystable}
    \centering
\begin{tabularx}{\textwidth}{|>{\centering\arraybackslash}m{5em}||>{\centering\arraybackslash}m{32em}|>{\centering\arraybackslash}m{25em}|}
    \cline{1-3}
Surface &  Fitted \(V_h(k)\) &  Fitted \(V_h(m_g)\) \\
    \cline{1-3}
    \(\mathbb{P}^1\times\mathbb{P}^1\times\mathbb{P}^1\) &\begin{equation*}\begin{aligned}
    V_h(k)&=\ln (k+\ln (0.4854 k)) \Big(8.374 \ln \left(0.4854 k-\ln ^2(0.4854 k)\right)\\&+\ln \left(\frac{7.19}{\ln k}\right)\Big) \ln
   (k+\ln k)
   \end{aligned}
\end{equation*}
with an \(R^2\) score of 1.0000 and a mean absolute error of 3.2689.  &\begin{equation*}\begin{aligned}
   V_h&=3.835 \big| m_g \big(7.968-\sqrt{m_g} (-9.743 m_g\\&+\ln (m_g-9.664)+\ln (0.170 m_g))\big)\big|
   \end{aligned}
\end{equation*}
with an \(R^2\) score of 1.0000 and a mean absolute error of 6.0229. \(k_c=6\) and \(m_g(k)=m(P(k))-m(P(6))\).\\
 \cline{1-3}
  \(\mathbb{P}^3\) &  \begin{equation*}\begin{aligned}
    &V_h=\bigg| -0.380 \bigg(-3.732 +\pi i+2.397\sqrt{0.174 k-1} \Big(3.724 k-73.933\\& \times\left| \ln \big| \sqrt{k}-314.956
   |(k-5.744 )(2.489 k-k^{-0.5})|(0.174 k-1)^2 k^{9.5} \big| \right| \Big)\\&\times(2.397 \sqrt{0.174
   k-1}-0.489 k)^{-1}-73.933 \sqrt{k}+1.489 k\bigg)\\&\times(\ln | 7.888 k-731.045| )^{-0.5}+244.509 \sqrt{| 0.174 k-1|}-0.186 k-1.074 \bigg|
   \end{aligned}
\end{equation*} with an \(R^2\) score of 1.0000 and a mean absolute error of 4.1666. & \begin{equation*}\begin{aligned}
    V_h &= \left(\sqrt{m_g}+7.95\right) (1.523 m_g+12.4201)\big(0.4794 m_g^2\\& +\Big(0.0264 \big(3.385+5.424/m_g +0.0028/\sqrt{m_g}\\&+0.8802 m_g-0.4794 m_g^2\big)\Big)/\sqrt{m_g}\big)
   \end{aligned}
\end{equation*}
with an \(R^2\) score of 0.9998 and a mean absolute error of 1.5644. \(k_c=4\) and \(m_g(k)=m(P(k))-m(P(4))\).\\
 \cline{1-3} 
 \(\mathbb{P}^2\times\mathbb{P}^1\) & \begin{equation*}\begin{aligned}
     V_h&= \bigg| \frac{1}{0.0068 \ln k-\frac{0.0068}{0.215 -0.0068 \sqrt{k}}}+(256.761 +5.049\pi i) \sqrt{0.118 k-1}-\sqrt{k}\\&+\sqrt{\frac{1}{0.0068 \ln k -0.0068 \sqrt{k}}+0.550 k}+4.174 \sqrt{\frac{k}{\ln k-\sqrt{k}}+\frac{k}{\ln k}}\\&+4.174 \sqrt{-\frac{k}{\ln
   \left(2 k-72.719 k \sqrt{\frac{k}{\ln k-\sqrt{k}}+\frac{k}{\ln k}+0.0061}\right)-\sqrt{k}}}\\&+\frac{1}{0.0045 \log
   (k-8.338)-0.0045 \sqrt{k}}-0.214 k+\ln k\bigg|
\end{aligned}
\end{equation*}
with an \(R^2\) score of 0.9999 and a mean absolute error of 8.0802. & \begin{equation*}\begin{aligned}
    &V_h= \bigg| 53.0889+\left(m_g^2-0.054 m_g-1.09316\right) \bigg(53.0889\\&+\Big(\big(-2 m_g+(\frac{0.146092}{m_g-8.175}-m_g-3.60814)\\&
   \times(\frac{9.966}{m_g-8.175}-7.324 m_g (m_g +2.04377 i))\big)\ln^{-1} m_g\Big)\bigg)\bigg|
   \end{aligned}
\end{equation*}
with an \(R^2\) score of 1.0000 and a mean absolute error of 3.1879. \(k_c=5\) and \(m_g(k)=m(P(k))-m(P(5))\).\\
 \cline{1-3}
\end{tabularx}
\caption{Summary of 3d results using symbolic regression.}
    \label{tab:3d results fitting}
\end{sidewaystable}

Moreover, since the results obtained using symbolic regression in the 3-dimensional case are too complicated to be useful, we also included the results obtained using NonlinearModelFit in \texttt{Mathematica} \cite{Mathematica} in Table \ref{tab:3d results Mfitting} to specifically test Conjecture 3.9 in \cite{Bao:2021fjd}. NonlinearModelFit requires an assumption of the structure of the fit model. In this case, a cubic equation in \(\ln k\) and a cubic equation in \(m_g\) were assumed based on the results in two dimensions. The results are relatively accurate based on their $R^2$ values and the mean prediction errors.

\begin{sidewaystable}
    \centering
\begin{tabularx}{\textwidth}{|>{\centering\arraybackslash}m{4em}||>{\centering\arraybackslash}m{21em}|>{\centering\arraybackslash}m{8em}|>{\centering\arraybackslash}m{19em}|>{\centering\arraybackslash}m{8em}|}
    \cline{1-5}
Surface &  Fitted \(V_h(k)\) & Plot \(V_h(k)\) &  Fitted \(V_h(m_g)\) & Plot  \(V_h(m_g)\) \\
    \cline{1-5}
    \(\mathbb{P}^1\times\mathbb{P}^1\times\mathbb{P}^1\) & \begin{equation*}
    V_h=7.981 \ln^3 k+0.292 \ln^2 k -26.439 \ln k -17.822
\end{equation*}
with an \(R^2\) score of 1.0000 and a mean prediction error of 0.0175.  & \raisebox{-.5\height}{ \includegraphics[width=\linewidth]{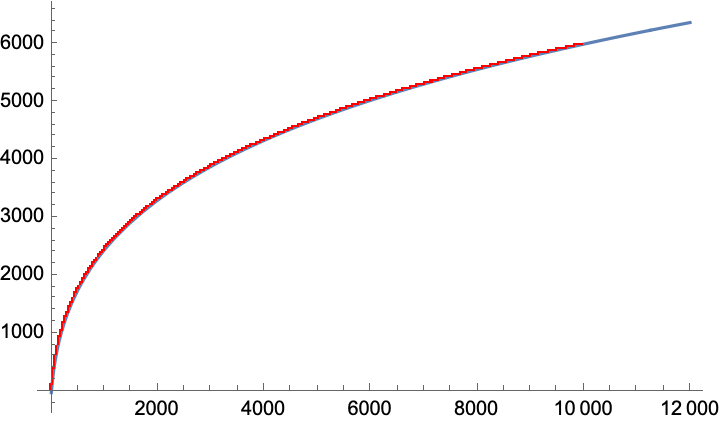}} &\begin{equation*}
    V_h=7.912 m_g^3+41.381 m_g^2+36.856 m_g-17.055
\end{equation*}
with an \(R^2\) score of 1.0000 and a mean prediction error of 0.0357.
& \raisebox{-.5\height}{ \includegraphics[width=\linewidth]{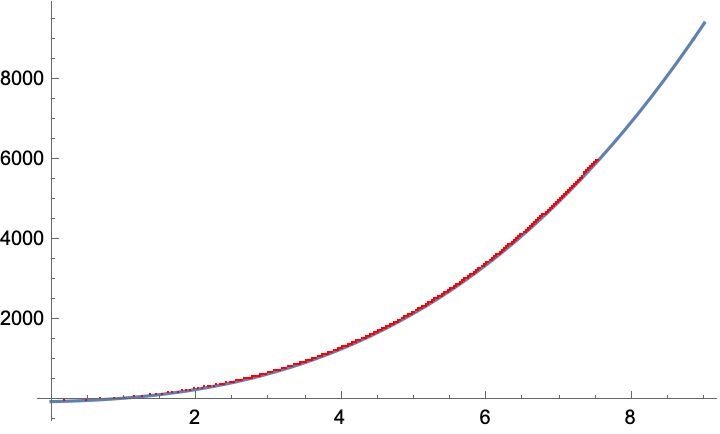}}\\
 \cline{1-5}
   \(\mathbb{P}^3\) &  \begin{equation*}
    V_h=10.299 \ln^3 k-0.941 \ln^2 k-25.871 \ln k +4.027
\end{equation*} with an \(R^2\) score of 1.0000 and a mean prediction error of 0.0051. & \raisebox{-.5\height}{ \includegraphics[width=\linewidth]{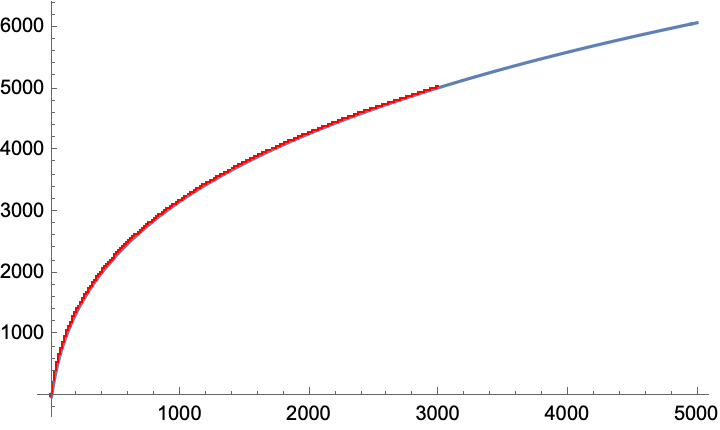}} & \begin{equation*}
    V_h = 10.294 m_g^3+40.936 m_g^2+27.867 m_g-6.787
\end{equation*}
with an \(R^2\) score of 1.0000 and a mean prediction error of 0.0055.
 & \raisebox{-.5\height}{ \includegraphics[width=\linewidth]{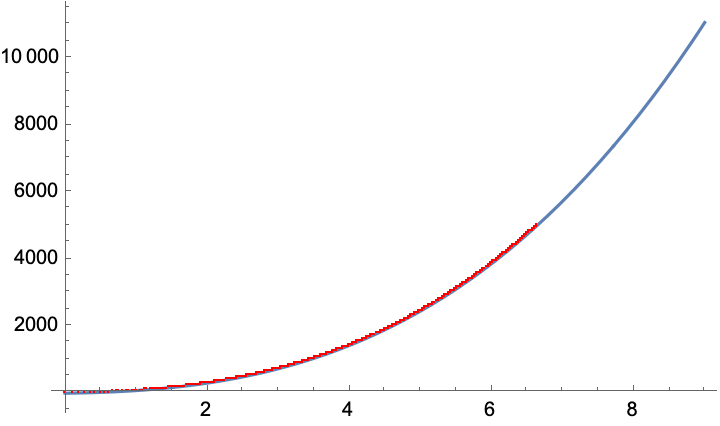}}\\
 \cline{1-5} 
 \(\mathbb{P}^2\times\mathbb{P}^1\) & \begin{equation*}
     V_h=8.196 \ln^3 k+1.986 \ln^2 k-26.103 \ln k -9.320
\end{equation*}
with an \(R^2\) score of 1.0000 and a mean prediction error of 0.0123. & \raisebox{-.5\height}{ \includegraphics[width=\linewidth]{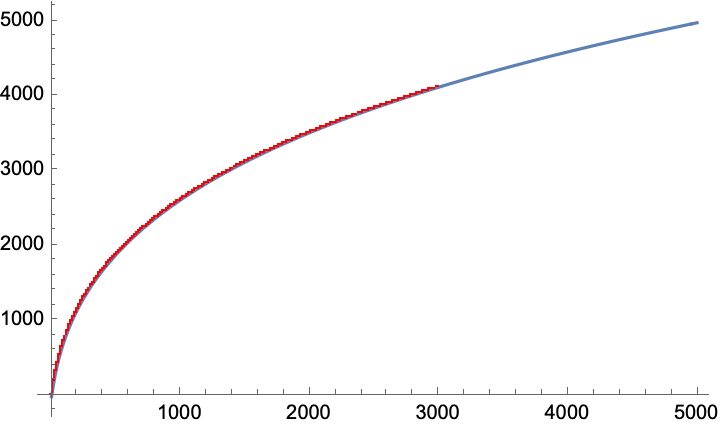}} &\begin{equation*}
    V_h= 8.169 m_g^3+40.011 m_g^2+35.901 m_g-12.671
\end{equation*}
with an \(R^2\) score of 1.0000 and a mean prediction error of 0.0131.
 & \raisebox{-.5\height}{ \includegraphics[width=\linewidth]{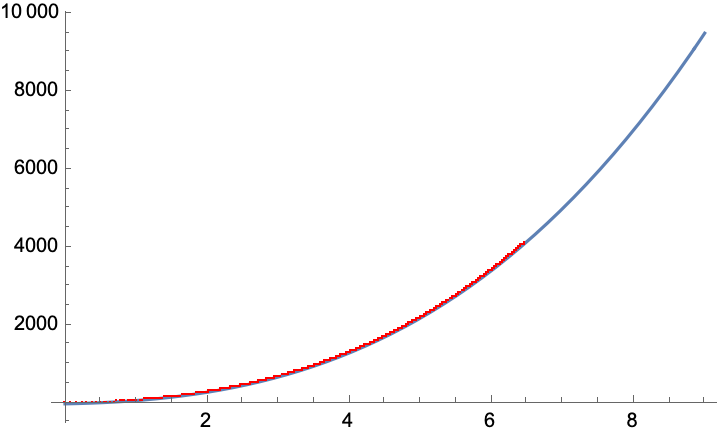}}\\
 \cline{1-5}
\end{tabularx}
\caption{Summary of 3d results using NonlinearModelFit in \texttt{Mathematica} \cite{Mathematica}.}
    \label{tab:3d results Mfitting}
\end{sidewaystable}

Results in Table \ref{tab:3d results Mfitting} agree with Conjecture 3.9 in \cite{Bao:2021fjd} in the \(n=3\) case: In the large $k$ limit, the volume of a bounded complementary region of the amoeba, \(V_h\), is cubic in \(\ln k\). However, we also notice that it is not possible to generalise an analytic expression for \(V_h\) analogous to the expression in Conjecture 3.8 in \cite{Bao:2021fjd}, because the volume of the 3-dimensional amoeba is almost always infinite whereas the area of the 2-dimensional amoeba is bounded from above \cite{pass2004}.

\subsection{Non-reflexive polytopes}
Following our consideration of the non-reflexive case in Section §\ref{sec:4}, it is interesting to also look at the relation between amoeba holes and the Mahler measure in this case. 
Amoebae for non-reflexive polytopes can have a geometric genus ranging from 1 to $n$, where $n$ is the number of interior points in the corresponding Newton polytope. It is noted in \cite{KenyonHarnack} that for amoebae with all holes open, a decrease in the size of one hole corresponds to an increase in the size of all others. 

Specifically, we are going to consider here the polytopes with two interior points, which corresponds to amoebae with a maximum of two bounded complementary regions. As we mention in Section §\ref{sec:4}, the  Mahler measure can now be represented as a function of two variables, $k_1$ and $k_2$, which correspond the two interior points. We can make a choice of which interior point we use as the origin. Where in the one dimensional case, we can generally find a critical value for $k$ at which the gas phase emerges in the amoeba, in two or more dimensions, we instead get a set of values for $(k_1,k_2)$ where gas phases emerge. We get another set of $(k_1,k_2)$ points where the genus of the amoeba changes from 1 to 2. Each of these gas phases appear and disappear individually depending on both the value of $k_1$ and of $k_2$.\par

\begin{figure}
    \centering

\tikzset{every picture/.style={line width=0.75pt}} 

\begin{tikzpicture}[x=0.75pt,y=0.75pt,yscale=-1,xscale=1]

\draw  (114.84,324.3) -- (403.95,324.3)(143.75,71.4) -- (143.75,352.4) (396.95,319.3) -- (403.95,324.3) -- (396.95,329.3) (138.75,78.4) -- (143.75,71.4) -- (148.75,78.4)  ;
\draw    (177.62,283.42) -- (165.16,324.65) ;
\draw    (143.71,302.45) -- (177.62,283.42) ;
\draw    (215.69,79.64) -- (177.62,283.42) ;
\draw    (177.62,283.42) .. controls (245.45,185.89) and (356.18,183.78) .. (397.71,181.4) ;

\draw (252.26,136.24) node [anchor=north west][inner sep=0.75pt]   [align=left] {Both holes open};
\draw (115.8,94.59) node [anchor=north west][inner sep=0.75pt]  [rotate=-0.37] [align=left] {Left hole only open};
\draw (250.56,267.86) node [anchor=north west][inner sep=0.75pt]   [align=left] {Right hole only open};
\draw (77.45,303.54) node [anchor=north west][inner sep=0.75pt]   [align=left] {No holes open};
\draw (324.02,327.62) node [anchor=north west][inner sep=0.75pt]    {$k_{2}$};
\draw (123.3,125.44) node [anchor=north west][inner sep=0.75pt]    {$k_{1}$};

\end{tikzpicture}

    \caption{The number of bounded amoeba complements present with respect to the value of \(k_1, k_2\) in the case of 2-dimensional non-reflexive polytopes with two interior points. 
    }
    \label{non-reflexive amoeba}
\end{figure}
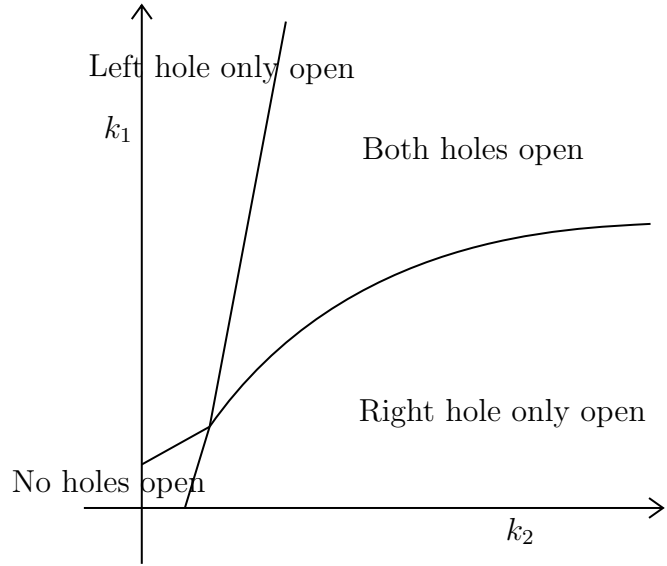

Based on our observation, for polytopes with two interior points, there are three ways that the bounded complementary regions of amoeba can evolve as we fix $k_1$ and move along $k_2$, or vice versa:
\begin{enumerate}
    \item There are initially no holes. At some critical value of \(k_2\) a hole opens up and continues to grow as we increase \(k_2 \rightarrow \infty\). This is similar to the reflexive case and only occurs when \(k_1\) is also small. 
    \item There is initially one hole. As we move along \(k_2\), the area of this hole decreases, until it closes. Another hole subsequently opens at the same or larger value of \(k_2\). This hole increases as we increase \(k_2 \rightarrow \infty\), like the reflexive case.
    \item There is initially one hole. As we move along \(k_2\), the area of this hole decreases. At some value of \(k_2\), a second hole opens. The area of this second hole continues to increase as the area of the first hole decreases. At some finite value of \(k_2\), the first hole closes, and the area of the second hole increases, as in the reflexive case.
\end{enumerate}
We get the same three cases if we instead fix $k_2$ and move along $k_1$. The values of $k_1$ and $k_2$ for which holes open up are not symmetrical, however. This is illustrated in Figure \ref{non-reflexive amoeba}.

\begin{figure}[!ht]
     \centering
     \includegraphics[width=.5\linewidth]{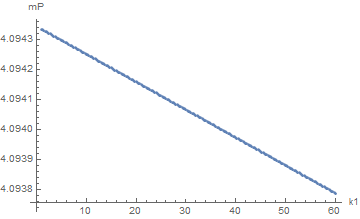}
     \caption{Mahler measure of polynomial associated with $\mathbb{C}^3/\mathbb{Z}_5$, with the origin being the left interior point. We set $k_2=60$ and varied $k_1$. We can see a clear decrease in the Mahler measure as we move along $k_1$}
     \label{decreasing mahler measure c3z5}
\end{figure}

\begin{figure}[!ht]
     \centering
     \includegraphics[width=.5\linewidth]{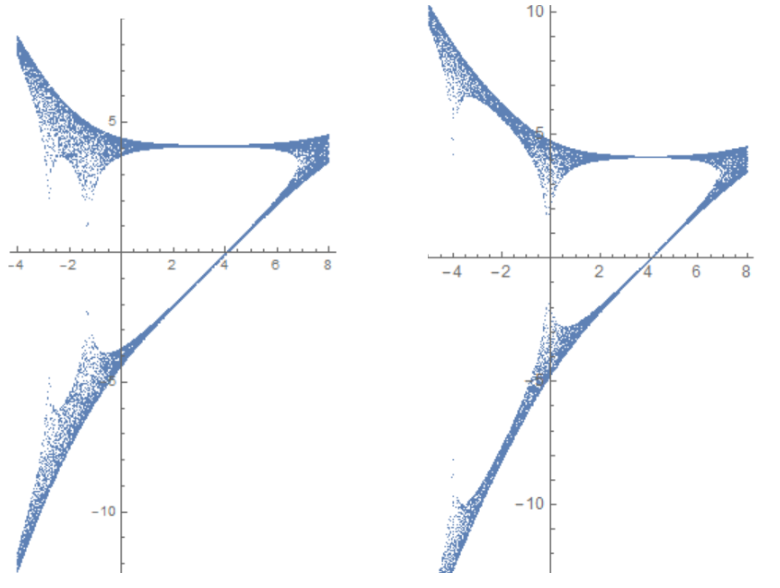}
     \caption{Amoebae of  polynomial associated with $\mathbb{C}^3/\mathbb{Z}_5$. In both cases we take the left hand interior point to be the origin and set $k_2=60$. In the left amoeba, we set $k_1=10$ and in the right we set $k_1=55$. There is a clear increase in hole size for larger $k_2$ }
     \label{c3z5 amoeba comparison}
\end{figure}

With respect to the relation between the amoeba holes and the Mahler measure, in general we expect a monotonic increase in the Mahler measure if we start at a point where no holes are open and move along either $k_1$ or $k_2$. This is very similar to the reflexive case, with there only ever being at most one hole open. This hole opens at some critical value of $k_1$, and its area continues to increase as $k_1\rightarrow\infty$. However, there are also instances where as the Mahler measure decreases, the area of the holes increase, or vice versa. An example is given in Figures \ref{decreasing mahler measure c3z5} and \ref{c3z5 amoeba comparison} where the value of the Mahler measure decreases as $k_1$ increases, but when compared with the evolution of the holes of the amoeba for coefficients in the same range, we see their area increases with increasing $k_1$.

We have however observed that as we move along the Mahler flow, as defined for the  two disconnected regions of the $(k_1,k_2)$ plane in Section §\ref{sec:4}, we do seem to get a monotonic increase in the area of the holes. This matches the monotonic behaviour we see in the Mahler measure in these regions.


It is evident that in the case of non-reflexive polytopes, the relations between the coefficients, the amoeba holes, and the Mahler measure are much more complicated. Nonetheless, we can still employ ML techniques, especially generic algorithm, to make their relations more precise. 

\subsubsection{2d Example: \texorpdfstring{$\mathbb{C}^3/\mathbb{Z}_5$}{C3Z5}}

\begin{figure}[!ht]
   \begin{minipage}{0.48\textwidth}
     \centering
     \includegraphics[width=.4\linewidth]{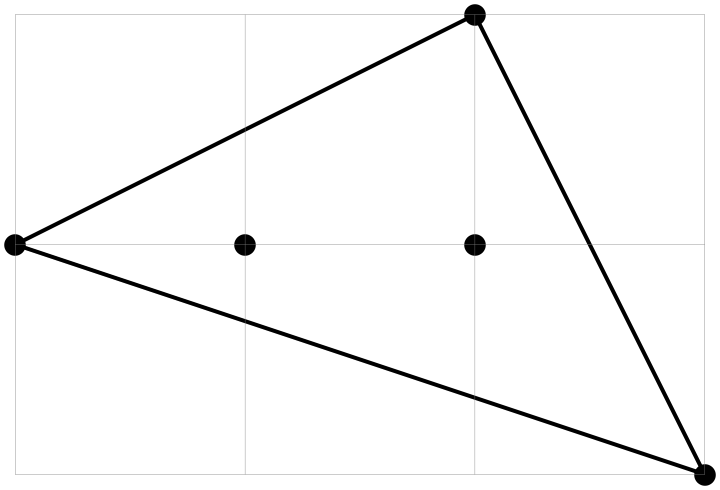}
     \caption{The toric diagram associated with $\mathbb{C}^3/\mathbb{Z}_5$\,.}
     \label{c3/z5 toric}
   \end{minipage}\hfill
   \begin{minipage}{0.48\textwidth}
     \centering
     \includegraphics[width=.4\linewidth]{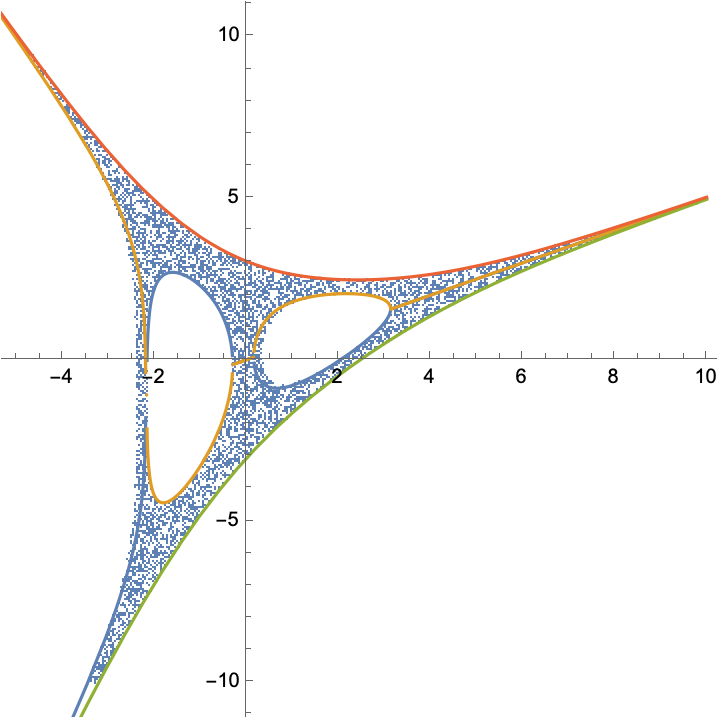}
     \caption{The boundary curves of the amoeba where \(k_1=10\) and \(k_2=10\)\,.}
     \label{c3/z5 boundary}
   \end{minipage}
\end{figure}

As a concrete example, we considered again the surface $\mathbb{C}^3/\mathbb{Z}_5$ whose associated toric diagram is given in Figure \ref{c3/z5 toric}, which has two interior points and is thus non-reflexive. Taking the left interior point as the origin, the Newton polynomial is \(P(z,w)=k_1+k_2z+z^{-1}+zw+z^2w^{-1}\). Following the same method in Section §\ref{sec:area_bhc}, the analytic boundary of the amoeba (Figure \ref{c3/z5 boundary}) is given by 
\begin{equation}
    y=\ln \left(\left|\frac{\pm k_1 e^{-x}-k_2-e^{-2x}}{2}\pm \sqrt{\left(\frac{\pm k_1e^{-x}-k_2-e^{-2x}}{2}\right)^2\pm e^x}\right|\right).
\end{equation}

We then explored the numerical relations between the areas of the bounded amoeba complements, the values of \(k_1\) and \(k_2\), and the Mahler measure using symbolic regression and NonLinearModelFit with an assumed form. Specifically, we restricted ourselves to the range of values where the two amoeba holes are both present.

We first machine-learned the relation between \(A_{1,2}\) and \(k_{1,2}\): From our discussion in the reflexive case and observation of the \texttt{gplearn} results, we expect the leading order dependence of the area on \(k_{1,2}\) should be second order in the logarithm of \(k_{1,2}\). Thus, we assumed the form of \(a\ln^2 k_1+b \ln k_1 \ln k_2+c\ln^2 k_2 +d \ln k_1+e\ln k_2 +h\) to use the NonLinearModelFit function in \texttt{Mathematica}. The ML and fitting results are presented in Table \ref{tab:A1,2(k1,2)}.

\begin{table}[!ht]
    \centering
    \addtolength{\leftskip} {-2cm}
    \addtolength{\rightskip}{-2cm}
    \begin{tabular}{|>{\centering\arraybackslash}m{21em}|>{\centering\arraybackslash}m{21em}|}
    \hline
    \texttt{gplearn} result & \texttt{Mathematica} result \\
    \hline
    \begin{equation*}\begin{aligned}
        A_1(k_1,k_2)&=-k_1^{0.25} + 9.62(k_1/\ln^{0.5} k_2)^{0.5} \\&- 2 k_2/k_1\end{aligned}
    \end{equation*} with an \(R^2\) score of 0.98054 and mean absolute error of 4.22914. & \begin{equation*}\begin{aligned}
        A_1(k_1,k_2)&=-20.0273+8.9449 \ln k_1 \\&+2.81341 \ln^2 k_1-4.3546 \ln k_2\\&+1.9363 \ln k_1 \ln k_2-1.6381\ln^2 k_2\end{aligned}
    \end{equation*} with an \(R^2\) score of 0.99997 and mean prediction error of 0.00378. \\
    \hline
    \begin{equation*}\begin{aligned}
        A_2(k_1,k_2)&=2.4692\big(-(0.1086k_1\ln (k_2^{0.5}/k_1) \\&+ 7.527k_2 - 
       69.3161)/\ln (7.527/k_1)\big)^{0.5}\end{aligned}
    \end{equation*} with an \(R^2\) score of 0.99431 and mean absolute error of 1.19120. & \begin{equation*}\begin{aligned}
        A_2(k_1,k_2)&=-0.7777-3.4303 \ln k_1\\&+0.9557 \ln^2 k_1+1.9742 \ln k_2\\&-3.3533 \ln k_1 \ln k_2+4.5230 \ln^2 k_2\end{aligned}
    \end{equation*} with an \(R^2\) score of 0.99998 and mean prediction error of 0.00170.\\
    \hline
    \end{tabular}
    \caption{Fits obtained from symbolic regression and NonLinearModelFit function}
    \label{tab:A1,2(k1,2)}
\end{table}

\begin{figure}[!ht]
  \centering
  \includegraphics[width=.4\linewidth]{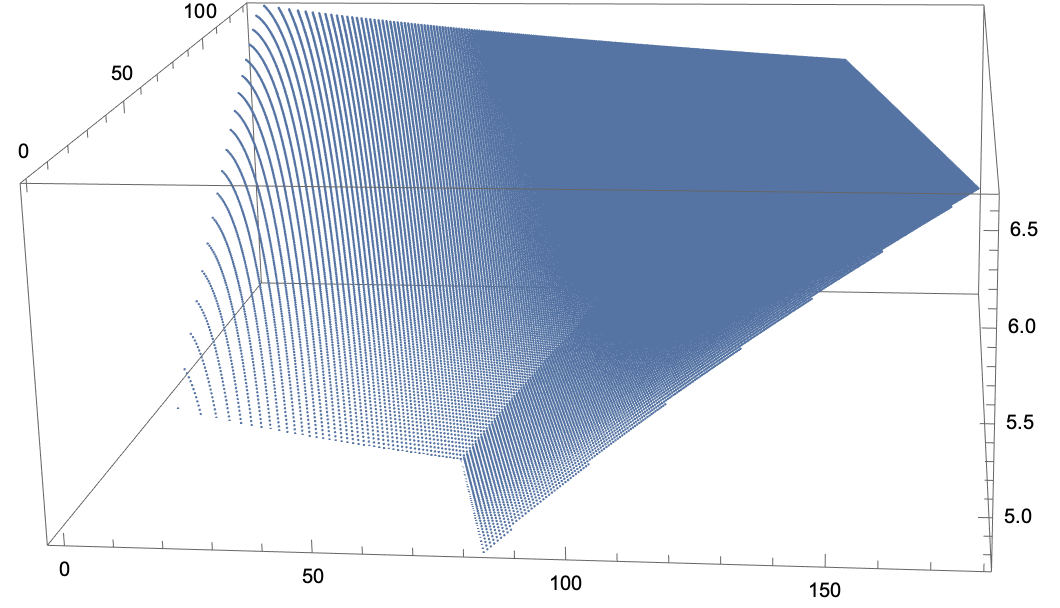}
  \caption{Plot of the Mahler measure against the areas of two amoeba holes}
  \label{a1a2m}
\end{figure}

To learn the relation between \(m(P)\) and \(A_{1,2}\), we computed the Mahler measure associated with the amoeba with two holes present in the range of \(0\leq k_{1,2}\leq 800\). The data points are plotted in Figure \ref{a1a2m}. There seems to be a discontinuous transition in the Mahler measure as we vary the sizes of the amoeba holes, and meaningful ML results using genetic symbolic regression can only be obtained if we fit two regions (left and right) separately. The numerical relations obtained are presented in Table \ref{tab:mL,R(A1,2)}.

\begin{table}[!ht]
    \centering
    \addtolength{\leftskip} {-2cm}
    \addtolength{\rightskip}{-2cm}
    \begin{tabular}{|>{\centering\arraybackslash}m{21em}|>{\centering\arraybackslash}m{21em}|}
    \hline
    \texttt{gplearn} result & \texttt{Mathematica} result \\
    \hline
    \begin{equation*}\begin{aligned}
       m_L(A_1,A_2)&=\bigg(\big(0.115314 A_2-0.0832856 A_1^{0.5}\big)\\& (A_1+A_2)\bigg)^{0.25}\end{aligned}
    \end{equation*} with an \(R^2\) score of 0.98803 and mean absolute error of 0.03493. & \begin{equation*}\begin{aligned}
        m_L(A_1,A_2)&=1.4397-0.1465 A_1^{0.25}+0.2205 A_1^{0.5}\\&-0.5654 A_2^{0.25}-0.1208 A_1^{0.25} A_2^{0.25}\\&+0.6588 A_2^{0.5}\end{aligned}
    \end{equation*} with an \(R^2\) score of 1.00000 and mean prediction error of 0.00002. \\
    \hline
    \begin{equation*}\begin{aligned}
        m_R(A_1,A_2)&=1.1745(-0.1761A_1 - 0.7250A_2^{0.5} \\&+ 
    1)^{0.5}\end{aligned}
    \end{equation*} with an \(R^2\) score of 0.99545 and mean absolute error of 0.01709. & \begin{equation*}\begin{aligned}
       m_R(A_1,A_2)&=1.0878 - 0.4489 A_1^{0.25} + 
 0.5654 A_1^{0.5} \\&- 0.0019 A_2^{0.25} - 
 0.1194 A_1^{0.25} A_2^{0.25}\\& + 0.1639 A_2^{0.5}\end{aligned}
    \end{equation*} with an \(R^2\) score of 0.99999 and mean prediction error of 0.00005.\\
    \hline
    \end{tabular}
    \caption{Fits obtained from symbolic regression and NonLinearModelFit function}
    \label{tab:mL,R(A1,2)}
\end{table}

\begin{figure}[!ht]
  \centering
  \includegraphics[width=.4\linewidth]{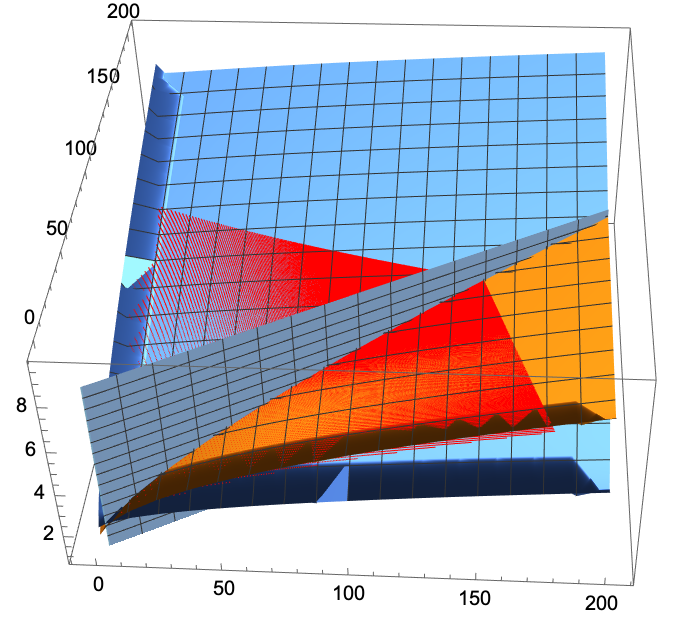}
  \caption{Plots of the two fitted surfaces (blue and orange), the plane that passes through the line of intersection (grey), and the data points (red).}
  \label{a1a2m fit}
\end{figure}

Moreover, we changed the parsimony coefficient in \texttt{gplearn} which controls the complexity of the equations from 0.02 to 0.002 in order for better learning result. The ansatz used in the NonLinearModelFit function is \(a +b A_1^{0.25} + c A_1^{0.5} +d A_2^{0.25} +e A_1^{0.25} A_2^{0.25}+h  A_2^{0.5}\) in both regions, by inverting the conjectured relation in 2d. Specifically, the line of intersection of two surfaces is found to be
\begin{equation}
\begin{aligned}
    A_2&=-0.4975 - 0.7157 A_1^{1/4} + 
 1.1806 A_1^{1/2} - 0.8423 A_1^{3/4} + 
 0.4857 A_1 \\&+ 
 (-0.0079 - 0.1660 A_1^{1/4} - 
   0.7928 A_1^{1/2} + 1.0819 A_1^{3/4} + 
   2.7237 A_1 \\&- 4.5800 A_1^{5/4} + 
   1.7311 A_1^{3/2} + 0.0090 A_1^{7/4} + 
   0.0000116A_1^2)^{1/2}.
  \end{aligned}
\end{equation}
The presence of this line of special values resemble the plots of Mahler measure in Section §\ref{sec:4}. The fitting results are plotted together with the data points in Figure \ref{a1a2m fit}.

The fitting results using NonLinearModelFit in Tables \ref{tab:A1,2(k1,2)} and \ref{tab:mL,R(A1,2)} have a rather high \(R^2\) value close to $1$. This provides further support for the adopted assumed forms based on previous \texttt{gplearn} results and conjectures, i.e. the degree of the polynomial relation equals to the dimension of the amoeba.

Given the extraordinary performance using  NonLinearModelFit in \texttt{Mathematica}, one is tempted to conjecture an exact formula. Converting the numerical coefficients to potential closed form \cite{wolframalpha}, an example for $m_L$ in Table \ref{tab:mL,R(A1,2)} is:
\begin{equation}
    \begin{aligned}
        m_L(A_1,A_2)&=\frac{11\pi}{24}-\frac{2\pi}{43} A_1^{1/4}+\frac{4\pi}{57} A_1^{1/2}-\frac{9\pi}{50} A_2^{1/4}-\frac{\pi}{26} A_1^{1/4} A_2^{1/4}+\frac{56}{85} A_2^{1/2}.
        \end{aligned}
\end{equation}
It would be interesting to prove results such as the above.

Additionally, we would like to note that the choice of origin would affect the areas of the corresponding amoeba holes given the same toric diagram. 
Specifically, if we set the coefficients to be of the form $P(z,w)=k_1-p(z,w)$, with all coefficients of $p(z,w)$ positive, the values of $(k_1,k_2)$ for which holes open up do not seem to be related. If we however set all coefficients in $P(z,w)$ to be the same sign, amoebae are equivalent to each other i.e. $A_l(k_1,k_2)=A_r(k_2,k_1)$, where $A_l$ is the amoeba when the left interior point is taken to be the origin. This is expected, as it is the same as multiplying the related polynomial by a factor of $z^aw^b$, while keeping all coefficients the same.

\section{Discussions and Outlook}\label{sec:6}

In this paper, we brought together amoebae in tropical geometry and the Mahler measure in number theory, in the context of brane configurations and dimer models.

First, we continued the study of applying machine learning techniques to the analysis of amoebae topology, initiated in \cite{Bao:2021olg}. We applied both MLP and CNN to examples of 3-dimensional reflexive amoebae and compared the results using data obtained from persistent homology and analytic conditions using lopsidedness. Although the analytic conditions always give clearer data separation shown with the MDS projection, it may not be available for complicated examples and persistent homology can be helpful in those cases. The ML performance on data from lopsidedness only improves marginally if the size of the ML data is increased, whilst the ML performance on data from persistent homology can be improved by increasing the data size at the cost of longer computation time. Similar to the 2-dimensional results in \cite{Bao:2021olg}, a simple MLP or CNN can predict the number of 2-dimensional cavities characterised by the second Betti number to a high accuracy.

Second, we extended the definition of the Mahler flow in \cite{Bao:2021fjd} to incorporate the extra degrees of freedom present in non-reflexive polytopes. We investigate the properties of the flow using a combination of analytical and numerical techniques, and discuss its relation to amoebae and dimers.

Finally and most importantly, we obtained a more precise relation between the amoeba and the Mahler measure which are closely but mysteriously related through dimer models and crystal melting models. To do so, we performed symbolic regression using genetic algorithms to machine learn the numerical relations between the volume of the bounded amoeba complement, coefficient \(k\) in the Newton polynomial, and the Mahler measure, which are conjectured in \cite{Bao:2021fjd}. We obtained the analytic expressions of the amoeba boundary by considering the poles of the gradient of the Ronkin function, which allowed computation of the volume of the bounded amoeba complement. Although the mean absolute error may be high in complicated examples such as the 3-dimensional amoebae or non-reflexive amoebae, the ML results are useful in making ans\"atzen required in the NonLinearModelFit function in \texttt{Mathematica} to obtain a better fit. At the end, we also considered an example 2-dimensional non-reflexive polytopes where the dynamics between the coefficients, the Mahler measure, and the areas of the amoeba holes is much more complicated. That said, we were able to find a numeric relation between these non-reflexive amoebae and the Mahler flow.

Our results from genetic symbolic regression in Section \S\ref{sec:5} provide numerical evidence for Conjecture 3.8 in \cite{Bao:2021fjd} in both two and three dimensions. Specifically, we found that the volume of the bounded complement of the amoeba is related to the gas phase contribution to the Mahler measure by a polynomial of degree of the dimension of the amoeba. 

In our discussion of the relation between Mahler measure and amoeba hole, we used the notion of gas phase contribution to the Mahler measure, but we also found that this notion needs to be refined in the case involving multiple coefficients or non-reflexive polytopes in 2-dimensional cases. In 3 dimensions, we defined an analogous notion of \(m_g(k)=m(k)-m(k_c)\), where \(k_c\) is the critical value at which the 2-dimensional cavity first appears. The interpretation of this \(m_g\) would require a 3-dimensional dimer model and can be a subject of future studies. We will leave the physical interpretation of the Mahler measure in these more complicated scenarios to future work.

Our analysis also implies the power of numerical analysis in this context, and we can continue in this direction to study concepts such as the closely related Ronkin functions and its Legendre dual.

\section*{Acknowledgments}
The authors would like to thank Jiakang Bao and Elli Heyes for insightful discussion.
YHH would like to thank STFC for grant ST/J00037X/2.
EH would like to thank STFC for a PhD studentship.
The research of AZ has been supported by the French “Investissements d’Avenir” program, project ISITE-BFC (No. ANR-15-IDEX-0003), and EIPHI Graduate School (No. ANR17-EURE-0002).

\appendix

\section{Additional examples of ML the Betti number of 3d amoebae}\label{appendix1}

\subsection{\texorpdfstring{\(\mathbb{P}^3\)}{P3}} 

\begin{figure}[!ht]
   \begin{minipage}{0.48\textwidth}
     \centering
     \includegraphics[width=.6\linewidth]{Fig/p3.png}
     \caption{Toric diagram for \(\mathbb{P}^3\)\,.}
     \label{p3}
   \end{minipage}\hfill
   \begin{minipage}{0.48\textwidth}
     \centering
     \includegraphics[width=.6\linewidth]{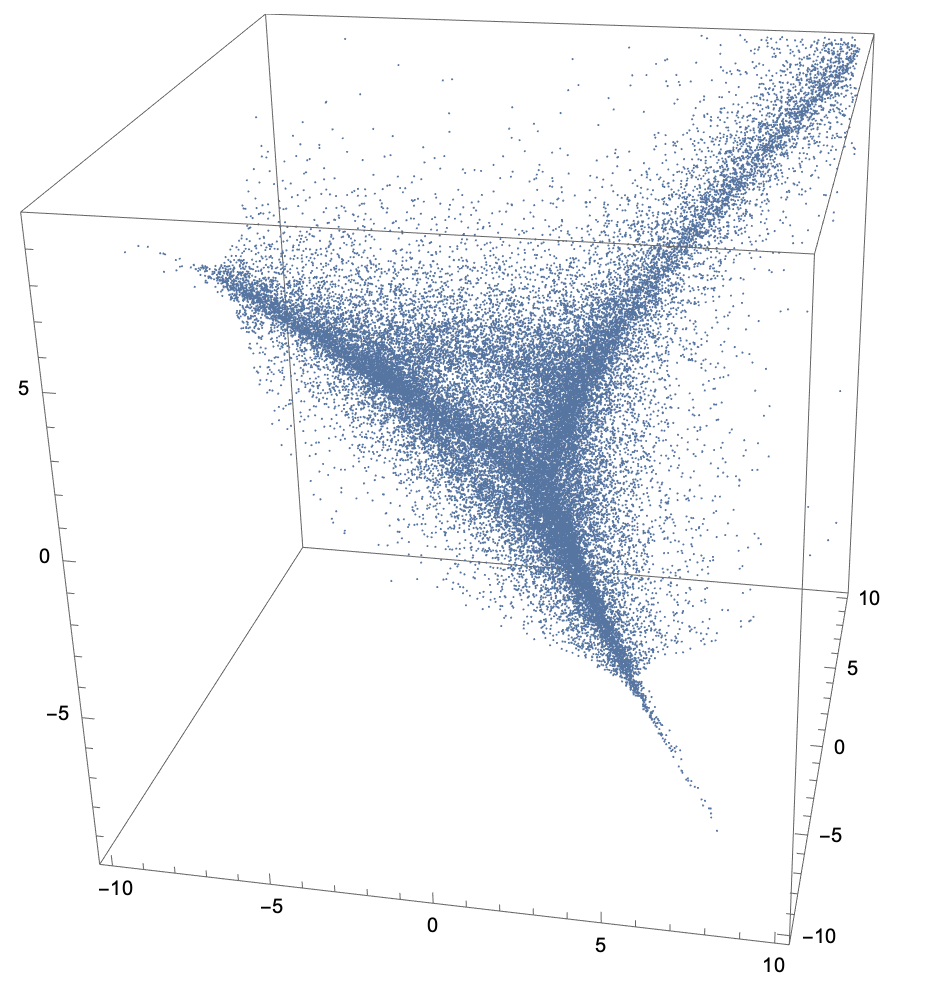}
     \caption{An example of the corresponding \(\mathbb{P}^3\) amoeba from Monte Carlo sampling.}
     \label{p3amb}
   \end{minipage}
\end{figure}

The Newton polynomial corresponding to \(\mathbb{P}^3\) is \(P(z_1,z_2,z_3)=c_1z_1+c_2z_2+c_3z_3+c_4z_1^{-1}z_2^{-1}z_3^{-1}+c_5\), whose toric diagram is given in Figure \ref{p3}, with an example Monte Carlo sampled amoebae in Figure \ref{p3amb}.

\subsubsection{Learning persistent homology \texorpdfstring{\(b_2\)}{b2}}
Using persistent homology to obtain the values of \(b_2\) for a set of 3000 coefficient lists. The values of \(b_2\) is determined as follows
\begin{equation}
    b_2 = \begin{cases} 0 \hspace{10pt} & \mathrm{No}\hspace{4pt} \mathrm{persistent}\hspace{4pt} \mathrm{pairs}\hspace{4pt} \mathrm{with}\hspace{4pt} q-p >0.24\,;\\
    1 \hspace{10pt} & \mathrm{Otherwise}\,.
    \end{cases}
\end{equation}
Then performing ML on this dataset achieves performance measures 
\begin{align}
    &\text{MLP:} \quad \text{ACC: } 0.840 \pm 0.015\,, \quad \text{MCC: } 0.699 \pm 0.022\,, \\
    &\text{CNN:} \quad \text{ACC: } 0.727 \pm 0.031 \,, \quad \text{MCC: } 0.457 \pm 0.073\,.
\end{align}

\subsubsection{Learning analytic lopsidedness \texorpdfstring{\(b_2\)}{b2}}
The analytic condition for \(b_2\) using lopsidedness is 
\begin{equation}
     b_2 = \begin{cases} 0 \hspace{10pt} & |c_5|\leq |c_1c_4|^{1/4}+|c_2c_4|^{1/4}+|c_3c_4|^{1/4}+|c_1c_2c_3|^{3/4}|c_4|^{1/4}\,;\\
    1 \hspace{10pt} & \mathrm{Otherwise}\,.
    \end{cases}
\end{equation}

For a balanced dataset of 7000 random samples with \(c_{1,2,3,4}\in [-5,5]\) and \(c_5\in [-10,10]\) using this analytic condition, the ML performance measures achieved over the 5-fold cross-validation were
\begin{align}
    &\text{MLP:} \quad \text{ACC: } 0.939 \pm 0.009\,, \quad \text{MCC: } 0.876 \pm 0.017\,, \\
    &\text{CNN:} \quad \text{ACC: } 0.910 \pm 0.010 \,, \quad \text{MCC: } 0.819 \pm 0.019\,.
\end{align}


\subsection{\texorpdfstring{\(\mathbb{P}^2\times\mathbb{P}^1\)}{P2P1}}
\begin{figure}[!ht]
   \begin{minipage}{0.48\textwidth}
     \centering
     \includegraphics[width=.6\linewidth]{Fig/p2p1.png}
     \caption{Toric diagram for \(\mathbb{P}^2\times\mathbb{P}^1\)\,.}
     \label{p2p1}
   \end{minipage}\hfill
   \begin{minipage}{0.48\textwidth}
     \centering
     \includegraphics[width=.6\linewidth]{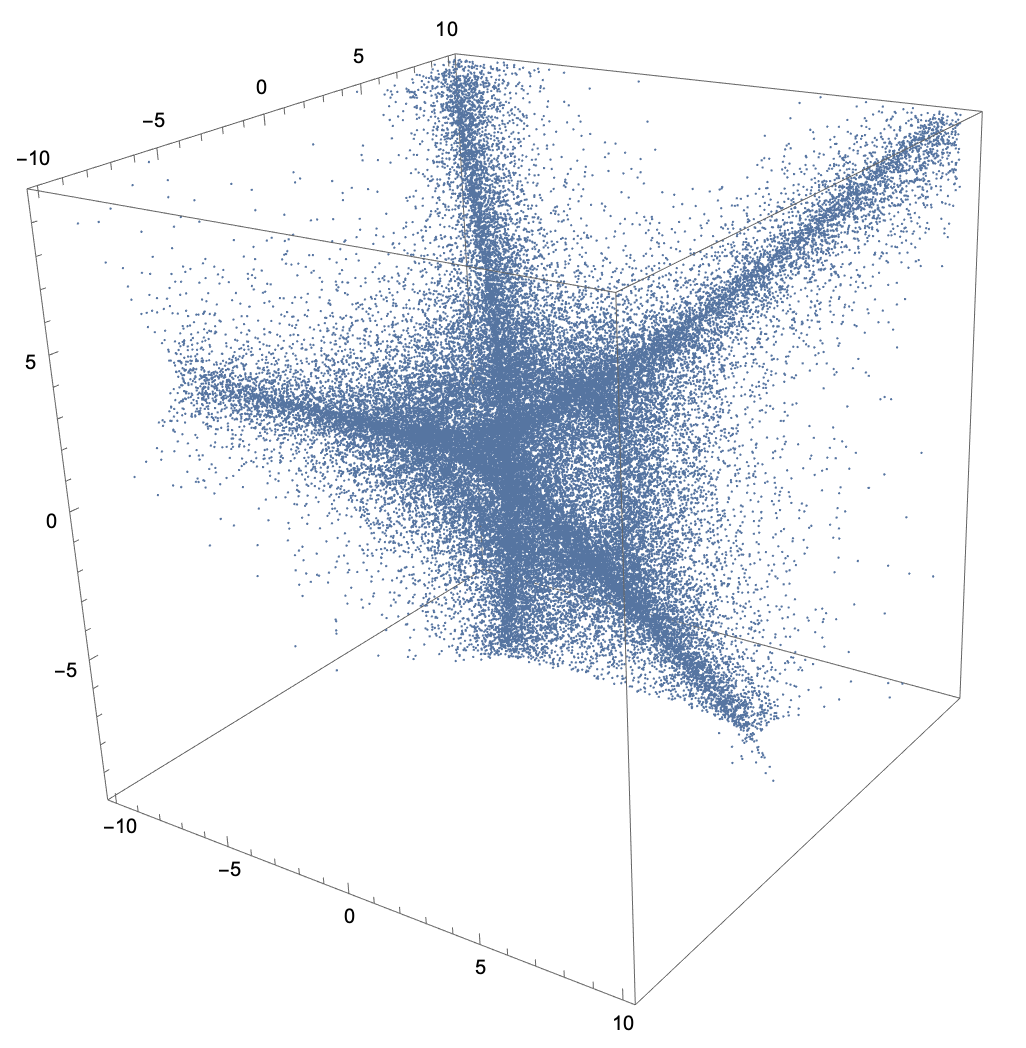}
     \caption{An example of the corresponding \(\mathbb{P}^2\times\mathbb{P}^1\) amoeba from Monte Carlo sampling.}
     \label{p2p1amb}
   \end{minipage}
\end{figure}

The Newton polynomial associated with \(\mathbb{P}^2\times\mathbb{P}^1\) is \(P(z_1,z_2,z_3)=c_1z_1+c_2z_2+c_3z_3+c_4z_1^{-1}+c_5z_2^{-1}z_3^{-1}+c_6\), and the toric diagram and example Monte Carlo amoeba are given in Figures \ref{p2p1} and \ref{p2p1amb}. This is also a reflexive polytope with only one interior point. Thus, \(b_2=0\) or \(1\).

\subsubsection{Learning persistent homology \texorpdfstring{\(b_2\)}{b2}}
A balanced data set of 4000 random samples is used with with \(c_{1,2,3,4,5}\in [-5,5]\) and \(c_6\in [-15,15]\). The values of \(b_2\) were determined as follows using persistent homology
\begin{equation}
    b_2 = \begin{cases} 0 \hspace{10pt} & \mathrm{No}\hspace{4pt} \mathrm{persistent}\hspace{4pt} \mathrm{pairs}\hspace{4pt} \mathrm{with}\hspace{4pt} q-p >0.28\,;\\
    1 \hspace{10pt} & \mathrm{Otherwise}\,,
    \end{cases}
\end{equation}
leading to ML results
\begin{align}
    &\text{MLP:} \quad \text{ACC: } 0.830 \pm 0.016\,, \quad \text{MCC: } 0.652 \pm 0.035\,, \\
    &\text{CNN:} \quad \text{ACC: } 0.825 \pm 0.027\,, \quad \text{MCC: } 0.630 \pm 0.063\,.
\end{align}

\subsubsection{Learning analytic lopsidedness \texorpdfstring{\(b_2\)}{b2}}
Following the same derivation methods, the $b_2$ values determined using lopsidedness used
\begin{equation}
     b_2 = \begin{cases} 0 \hspace{10pt} & |c_6|\leq 2|c_1c_4|^{1/2}+3|c_2c_3c_5|^{1/3}\,;\\
    1 \hspace{10pt} & \mathrm{Otherwise}\,,
    \end{cases}
\end{equation}
equivalently leading to ML results
\begin{align}
    &\text{MLP:} \quad \text{ACC: } 0.947 \pm 0.007\,, \quad \text{MCC: } 0.893 \pm 0.014\,, \\
    &\text{CNN:} \quad \text{ACC: } 0.920 \pm 0.011 \,, \quad \text{MCC: } 0.841 \pm 0.023\,.
\end{align}

\section{Additional example of ML 2d amoebae and Mahler measure}\label{appendix2}

\subsection{\texorpdfstring{\(\mathbb{P}^2\)}{P2}}

\begin{figure}[!ht]
  \centering
  \includegraphics[width=.4\linewidth]{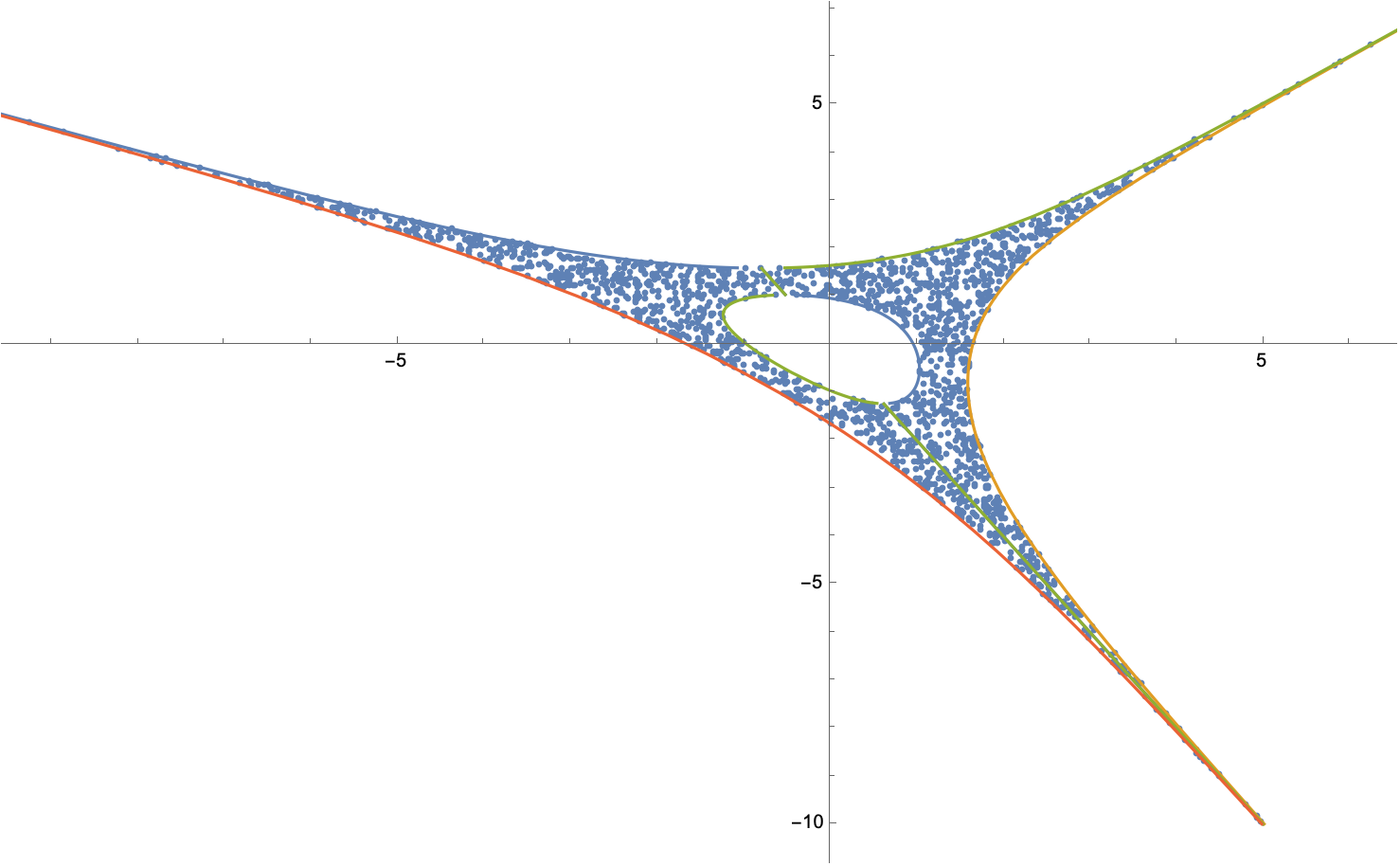}
  \caption{The analytic boundary of amoeba with \(k=4\).}
  \label{p2boundary}
\end{figure}

The Newton Polynomial in this case is \(P(z,w)=k-z-w-z^{-1}w^{-1}\). The analytic boundary of the amoeba is 
\begin{equation}
    x= \ln \left(\left| \frac{k}{2} \pm\frac{e^y}{2}\pm\sqrt{\frac{1}{4} \left(k\pm e^y\right)^2\pm e^{-y}}\right| \right),
\end{equation}
for \(k\geq 3\) (Figure \ref{p2boundary}).

The Mahler measure for \(P(z,w)=k-z-w-z^{-1}w^{-1}\) as a function of \(k\) obtained using Taylor expansion and Cauchy residue theorem is
\begin{equation}
    m(P)=\ln k -2k^{-3} {}_{4}F_{3}\left(1,1,\frac{4}{3},\frac{5}{3};2,2,2;27k^{-3}\right),
\end{equation}

\begin{figure}[!ht]
  \centering
  \includegraphics[width=.4\linewidth]{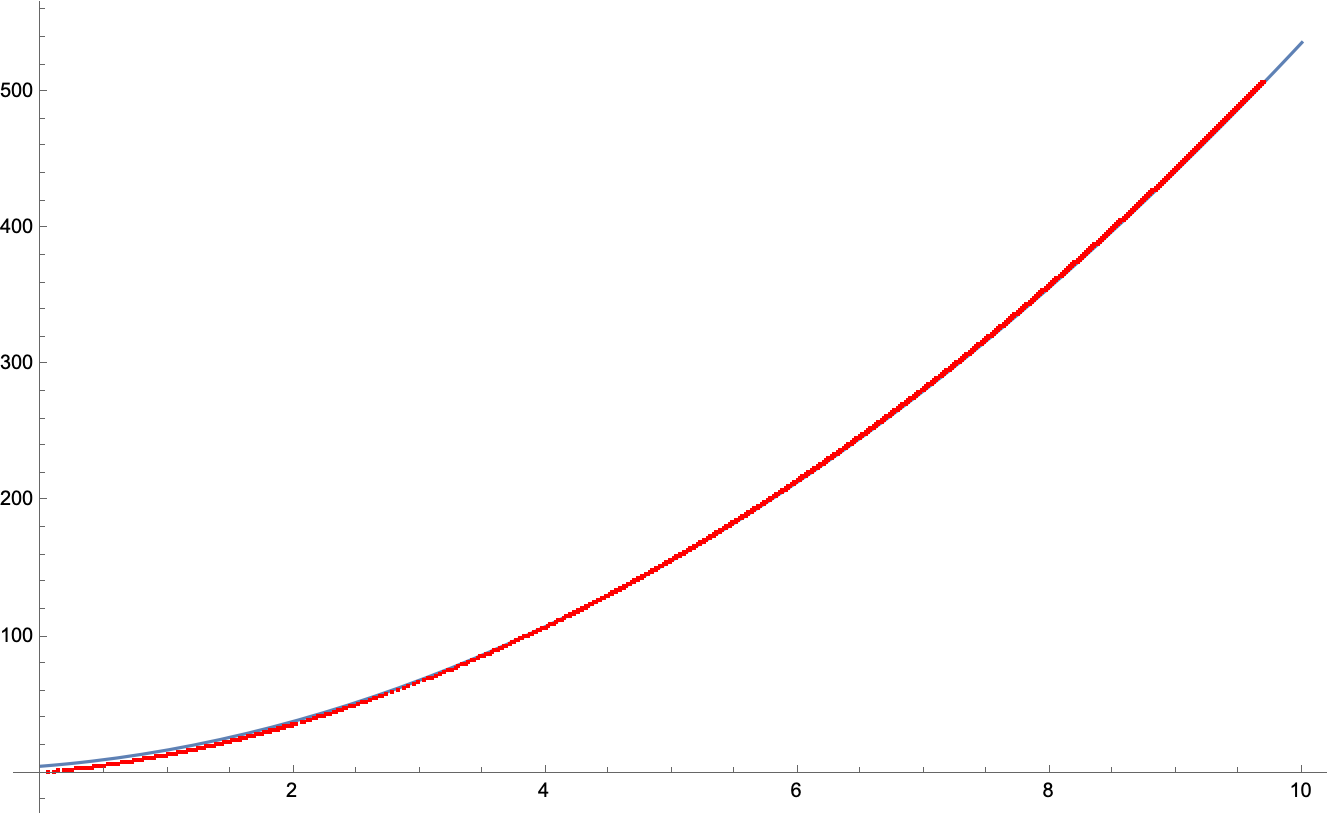}
  \caption{Data points (red) and the fitted relation (blue) between \(A_h\) and \(m_g\).}
  \label{p2 a vs mg}
\end{figure}

The relation between the gas phase contribution \(m_g(P)\) and the amoeba hole area is fitted with 20000 data pairs and found to be
\begin{equation}
    A_h = 4.59038 m_g^2+7.24861m_g+5.09803\,,
\end{equation}
with an \(R^2\) score of 1.0000 and mean absolute error of 0.1340 (plotted in Figure \ref{p2 a vs mg}).

\begin{figure}[!ht]
  \centering
  \includegraphics[width=0.4\linewidth]{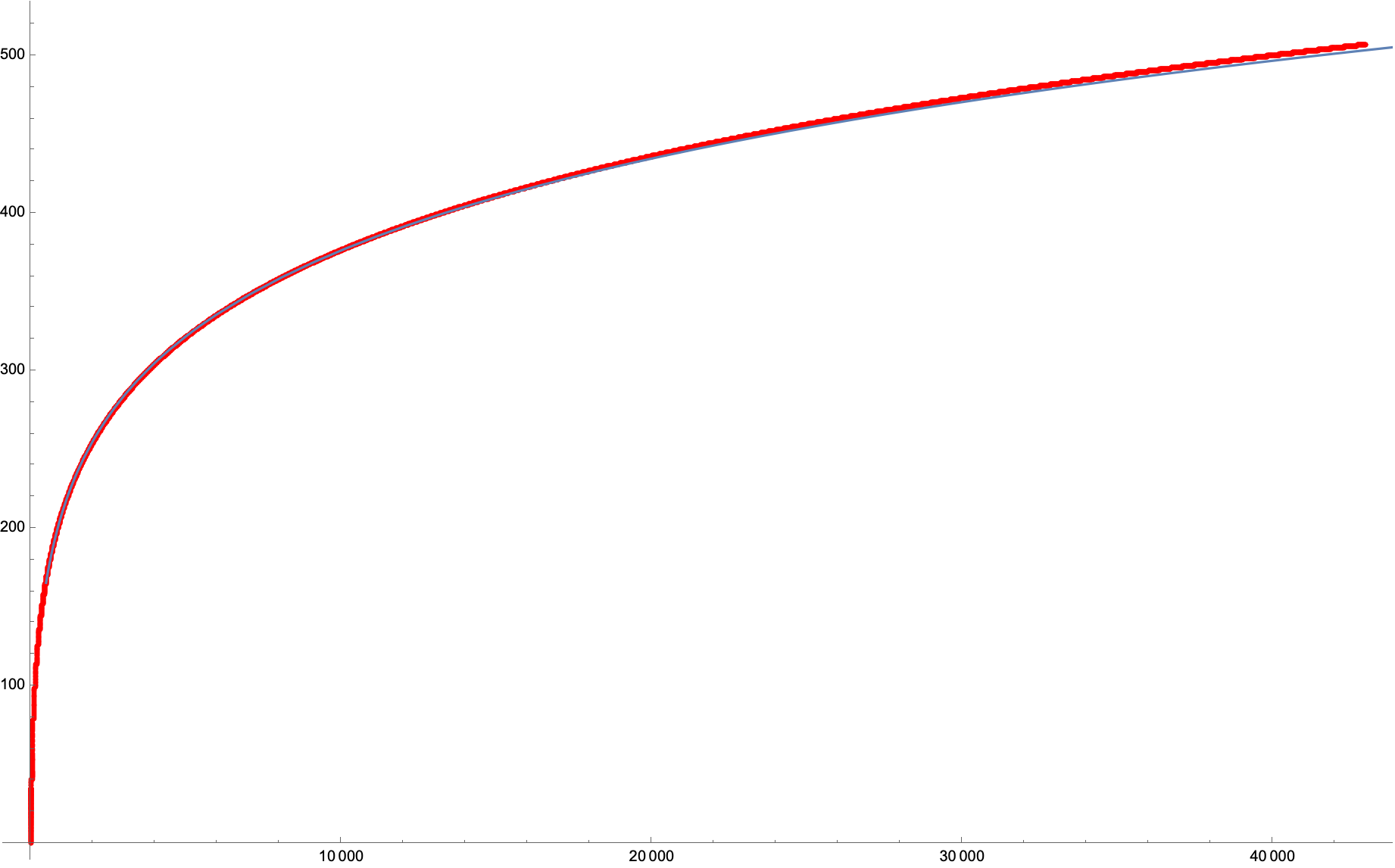}
  \caption{Data points (red) and the fitted relation (blue) between \(A_h\) and \(k\).}
  \label{p2 a vs k}
\end{figure}

The relation between the amoeba hole area and the value of k for \(k\geq 3\) is fitted with 5000 data pairs, and is found to be
\begin{equation*}
   A_h=2 \ln^2 k+\ln (k \ln k)-5.49+\ln k^2 \ln ((k-\ln k^2 ) (\ln (k \ln k)-5.49))\,,
\end{equation*}
with an \(R^2\) score of 0.9998 and mean absolute error of 1.0983 (plotted in Figure \ref{p2 a vs k}). 

\section{Explicit example of the expansion method}\label{appendix3}

In this section we outline the method used to calculate expressions for the Mahler measure, and present an explicit example. This method can be used to calculate the Mahler measure of any polynomial. In this example, we will derive Eq.(\ref{eqc3z5}), which corresponds to the $\mathbb{C}^3/\mathbb{Z}_5$ polygon with $s$ as the origin:
\begin{equation}
    m(P_s(z,w)) = m\left({k_1-k_2z-\frac{1}{z}-zw-\frac{z^2}{w}}\right).
\end{equation}

\noindent In order to expand, we write $P_s(z,w)=k_1-p_s(z,w)$, where $p_s(z,w) = k_2z+z^{-1}+zw+z^2w^{-1}$. From Eq.(\ref{mexpansionequation}), we expand as:

\begin{equation}\label{expandedm}
 m(P_s(z,w)) = \log k_1 - \sum_{n=1}^\infty\frac{[p_s^n(z,w)]_0}{nk_1^n},
\end{equation}

\noindent where $[p_s^n(z,w)]_0$ is the constant term of the $n^{\mathrm{th}}$ power of $p_s(z,w)$. To calculate these constant terms, we use a binomial expansion as follows:
\begin{align}
    p_s^n(z,w) &= \left(k_2z+z^{-1}+zw+z^2w^{-1}\right)^n\\
    &= \sum_{i=0}^n\binom{n}{i}(k_2z+z^{-1})^i(zw+z^2w^{-1})^{n-i}\\\label{fullexpansion}
    &= \sum_{i=0}^n\binom{n}{i}\left[\sum_{l=0}^i\binom{i}{l}k_2^lz^lz^{l-i}\right]\left[\sum_{j=0}^{n-i}\binom{n-i}{j}z^{2(n-i-j)}w^{-(n-i-j)}z^jw^j\right].
\end{align}

\noindent We are looking for constant terms only, so the sum of the powers of both $z$ and $w$ should be equal to zero. Grouping $w$ and $z$ terms individually, we get:

\begin{equation}
    2j+i-n=0 \Rightarrow j=\frac{n-i}{2},
\end{equation}

\begin{equation}
    2l+2n-3i-j=0 \Rightarrow l= \frac{5i-3n}{4}.
\end{equation}

\noindent Subbing this into Eq.(\ref{fullexpansion}), we arrive at

\begin{equation}
    [p_s^n(z,w)]_0=\sum_{i=0}^n\binom{n}{i}\binom{n-i}{\frac{n-i}{2}}\binom{i}{\frac{5i-3n}{4}}k_2^{\frac{5i-3n}{4}}.
\end{equation}

\noindent Finally, inserting this into Eq.(\ref{expandedm}), we arrive at our final result:

\begin{equation}
    m\left(P_s(z,w)\right) = \log k_1 - \sum_{n=1}^{\infty}\sum_{i=0}^{n}\binom{n}{i}\binom{n-i}{\frac{n-i}{2}}\binom{i}{\frac{5i-3n}{4}}\frac{k_2^{(\frac{5i-3n}{4})}}{k_1^nn},
\end{equation}

\noindent which is valid for all $k_1\geq \underset{|z|,|w|=1}{\mathrm{max}}|p_s(z,w)|$. This expression comes with some constraints, which ensures all entries in the binomials are positive integers, and for $\binom{n}{r}$, we always have $n\geq r$. We require that $i\geq 3n/5$ and that $(3i-5n)\!\!\!\mod4=0$. This can greatly decrease the number of terms in the series.\par

This method can be used to calculate the Mahler measure of any polynomial. In cases where the number of terms in the polynomial becomes large, we may have to sum over a large number of indices. In general, the number of indices we sum over is equal to (excluding $n$): Number of indices summed over = ((Number of non-constant terms in the polynomial)$-$1)$-$(Number of variables). Because of this, for polynomials with a large number of variables, this expansion method is often much more efficient than numerical integration method, where we would have to integrate over each variable.

\printbibliography

\end{document}